\documentclass[aip,pof,preprint]{revtex4-1}

\usepackage{amsfonts,amsmath,amssymb,amsthm}

\usepackage[usenames,dvipsnames,svgnames,table]{xcolor}
\usepackage{color}

\usepackage{accents}
\usepackage{calc}
\usepackage{enumerate}
\usepackage{enumitem}
\usepackage[T1]{fontenc}
\usepackage{helvet}
\usepackage{graphicx}
\usepackage{ifdraft}
\usepackage{ifthen}
\usepackage{listings}
\usepackage{mathtools}
\usepackage{multirow}
\usepackage{natbib}
\usepackage{rotating}
\usepackage{setspace}
\usepackage{tabularx}
\usepackage{textcomp}
\usepackage{tikz}
\usepackage{url}
\usepackage{varioref}
\usepackage{yhmath}
\usepackage{xkeyval}
\usepackage{xspace}

\usepackage[textsize=scriptsize]{todonotes}

\usepackage[colorlinks=true,
            linkcolor=blue,
            urlcolor=blue,
            citecolor=blue,
            final,
            hypertexnames=false]{hyperref}
\usepackage{cleveref}

\usepackage{subfig}

\usepackage{algorithm}
\usepackage{algorithmic}

\graphicspath{{figs/}}

\DeclareMathOperator{\variance}{Var}

\newcommand{\avg}[1]{\ensuremath{\langle #1 \rangle}}
\newcommand{\rarrow}{\ensuremath{\rightarrow}}
\newcommand{\mcal}[1]{\ensuremath{\mathcal{#1}}}
\newcommand{\bq}{\ensuremath{\bar{q}}}
\newcommand{\dd}[2]{\ensuremath{\frac{d #1}{d #2}}}

\begin{document}

% title and author
\title{Estimating Uncertainties in Statistics Computed from DNS}

\author{Todd A.~Oliver}
\email{oliver@ices.utexas.edu}
\affiliation{Center for Predictive Engineering and Computational Sciences,  \\ Institute for Computational Engineering and Sciences, \\ The University of Texas at Austin, \\ Austin, TX 78712, USA}

\author{Nicholas Malaya}
\email{nick@ices.utexas.edu}
\affiliation{Center for Predictive Engineering and Computational Sciences,  \\ Institute for Computational Engineering and Sciences, \\ The University of Texas at Austin, \\ Austin, TX 78712, USA}

\author{Rhys Ulerich}
\email{rhys@ices.utexas.edu}
\affiliation{Center for Predictive Engineering and Computational Sciences,  \\ Institute for Computational Engineering and Sciences, \\ The University of Texas at Austin, \\ Austin, TX 78712, USA}

\author{Robert D.~Moser}
\email{rmoser@ices.utexas.edu}
\affiliation{Center for Predictive Engineering and Computational Sciences, \\ Institute for Computational Engineering and Sciences, \\ The University of Texas at Austin, \\ Austin, TX 78712, USA}
\affiliation{Department of Mechanical Engineering, \\ The University of Texas at Austin, \\ Austin, TX 78712, USA}

% the abstract
\begin{abstract}
Rigorous assessment of uncertainty is crucial to the utility of DNS
results.  Uncertainties in the computed statistics arise from two
sources: finite statistical sampling and the discretization of the Navier--Stokes
equations.  Due to the presence of non-trivial sampling error,
standard techniques for estimating discretization error (such as
Richardson extrapolation) fail or are unreliable.  This work provides
a systematic and unified approach for estimating these errors.  First,
a sampling error estimator that accounts for correlation in the input
data is developed.  Then, this sampling error estimate is used as part
of a Bayesian extension of Richardson extrapolation in order to
characterize the discretization error.  These methods are tested using
the Lorenz equations and are shown to perform well. These techniques
are then used to investigate the sampling and discretization errors in
the DNS of a wall-bounded turbulent flow at $Re_{\tau} \approx 180$.
Both small ($L_x/\delta \times L_z/\delta = 4 \pi \times 2 \pi$) and large
($L_x/\delta \times L_z/\delta = 12 \pi \times 4 \pi$) domain sizes are
investigated.  For each case, a sequence of meshes was generated by
first designing a ``nominal'' mesh using standard heuristics for
wall-bounded simulations.  These nominal meshes were then coarsened to
generate a sequence of grid resolutions appropriate for the
Bayesian Richardson extrapolation method.  In addition, the small
box case is computationally inexpensive enough to allow simulation on a
finer mesh, enabling the results of the extrapolation to be validated
in a weak sense.  For both cases, it is found that while the sampling
uncertainty is large enough to make the order of accuracy
difficult to determine, the estimated discretization errors are quite
small. This indicates that the commonly used heuristics provide adequate
resolution for this class of problems.  However, it is also found that,
for some quantities, the discretization error is not small relative to
sampling error, indicating that the conventional wisdom that sampling
error dominates discretization error for this class of simulations
needs to be reevaluated.
\end{abstract}

\maketitle

\section{Introduction}
Direct numerical simulation (DNS) of turbulence is a valuable tool for
the study of turbulent flows. Statistical quantities
computed from DNS results are commonly used both to further
understanding of flow physics and test hypotheses regarding
turbulence~\citep{Moin1998Tool, Jimenez2007Learning, Alfonsi2011} as
well as to calibrate and validate engineering turbulence
models~\citep{Pope2000Turbulent, Wilcox2006Turbulence,
Durbin2011Statistical, RESS-2010, ETC13-2011}.  DNS data are thus
commonly used like experimental data.  Therefore, as with experimental
data, to have confidence in the interpretation of a DNS or in the meaning
of any comparison with DNS data, one must understand the
uncertainty in that data.  However, it is not common in the DNS
literature to report these uncertainties because uncertainties in the
data are generally not systematically evaluated.  Instead, it is
common for expert practitioners to determine grid spacing
requirements, required simulation time, etc. based on a combination of
knowledge gained from previous experience and observations of simulation outputs.
%sanity checks\todo{Too informal? ``observing anticipated, non-trivial behavior of''} on simulation results.  
The goal of this work is to improve upon this practice by providing a systematic
method for estimating uncertainty in the statistics computed from DNS
data.

Uncertainty estimation for DNS is an example of solution verification
for the DNS statistical quantities.  The goal of solution verification
is to ensure that numerical solutions of a mathematical model are
sufficiently accurate approximations to the exact solution of the
model~\cite{AIAA_Guide_1998, ASME_VV10_2006}.  Solution verification
techniques for computational fluid dynamics (CFD) have been the topic
of a large body of research~\cite{Roache1998Verification, Roy2005,
Oberkampf2010}.  The simplest techniques in this domain are based on
Richardson extrapolation for estimating discretization error given a
sequence of simulations on successively finer meshes.  However, these
developments have had little impact on the practice of DNS due to the
fact that the outputs are generally statistical quantities that
are contaminated not only by discretization error but also by sampling
error.  Since the
goal of DNS is to resolve all relevant physical scales, it is generally
expected that errors due to finite sampling are significant relative to
discretization errors.  Thus, simple methods for estimating
discretization error that are common for other CFD calculations, like
Richardson extrapolation, are not directly applicable to DNS results,
because the estimated discretization error is greatly affected by
sampling error.  The result is that, while systematic mesh resolution
studies have been performed\citep{Donzis2008}, it is not common to
actually estimate discretization error.

To address this issue, it is of primary importance to estimate
sampling errors.  Of course, if the data used to compute the
statistics were samples from independent, identically distributed
random variables, the central limit theorem allows easy estimation of
the sampling error.
%application of the central limit theorem implies that, as the number of
%samples $N$ goes to infinity, the error of the sample average goes to a
%zero-mean Gaussian random variable with variance $\sigma^2/N$, where
%$\sigma^2$ is the variance of the random variables.  
However, the samples used to generate DNS statistics are drawn from a
time history and/or spatial field and are generally not independent.
To reduce the correlation, the samples used to compute statistics are
sometimes taken ``far'' apart in time and then treated as
independent~\citep{Donzis2008}.  While this procedure has intuitive
appeal, it can lead to underestimated uncertainty if the snapshots are
not sufficiently separated.  Alternatively, if the snapshots are taken
too far apart, it leads to fewer samples and overestimates of
sampling error.

Instead of restricting the samples in this way, it is preferable to use
all the available data and account for correlations. One approach to
accounting for the correlations in DNS statistics, which was proposed by Hoyas and
Jim\'{e}nez\citep{Hoyas2008Reynolds}, uses a sequence of ``coarse
grainings'' of the data.  However, our experience has been that it is
difficult to automate this procedure because the presence of noise often
requires both user intervention and interpretation. 
%\todo{RDM greatly shortened this. Nick, is it OK}

%To the best of the
%authors' knowledge, there is only one existing method for estimating
%sampling error in the presence of correlation that has been applied
%to DNS results.  This method, due to Hoyas and
%Jim\'{e}nez~\citep{Hoyas2008Reynolds}, estimates the standard deviation
%of the sample average by ``coarse-graining'' the data.  Specifically, the
%method estimates the variance of the sample average by finding a sequence
%of sampling error estimates by taking every $k$\textsuperscript{th}
%sample for a range of coarsening parameters $k$.  When it exists, the
%convergent plateau of these estimates is reported as the sampling
%error.  However, our experience has shown that due to the 
%presence of noise, it is difficult to automate the detection of this 
%limiting behavior, with the resulting uncertainty estimate requiring
%both user intervention and interpretation.  \todo{Nick, is this a fair
%description of your experience?  Either way, please alter as you see
%fit. nick: only slightly updated, I think this is solid.}

A more promising approach based on direct estimation of the
correlations in the data has been used to estimate sampling errors in many
fields, including the weather and climate
communities~\citep{Trenberth1984Some, Zwiers1995Taking}.
In this approach, the autocorrelation of the data, which is not known
\emph{a priori}, must be estimated from the data, which presents its own
challenges.  Here, we follow the work
of \citet{Broersen2002Automatic,Broersen2006Automatic} and fit autoregressive models
from which the autocorrelation function is then computed.

Given an estimate of the sampling error, discretization errors are
estimated using data from simulations with different resolution
levels.  As noted earlier, because sampling uncertainty is generally
expected to be of the same magnitude as the discretization error, at
least for grid spacing and time steps used for production DNS,
standard Richardson extrapolation generally fails to correctly
estimate the discretization error.  Here, a Bayesian extension of the
standard Richardson extrapolation that accounts for both statistical
uncertainty and prior information (e.g., the expected asymptotic order
of accuracy) is formulated.  This Bayesian statistical formulation
effectively regularizes the Richardson extrapolation
problem to decrease the sensitivity of the estimated discretization
error to finite sampling effects.

The performance of these estimators is tested using the Lorenz
equations.  They are then applied to the problem of assessing
uncertainties in statistics from the DNS of incompressible, turbulent
channel flow at $Re_{\tau} \approx 180$.  The resulting discretization
error estimates are assessed using a small domain case where it
is feasible to run a simulation with twice the resolution of the
nominal simulation, which is designed according to typical DNS heuristics.

For many quantities, including the mean velocity, Reynolds shear stress,
and skin friction coefficient, the discretization error model is
validated, meaning that its predictions agree with the observations at
higher resolution.  For these quantities, the model is then used to
predict the discretization error present in a large domain simulation
with resolution again set by the usual heuristics.  The results
demonstrate that, for these quantities, the discretization errors are
small, generally much less than one percent.  Thus, the usual mesh
heuristics appear to be adequate.  It should be pointed out however
that the estimated discretization error is often similar to or larger
than the estimated sampling error.  
%For example, for the skin friction
%coefficient, the sampling error is estimated to be smaller than the
%discretization error.  
This result violates the conventional wisdom that
sampling error dominates, indicating that it is important to
systematically estimate discretization error effects as well.

Unfortunately, for other quantities, including the streamwise velocity
variance and the vorticity variances, our simple discretization error
model is invalidated by the high resolution small domain simulation
results.  While the observed changes between the nominal and high
resolution simulation are small, and so there is no indication that the
nominal resolution is inadequate, this invalidation precludes the use
of the model to predict the discretization error with any confidence.
Thus, no discretization error estimates are presented for these
quantities.

The remainder of the paper is organized as follows.  The full error
estimation methodology is presented in \S\ref{sec:method}, including
the sampling error estimation (\S\ref{sec:sampling_error}), the
Bayesian Richardson extrapolation procedure
(\S\ref{sec:discretization_error}), and the illustrative Lorenz
example (\S\ref{sec:lorenz}).  Results for DNS of $Re_{\tau} = 180$
channel flow are given in \S\ref{sec:channel},
and \S\ref{sec:conclusions} provides conclusions.

\section{Methodology} \label{sec:method}
This work addresses two major sources of uncertainty in statistics
computed from DNS: finite sampling error and discretization error.
The sampling error estimator is described briefly in
\S\ref{sec:sampling_error}.  This estimate is then used in a Bayesian
extension of Richardson extrapolation to determine probabilistic
estimates of the discretization error and the exact value of the
statistic of interest, as described in
\S\ref{sec:discretization_error}.  To assess the characteristics of these
procedures in a simple setting where different regimes can easily be
explored, both estimators are used to evaluate simulations of the
Lorenz equations in \S\ref{sec:lorenz}.

\subsection{Sampling Error} \label{sec:sampling_error}

This section outlines a method for estimating the variance of a sample
average computed from correlated data.  To fix notation, let $X$ denote a
scalar flow quantity (e.g., a velocity component).  Assume that the
DNS produces a sample from a statistically stationary sequence of
random variables $\{X_i\}$ for $i= 0, 1, \ldots$.  Of course, the
simulation can only run for finite time, so only the
first $N$ components of this sequence are known.  The average of the $N$
available samples,
\begin{equation*}
\avg{X}_N = \frac{1}{N} \sum_{i=0}^{N-1} X_i,
\end{equation*}
is then an approximation of the true mean $\mu = E[X] = E[X_0]$, where
here $E[\cdot]$ is the expected value.  Then, the sampling error $e_N$
is simply the difference between the sample average and the true mean:
\begin{equation*}
e_N \equiv \avg{X}_N - E[X].
\end{equation*}
Extensions of the central limit theorem (CLT) valid for sequences in
which independence is approached for large separations, as is
expected for turbulence time series, imply that for large $N$, $e_N$
converges to a normal distribution with zero mean (see
Appendix~\ref{app:background_sample_error}). The variance of $e_N$ is
thus all that is required to completely characterize the sampling
error. The estimator for the variance used here is motivated by this
same generalization of the CLT, as described in
Appendix~\ref{app:background_sample_error}.

%
%\subsection{A Computable Estimate} \label{sec:compute_sample_error}
%Of course, neither the typical version of the CLT nor the extension
%for weak dependence is useful in the form given
%in \S\ref{sec:background_sample_error} because the inputs, such as the
%true variance $\sigma^2$, cannot be computed from the available data.
%Instead, we must estimate the desired quantities from the samples.
%For instance, for independent samples, it is standard to estimate the
%true variance using
%%
%\begin{equation*}
%\sigma^2 \approx \sigma_N^2 \equiv \frac{1}{N-1} \sum_{i=0}^{N-1} (X_i - \avg{X}_N)^2.
%\end{equation*}
%%
%For correlated samples, more in-depth considerations are required
%because we must also estimate the autocorrelation function.
Following \citet{Trenberth1984Some}, the sampling error is estimated as
\begin{equation}
%  \operatorname{Var} \bar{X}
%  &\approx
%  \frac{\frac{N}{N-T_0} s^2}{\frac{N}{T_0}}
%  =
%  \frac{T_0}{N-T_0} \sum_{i=1}^{N} \left(X_i - \bar{X}\right)^2
\variance e_N \approx \frac{\hat{\sigma}_N^2 T_0}{N},\label{eqn:Neff}
\end{equation}
where
%
%\begin{gather}
\begin{equation}
\hat{\sigma}_N^2 = \frac{1}{N - T_0} \sum_{i=0}^{N-1} ( X_i - \avg{X}_N
 )^2, \label{eqn:sigN}
\end{equation}
%\hat{N}_\text{eff} = \frac{N}{T_0}. \label{eqn:Nhat}
%\end{gather}
%
and $T_0$ is the
decorrelation separation distance. 
%$\hat{N}_\text{eff}$ is the effective sample size and $T_0$ is the
%decorrelation separation distance.  
Specifically,
%In~\eqref{eqn:sigN} and~\eqref{eqn:Nhat}, $T_0$ is the estimated
%decorrelation separation:
%
\begin{equation}
T_0 = 1+2\sum_{k=1}^{N-1} \left(1-\frac{k}{N}\right) \hat{\rho}(k),
\label{eq:meanT0}
\end{equation}
where $\hat{\rho}$ is an estimate of the unknown true autocorrelation
function $\rho$.  The possibly unexpected $1-k/N$ factor is a common
artifact~\citep{Thiebaux1984Interpretation,Zwiers1995Taking} of
choosing a biased estimator, which is used here because it possesses
more desirable properties than the ``unbiased'' version in this
context~\citep{Trenberth1984Some,Percival1993Three}. The expression
(\ref{eqn:Neff}) for $\variance e_n$ is the same as the estimate that
would be obtained if $N_{\text{eff}} = N/T_0$ independent samples were
used, making $N_\text{eff}$ a measure of the effective size of a
sample.

%The autocorrelation $\rho$ is a function of the sample separation or ``lag'' $k$.
%$T_0$ is called the ``decorrelation time'' but is more accurately a
%decorrelation separation---multiplying $T_0$ by the mean sampling period
%produces a genuine timescale.  The ratio $N/T_0$ is often called the
%``effective sample size'' as it gauges the number of effectively independent
%samples within the data~\citep{Thiebaux1984Interpretation,Lee2008Equivalent}.

The fundamental challenge in estimating the variance of the sample average is the
approximation of the autocorrelation $\rho$.  While $\rho$ can be
approximated directly from the definition, such a naive approximation
tends to be noisy, which can lead to bad estimates of
$T_0$~\citep{Percival1993Three}. Obtaining a useful estimate of $\rho$
requires more sophisticated techniques, especially for modest sample
sizes.
%Specifically, direct calculations of
%$\rho$ tend to be noisy for separations $k$ that are a significant
%fraction of $N$ since the sample set contains few samples separated by
%such large $k$.  This noise contaminates the calculation of $T_0$,
%which requires information about $\rho$ for large separation $k$.
%
%Thus, obtaining a useful estimate of $\rho$ given only small-to-moderate data
%requires more sophisticated techniques.  
%%
%Given the duality between the power
%spectral density and the autocorrelation function, segmented averaging or other
%classical ``periodogram'' estimation techniques may be used
%\citep{Bartlett1948Smoothing,Welch1967Use}.  However, their application to the
%constant Fourier mode used by the proofs in the appendix of
%\citet{Trenberth1984Some} and often used to motivate effective sample
%sizes~\citep{Thiebaux1984Interpretation} is considered
%problematic~\citep[\textsection{}12.3.7]{Storch2001Statistical} and requires
%special care~\citep{Madden1976Estimates}.  Least squares approaches related to
%the work of \citet{Lomb1976Leastsquares} and \citet{Scargle1982Studies}
%initially seem applicable but these often fail to accurately capture the slopes
%in spectral densities\todo{Citation?}. Therefore, their use is expected to
%incur errors within the summation appearing in equation~\eqref{eq:meanT0}.
%
Here, we follow \citet[\textsection{}17.1.3]{Storch2001Statistical}
and fit an autoregressive (AR) time series model \citep[see,
e.g.,][]{Box2008Time,Priestley1981SpectralI} to the observed sequence
$X_i$.  An AR process of order $p$ takes the following form:
\begin{equation}
X_n + a_1 X_{n - 1} + \dots + a_p X_{n - p} = \epsilon_n, \quad \epsilon_n \sim{} \mcal{N}\left(0, \sigma^2_\epsilon\right).
\label{eq:ar_model}
\end{equation}
where $\epsilon_n \sim \mcal{N}(m,s^2)$ indicates that $\epsilon_n$ is a
Gaussian random variable with mean $m$ and variance $s^2$.  The process
parameters $a_1, \ldots, a_p$ and noise variance $\sigma_{\epsilon}^2$
completely define the process, and thus, given these parameters, the
exact autocorrelation function of the AR process may be computed.
This autocorrelation function is then used as $\hat{\rho}$ to compute
$T_0$ according to~\eqref{eq:meanT0}.

Thus, estimating the autocorrelation reduces to estimating the
parameters of an AR model.  However, because the ``true''
process order is unknown, a hierarchy of models with increasing order
$p$ are simultaneously
estimated~\citep{Broersen2002Automatic,Broersen2006Automatic}.  From
these candidates, the best model is chosen using an
information-theoretic, finite sampling model selection
criterion~\citep{Broersen2000Finite}.

Fitting such models to observed data has been studied
extensively~\cite{Burg1975, Andersen1978Comments,Faber1986Commentary,
Andersen1974Calculation,Broersen2002Automatic,Broersen2006Automatic},
and there are a number of available algorithms.  Here, classical Burg
recursion \citep{Andersen1974Calculation} is used to compute the
parameters because it is less susceptible to round-off error
accumulation than the more efficient
recursive denominator variant~\citep{Andersen1978Comments,Faber1986Commentary}.  An open
source, header-only C++ reference implementation is
available at \url{http://rhysu.github.com/ar/}.
% available\citep{ARweb}.  % Reference with URL
Convenient wrappers for GNU
Octave~\citep{Eaton2008GNU} and Python~\citep{Drake2011Python} are
also provided.  While this implementation is sufficient for the
results shown in~\S\ref{sec:lorenz} and~\S\ref{sec:channel}, it can fail
in some circumstances. The
authors have observed stability-related problems when processing large
data sets
that have extremely low noise or 
very long decorrelation times relative to the sampling rate.
These issues are related to accumulation of round-off 
error~\citep{Andersen1974Calculation}.  

\subsection{Discretization Error} \label{sec:discretization_error}
In addition to the sampling error, discretization error contributes to
the error in statistics computed from DNS.  As part of a typical
calculation, statistics computed from multiple levels of mesh
resolution are available because course meshes are often used to speed
convergence to a statistically stationary state.  In principle, this
information can be used to estimate discretization error.  However,
the standard procedure, Richardson extrapolation, does not account for
sampling error, which can lead to misleading results.  This procedure
and issues introduced by sampling error are described
in \S\ref{sec:std_richardson}.  An extension of this method that
accounts for the sampling error through a Bayesian calibration
procedure is described in \S\ref{sec:bayes_richardson}.

\subsubsection{Assessing Order of Accuracy without Sampling Error} \label{sec:std_richardson}

Given simulations using at least three distinct resolutions, the
convergence rate of a discrete approximation to an unknown continuum
value may be assessed~\cite{Roache1998Verification, Roy2005}, assuming
that all three resolutions are in the asymptotic convergence range.  Let
$q$ denote the exact value of some output quantity and $q_h$ denote
the discrete approximation of $q$ at resolution level $h$.  Assuming
that
%an approximation $\hat{q}(h)$ shows an $h$-dependent truncation error
%compared to an exact value $q$, \textit{viz.}
\begin{align}
\label{eq:taylorbasedtruncationmodel}
q - q_h &= C_0h^{p} + C_1h^{p+1} + \cdots
,
\end{align}
gives rise to the classical Richardson extrapolation procedure.  The
input data are a sequence of outputs $q_{h_0}$, $q_{h_1}$, and
$q_{h_2}$ resulting from computations for successively finer discrete
approximations $h_0$, $h_1$, and $h_2$.
%Specifically, a sequence of successively finer discrete
%approximations $h_0$, $h_1$, and $h_2$ gives a sequence of results
%$q_{h_0}$, $q_{h_1}$, and $q_{h_2}$.  Then, 
Given this data and neglecting $O(h^{p+1})$
contributions, one can estimate the leading error order $p$ by
solving
\begin{align}
\label{eq:orderest}
\frac{q_{h_2} - q_{h_1}}{q_{h_1} - q_{h_0}} = r_1^p \frac{(r_2^p - 1)}{(r_1^p -1)}
%  q
%&=
%  \frac{t^{p} q_{h/t} - q_h}
%       {t^{p}-1}
%%  + O(h^{p_1}) % neglected
%=
%  \frac{s^{p} q_{h/s} - q_h}
%       {s^{p}-1}
%%  + O(h^{p_1}) %neglected
\end{align}
for $p$, where $r_1 = h_1/h_0$ and $r_2 = h_2/h_1$. 

Unfortunately, when the computed discrete approximation is a
statistical quantity that is contaminated by sampling error, this
procedure can give misleading results.  For instance, when the
sampling error is large, the computed order $p$ may be very far from
the true $p$ that would be obtained if sampling error were eliminated,
making it appear that the discretization error is either much larger
or much smaller than the true error.  If the sampling error is large
enough, it can make the implied $p$ negative, making it appear that
the solution is diverging.  Or, \eqref{eq:orderest} may have no
solution at all, making it impossible to assess $p$ or the
discretization error.  Thus, this procedure is insufficient when
significant sampling error is expected.

\subsubsection{Accounting for Sampling Error} \label{sec:bayes_richardson}
To account for sampling error, a probabilistic model of the true mean
that includes both the discretization error described in
\S\ref{sec:std_richardson} and the sampling error estimate described in
\S\ref{sec:sampling_error} is needed.  Using this model, the parameters of the
discretization error model (e.g., the constants $C_0$ and $p$) are then
estimated using Bayesian inference.  This formulation is advantageous
relative to a deterministic procedure (e.g., least-squares or maximum
likelihood estimation) in the current context because it naturally
assesses the uncertainty in the discretization error estimate,
eliminating the flaw in the standard procedure described
in \S\ref{sec:std_richardson}.  Since the Bayesian approach to inverse
problems is described in more detail by many authors~\cite{Cox_1961, Christian_2001, Jaynes_2003, Kaipo_2005,
  Calvetti_2007},
% Beck_1991, Beck_1998, Kennedy_2001,
%  Higdon_2004, Muto_2008, Oliver_POF-2012},
additional background information is omitted here.%\todo{Do we need all these Bayesian refs?}

%While effective in the deterministic , the three-sample technique does not account for the presence
%of sampling error within the $q_h$.  A straightforward extension of the
%technique
%%, initially presented by \citet{Malaya2012Estimating}, 
%permits both
%accounting for the sampling error and assessing its impact relative to the
%estimated discretization error.

To develop a probabilistic model for the true mean $E[q]$, let
$e_{h,N}$ denote the sampling error for the sample average computed
from $N$ correlated samples at resolution $h$.  That is,
\begin{equation*}
e_{h,N} = E[q_h] - \avg{q_h}_N.
\end{equation*}
where $E[q_h]$ is the true mean at resolution $h$ and $\avg{q_h}_N$
is the sample average computed from $N$ samples.  Further, letting
$\epsilon_h = E[q] - E[q_h]$ denote the discretization error, we have
\begin{equation}
E[q] = \avg{q_h}_N + e_{h,N} + \epsilon_h.
\label{eqn:error_model}
\end{equation}
Using the sampling error estimator from \S\ref{sec:sampling_error} for
$e_{h,N}$ and the form of $\epsilon_h$
from (\ref{eq:taylorbasedtruncationmodel}), one has a complete probabilistic
model of the true mean $E[q]$.  Specifically,
\begin{equation*}
E[q] = \avg{q_h}_N + e_{h,N} + C_0 h^p + C_1 h^{p+1} + \ldots,
\end{equation*}
where $e_{h,N} \sim \mcal{N}(0, \hat{\sigma}_{h,N}^2)$ and
$\hat{\sigma}_{h,N}$ is the estimate from \eqref{eqn:sigN} computed at
resolution $h$.  Note that, while $E[q]$ is a deterministic quantity,
our knowledge of $E[q]$ is incomplete.  Since Bayesian probability is
a representation of incomplete knowledge, it is appropriate that
$E[q]$ is represented by a probabilistic model.  Neglecting the
$O(h^{p+1})$ terms gives
\begin{equation}
E[q] - \avg{q_h}_N - C_0 h^p =  e_{h,N} \sim \mcal{N}(0, \hat{\sigma}_{h,N}^2).
\label{eqn:likelihood_model}
\end{equation}
This model forms the basis of the Bayesian inverse problem formulated
later in this section, and we use it exclusively in this work.
However, with appropriate modifications of the likelihood
function defined below, any discretization error model may be used
here in place of $C_0 h^p$.
% Not anymore
%  We make use of this
%fact in the channel flow results shown in \S\ref{sec:channel} to
%invesigate different descriptions of the error.

For brevity, let $\bq = E[q]$ from now forward.  Then, given $M$
sample averages $\hat{q}_i = \avg{q_{h_i}}_{N_i}$ $i = 1, \ldots, M$
computed using distinct mesh sizes $h_i$, Bayes' theorem implies that
\begin{equation}
\pi(\bq, C_0, p | \hat{q}_1, \ldots \hat{q}_M) \propto 
\pi (\bq, C_0, p) \, \pi(\hat{q}_1, \ldots, \hat{q}_M | \bq, C_0, p),
\label{eqn:bayes}
\end{equation}
where $\pi(a | b)$ denotes the probability density function (PDF) for
$a$ conditioned on $b$.  The right hand side of \eqref{eqn:bayes} is
composed of two factors: the prior PDF and the likelihood function.  The
prior PDF $\pi(\bq, C_0, p)$ encodes any available information about the
parameters $\bq$, $C_0$, and $p$ that is independent of the observations
$\hat{q}_i$.  For instance, one may have strong prior information
regarding $p$ because the formal order of accuracy of the numerical
scheme is known.  The likelihood function assesses the consistency of
the model with particular values of the parameters $\bq$, $C_0$, and $p$
and the computed values $\hat{q}_1, \ldots, \hat{q}_M$.  It is derived
from the probabilistic model \eqref{eqn:error_model}.  Assuming that
sampling errors for different resolutions $h_i$ are independent,
\begin{equation*}
\pi(\hat{q}_1, \ldots, \hat{q}_M | \bq, C, p) = \prod_{i=1}^{M} \pi(\hat{q}_i | \bq, C, p).
\end{equation*}
Then, from \eqref{eqn:likelihood_model}, it is clear that
\begin{equation*}
\pi(\hat{q}_i | \bq, C, p) = \frac{1}{\sigma_i} \phi\!\left(\frac{\bq - \hat{q}_i - C h_i^p}{\sigma_i}\right)
\end{equation*}
where $\phi$ is the standard normal density
$
  \phi(x) = \frac{1}{\sqrt{2\pi}}\exp\left(-\frac{1}{2}x^2\right)
$, 
and $\sigma_i = \hat{\sigma}_{h_i,N_i}$.

Note that, for $M=3$, as $\sigma_i \to 0$ the likelihood PDF approaches
the $\delta$ distribution centered at the observed values, and thus,
this Bayesian procedure reduces to the deterministic Richardson
extrapolation approach described in \S\ref{sec:std_richardson}.
%However, for significant sampling error, the Bayesian procedure is
%superior because it systematically accounts for the uncertainty in the
%data.

To complete the specification of the Bayesian inverse problem, one
must set priors on $\bq$, $C_0$, and $p$.  
% Without loss of generality,
% we
%take $\hat{q}_M$ to be the observation from the finest mesh with the
%smallest sampling error variance.  As $\hat{q}_N$ should be the best
%approximation to $q$ we know \textit{a priori}, we choose $q \sim
%N\left(\hat{q}_N,\sigma_q^2\right)$ for some moderate $\sigma_q$ from
For simplicity, we take $\bq$, $C_0$, and $p$ to be independent in the
prior.  Further, we choose
$\bq \sim \mcal{N}\left(\hat{q}_M,\sigma_q^2\right)$, where
$\hat{q}_M$ is the result at the finest resolution, for some moderate
$\sigma_q$.  Then,
\begin{align}
  \pi\left(\bq\right)
  &=
  \frac{1}{\sigma_q}
  \phi\!\left(\frac{\bq - \hat{q}_M}{\sigma_q}\right)
  .
\end{align}
In principle, $C_0$ may take any real value, but it is algorithmically convenient
to limit the probable range of $C_0$ by choosing
%RDM there is nothing wrong with an improper prior in this context
%we wish to avoid using an
%improper uniform prior over $\mathbb{R}$, 
%we choose 
$C \sim \mcal{N}\left(0,\sigma_C^2\right)$ for some large $\sigma_C$
from which
\begin{align}
  \pi\left(C_0\right)
  &=
  \frac{1}{\sigma_C}
  \phi\!\left(\frac{C_0}{\sigma_C}\right)
  .
\end{align}
Because $p\geq 1$ is expected for most convergent numerical schemes but
detecting pathologically-slow convergence when $p>0$ is desirable, we
select a prior distribution that goes to zero at $p=0$, is maximum near
the expected convergence order (if known), and has a broad range of
plausible $p$. The Gamma distribution with $\alpha> 1$ meets these
requirements for suitable values of the parameters $\alpha$ and $\beta$, so the prior
on $p$ is given by
\begin{equation}
  \pi\left(p\right)
  =
  \frac{\beta^\alpha p^{\alpha-1}}{\Gamma(\alpha)}
  \exp\left(-\beta{}p\right)
  =
  \frac{\sqrt{2\pi} \beta^\alpha p^{\alpha-1}}{\Gamma(\alpha)}
  \phi\!\left(\sqrt{2\beta{}p}\right)
  .
\end{equation}
Substituting these priors into \eqref{eqn:bayes} gives
%Using these prior selections within our last Bayesian proportionality gives
%
\begin{multline}
  \pi\left(
    \bq,C,p\mid\hat{q}_1,\ldots,\hat{q}_N;
    \sigma_q,\sigma_C,\alpha,\beta
  \right)
\\
\propto
  \frac{\sqrt{2\pi} \beta^\alpha p^{\alpha-1}}
       {\sigma_q\,\sigma_C\,\Gamma(\alpha)}
  \phi\!\left(\frac{\bq - \hat{q}_N}{\sigma_q}\right)
  \phi\!\left(\frac{C}{\sigma_C}\right)
  \phi\!\left(\sqrt{2\beta{}p}\right)
  \prod_{i=1}^{M}
  \frac{1}{\sigma_i}
  \phi\!\left(\frac{\bq - \hat{q}_i - C h_i^p}{\sigma_i}\right),
\label{eqn:posterior}
\end{multline}
where the dependence on prior parameters $\sigma_q$, $\sigma_C$,
$\alpha$, and $\beta$ has been noted.  While the posterior PDF is
simple to write down according the Bayes' theorem, working with this
PDF can be difficult.  In general it is not possible to compute
statistics for the posterior analytically because the necessary
integrals cannot be evaluated in closed form.  Instead, it is common
to use Markov chain Monte Carlo (MCMC) algorithms to sample the
posterior~\cite{Robert2010MC}.  In this work, we use a
Python~\citep{Drake2011Python} implementation relying on
the \texttt{emcee} implementation~\citep{ForemanMackey2012Emcee} of
\citeauthor{Goodman2010Ensemble}'s affine invariant MCMC sampling technique~\citep{Goodman2010Ensemble}.

% Settled on don't provide implementation then? 
%
%% A reference
%% implementation similar to that used to produce the results shown
%% in \S\ref{sec:lorenz} and \S\ref{sec:channel} is provided at ??.
%
%%A simple Python implementation for sampling the
%%posterior is given in \eqref{eqn:posterior} is provided at ??.
%\todo{RDM: don't want to publish code in the Journal, instead, let's
%make available through the auxiliary data repository}
%\todo{I entirely pulled the reference implementation comment --Rhys}
%%shown in Appendix~\ref{app:code}.

\subsection{Illustrative Example: The Lorenz Equations} \label{sec:lorenz}

To illustrate the application of the sampling and discretization error
estimation techniques discussed here, they are applied to
estimates of the means computed from solutions of the Lorenz equations.
The Lorenz equations are a system of three ordinary differential equations:
\begin{subequations}
\begin{align}
\dd{x}{t} & = \sigma (y - x), \label{eqn:lorenz_x} \\
\dd{y}{t} & = x (\rho - z) - y, \label{eqn:lorenz_y} \\
\dd{z}{t} & = x y - \beta z. \label{eqn:lorenz_z}
\end{align}\
\label{eqn:lorenz}
\end{subequations}
Depending on the values of the parameters $\sigma$, $\beta$, and
$\rho$, the system exhibits chaotic behavior.  The
methods described in \S\ref{sec:sampling_error}
and \S\ref{sec:discretization_error} are therefore applicable to 
estimating errors
in statistical quantities, such as the mean of $z$, computed from
discrete approximations.  In the results presented here,  the
parameters are set to their typical values: $\sigma = 10$, 
$\beta=\frac{8}{3}$, $\rho = 28$.
Further, \eqref{eqn:lorenz} are discretized
using fourth-order Runge-Kutta time discretization (RK4).

\subsubsection{Sampling Error Estimator Performance}
\label{sec:estimator_performance} First, we examine the performance of
the sampling error estimator.  Estimates of the standard deviation
$\sigma_z$ of
$\langle z\rangle_T$, the average of $z$ over a time period $T$, were
determined using the techniques in \S\ref{sec:sampling_error} for
several averaging periods $T$. To assess the reliability of these
estimates, they were repeated for each of a set of 10,085 different
Lorenz simulations, which were started with randomly selected initial
conditions 
%RDM OK, I guess we don't need to quote the burn in period. Doesn't fit
%well into the prose.
%\todo{RDM: is this true, if so what was the burn-in period?}
so that the variability in these estimates could be assessed.  In each
simulation, $\avg{z_h}_T$ was computed by sampling the solution
every $\Delta t_s=0.075$ time units, which was every third RK4 time step 
($\Delta t=0.025$). 

%
% for calculations below: 
%
% T0 \aprox 103.5 (samples)
% however, 1 sample per .075 T
% thus, .075*103.5 = 7.76 T
%
% number of samples per T = 1/.075 = 13.333
% N_eff = N/T0 
% N_eff = 13.33/103.5 = .1288
%

\begin{figure}[htb]
\begin{center}
\subfloat[$T \approx 1 \times 10^6$]{\includegraphics[width=0.49\linewidth]{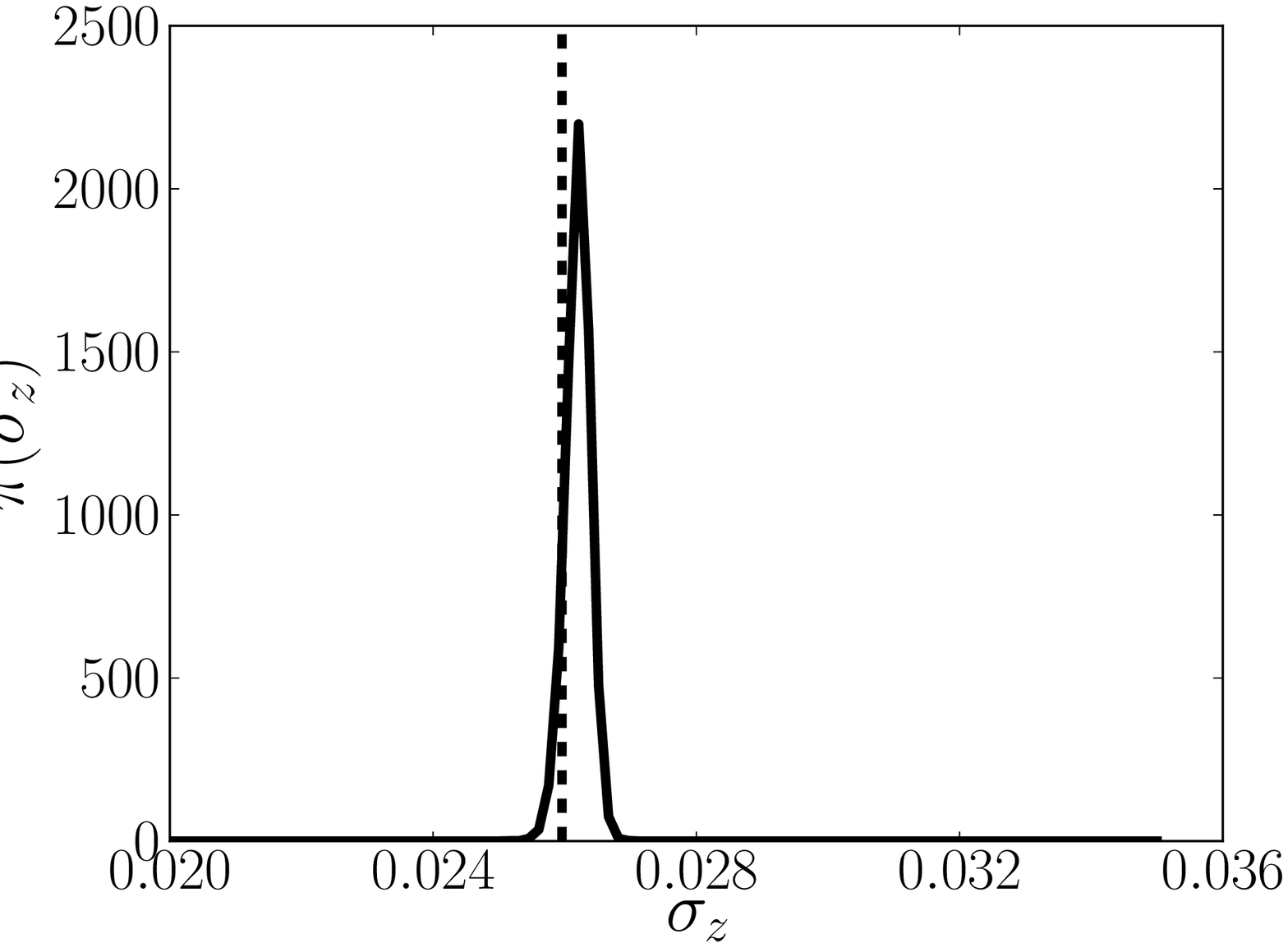}}
\label{fig:sigma_estimate_large}
\subfloat[$T \approx 5 \times 10^5$]{\includegraphics[width=0.49\linewidth]{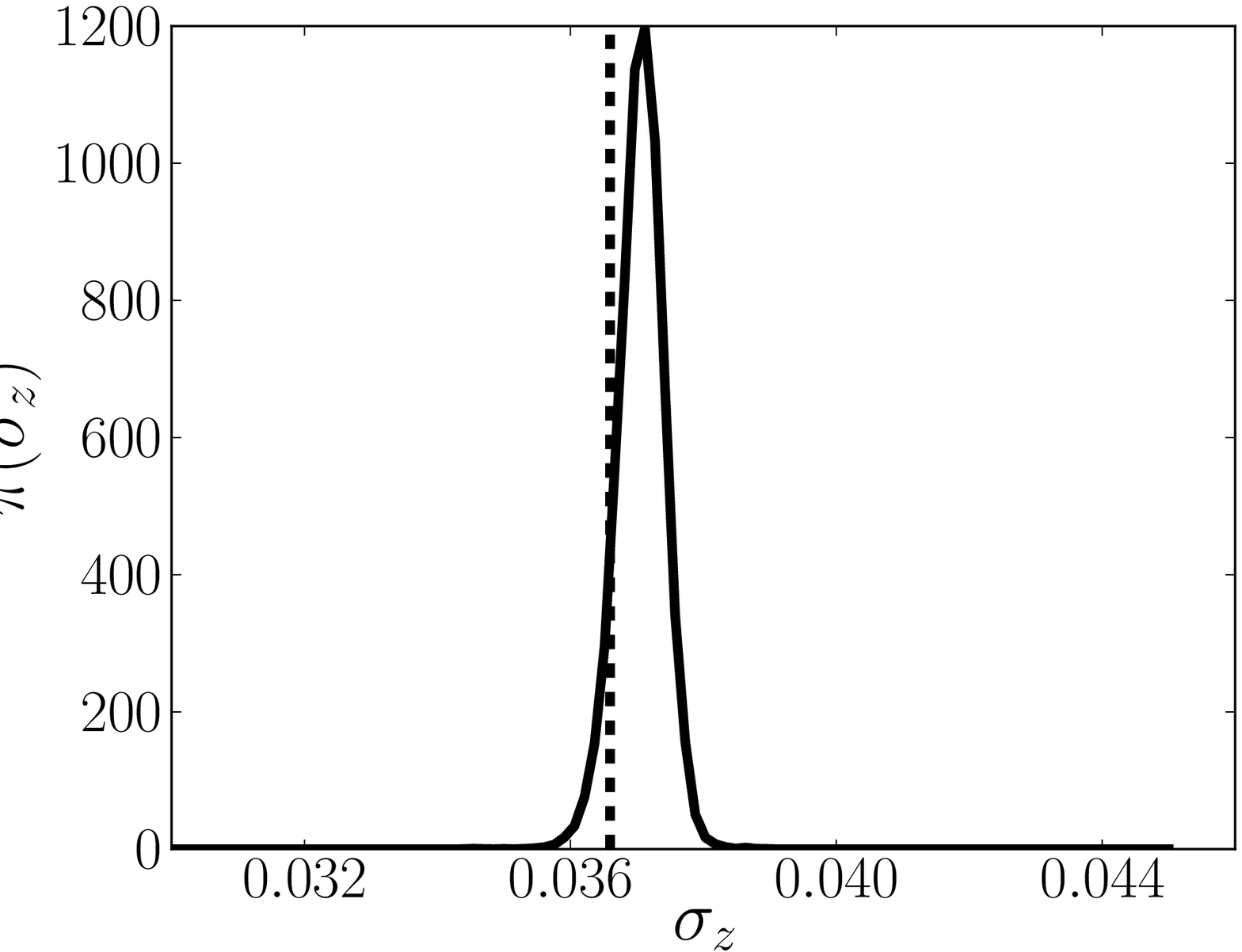}}
\label{fig:sigma_estimate_mid}
\subfloat[$T \approx 2.5 \times 10^5$]{\includegraphics[width=0.49\linewidth]{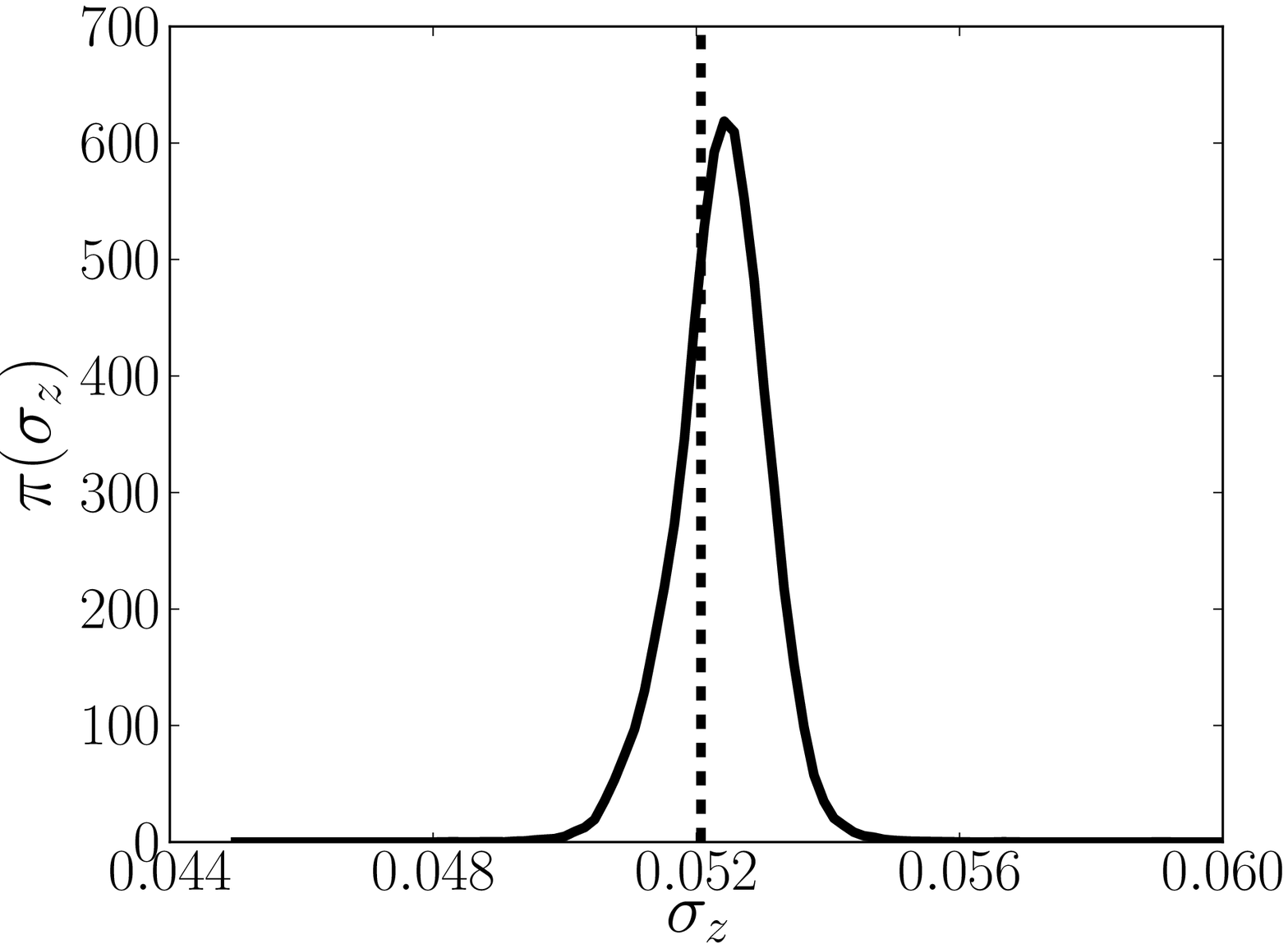}}
\label{fig:sigma_estimate_small}
\subfloat[$T \approx 1.25 \times 10^5$]{\includegraphics[width=0.49\linewidth]{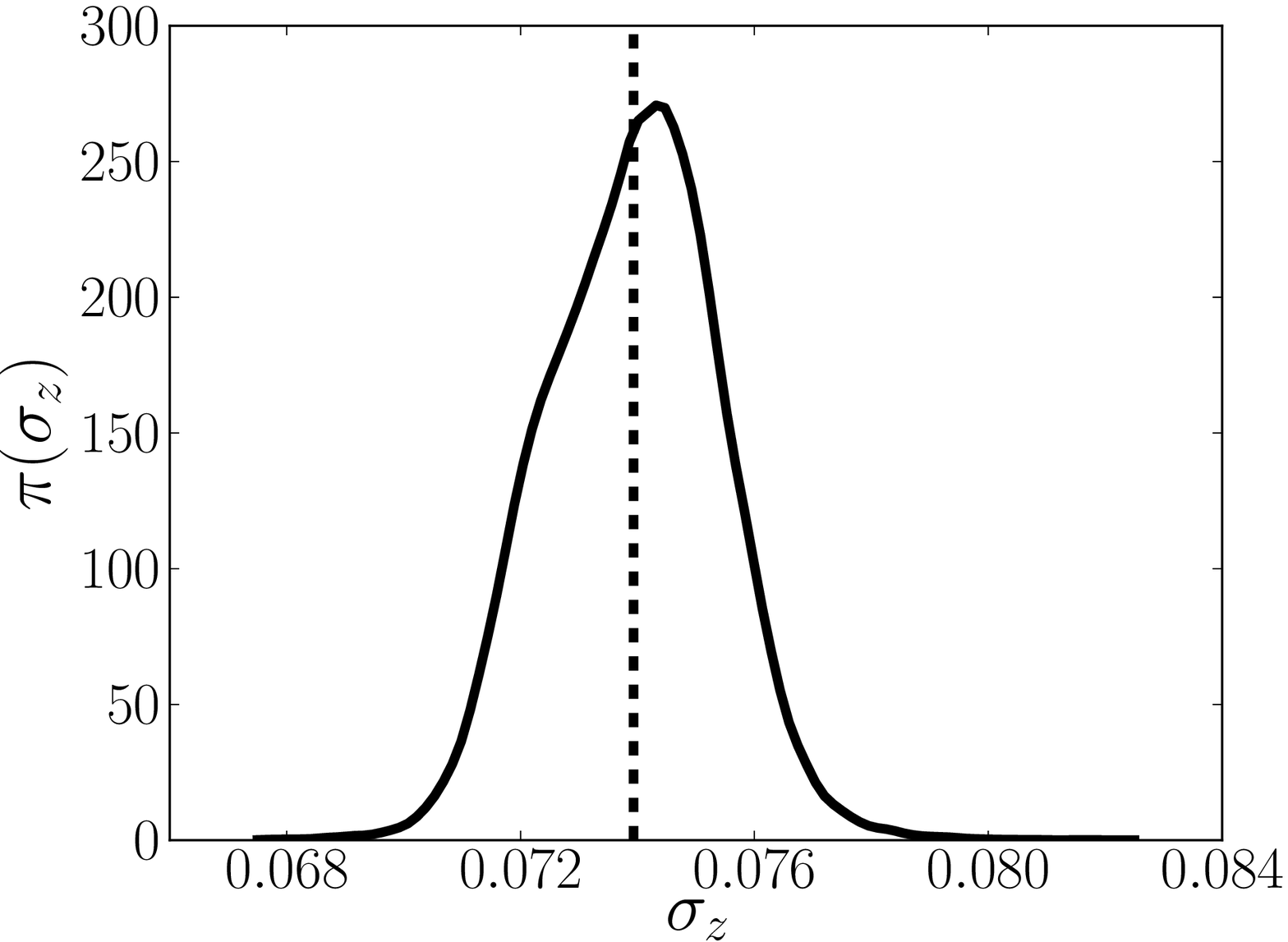}}
\label{fig:sigma_estimate_smallest}
\end{center}
\caption{PDFs of the estimate of the standard deviation of the sample
  average of $z$ computed according to \eqref{eqn:sigN} for varying
  averaging periods $T$ computed from an ensemble of 10,085
  Lorenz simulations.  For comparison, the vertical dashed lines
  indicate the standard deviation of the sample average computed directly
  from the ensemble according to
  \eqref{eqn:sigma_true}.} 
\label{fig:sigma_estimate}
\end{figure}

The decorrelation separation distance $T_0$ computed as in
\S\ref{sec:sampling_error} varied somewhat but was approximately 7.76
time units (103.5 samples), making the effective sample size
$N_{\mathrm{eff}}\approx{}.128\,T$.
The distributions of $\sigma_z$ obtained from the ensembles of Lorenz
simulations are shown in Fig.~\ref{fig:sigma_estimate} for four different
averaging periods $T$. An estimate of the ``true'' value of $\sigma_z$
is also shown in the figure. It is determined directly from the sample
variance of the ensemble of estimates $\avg{z_h}_T$:
\begin{equation}
\sigma^2_\text{true} = \frac{1}{S-1} \sum_{i=1}^{S} \left( \mu_\text{true} - \avg{z_h}_{T,i} \right)^2,
\label{eqn:sigma_true}
\end{equation}
where 
\begin{equation*}
\mu_\text{true} = \frac{1}{S} \sum_{i=1}^{S} \avg{z_h}_{T,i},
\end{equation*}
and $\avg{z_h}_{T,i}$ denotes the sample average of $z$ for the $i$th
simulation.

%
%obtained 
%
%Figure~\ref{fig:sigma_estimate} shows the estimate of the standard
%deviation of the sample average of $z$, denoted by $\sigma_z$, for
%time averages over periods of various durations.
%%\todo{``durations''?}, 
%which lead to different $N_{\mathrm{eff}}$
%values.
% 
%
%
% sigma estimate (should this come before the effectiveness of the sampling?)
%
% we also have sigma_estimate_mid == 50k $N_eff$
%
% these data are available in: 
%   sigma_estimate/ 
%
%

% 
%\todo{I'm finding the repeat use of ensemble confusing here. perhaps
%ensemble for simulations, and collections for estimates of standard deviation?}
%For each $N_{\mathrm{eff}}$, an ensemble of $S = 10085$ different
%Lorenz simulations was conducted.  From this ensemble, we generate a
%ensemble of estimates of the standard deviation of the sample average 
%$\sigma_z$.  The PDF of $\sigma_z$ is approximated from these samples
%using a kernel density estimator (KDE).  Further, by computing the
%standard deviation of the ensemble of sample averages computed from the
%ensemble of simulations, we have a sample-based estimate of the true
%standard deviation of the sample average.  Specifically, the true
%standard deviation is estimated as follows:
%
%\begin{equation}
%\sigma^2_\text{true} = \frac{1}{S-1} \sum_{i=1}^{S} \left( \mu_\text{true} - \avg{z_h}_{N,i} \right)^2,
%\label{eqn:sigma_true}
%\end{equation}
%
%where 
%%
%\begin{equation*}
%\mu_\text{true} = \frac{1}{S} \sum_{i=1}^{S} \avg{z_h}_{N,i},
%\end{equation*}
%% 
%and $\avg{z_h}_{N,i}$ denotes the sample average of $z$ for the $i$th
%simulation.  The estimated true values of the standard deviation are
%indicated in Figure~\ref{fig:sigma_estimate} by dashed vertical lines.

In all cases, the estimated true value is well
within the support of the distribution of $\sigma_z$, indicating that
the estimate is consistent with the true value.  Additional details
for comparison are provided in Table~\ref{tbl:sigma_numbers}.
\begin{table}[tbp]
\caption{Statistics of the estimator $\sigma_z$ for varying simulation duration.}
\label{tbl:sigma_numbers}
\begin{center}
\begin{tabular}{|c|c|c|c|c|c|c|}
\hline
$T$ & $\sigma_\text{true}$ & $\mathrm{mean}(\sigma_z)$ & $| \sigma_\text{true} - \mathrm{mean}(\sigma_z)|$  & $\mathrm{stddev}(\sigma_z)$ & $\min(\sigma_z)$ & $\max(\sigma_z)$ \\
\hline
$1.25 \times 10^5$ & 7.393e-02 & 7.389e-02 & 4.232e-05 & 1.491e-03 & 6.565e-02 & 8.686e-02 \\ 
$2.5 \times 10^5$ & 5.207e-02 & 5.236e-02 & 2.926e-04 & 7.079e-04 & 4.843e-02 & 5.550e-02 \\ 
$5 \times 10^5$ & 3.660e-02 & 3.705e-02 & 4.503e-04 & 3.491e-04 & 3.454e-02 & 3.877e-02 \\ 
$1 \times 10^6$ & 2.595e-02 & 2.620e-02 & 2.453e-04 & 1.813e-04 & 2.515e-02 & 2.695e-02 \\ 
\hline
\end{tabular}
\end{center}
\end{table}
The table shows that the error estimate $\sigma_z$ is quite
consistent with the standard deviation of the sample average, indicating
that it is a good estimator.  For instance, the difference between
$\sigma_{\mathrm{true}}$ and the sample average of $\sigma_z$ is
never more than $1.2 \%$ of $\sigma_{\mathrm{true}}$.  Further, even
for the maximum and minimum $\sigma_z$ the errors are generally around
$10\%$ or less, and the maximum error is $17.5\%$ of
$\sigma_{\mathrm{true}}$.  Finally, note that the results show the
correct $\sigma_z \propto 1/\sqrt{T}$ scaling, as
expected.

\begin{figure}[htb]
  \includegraphics[width=\linewidth]{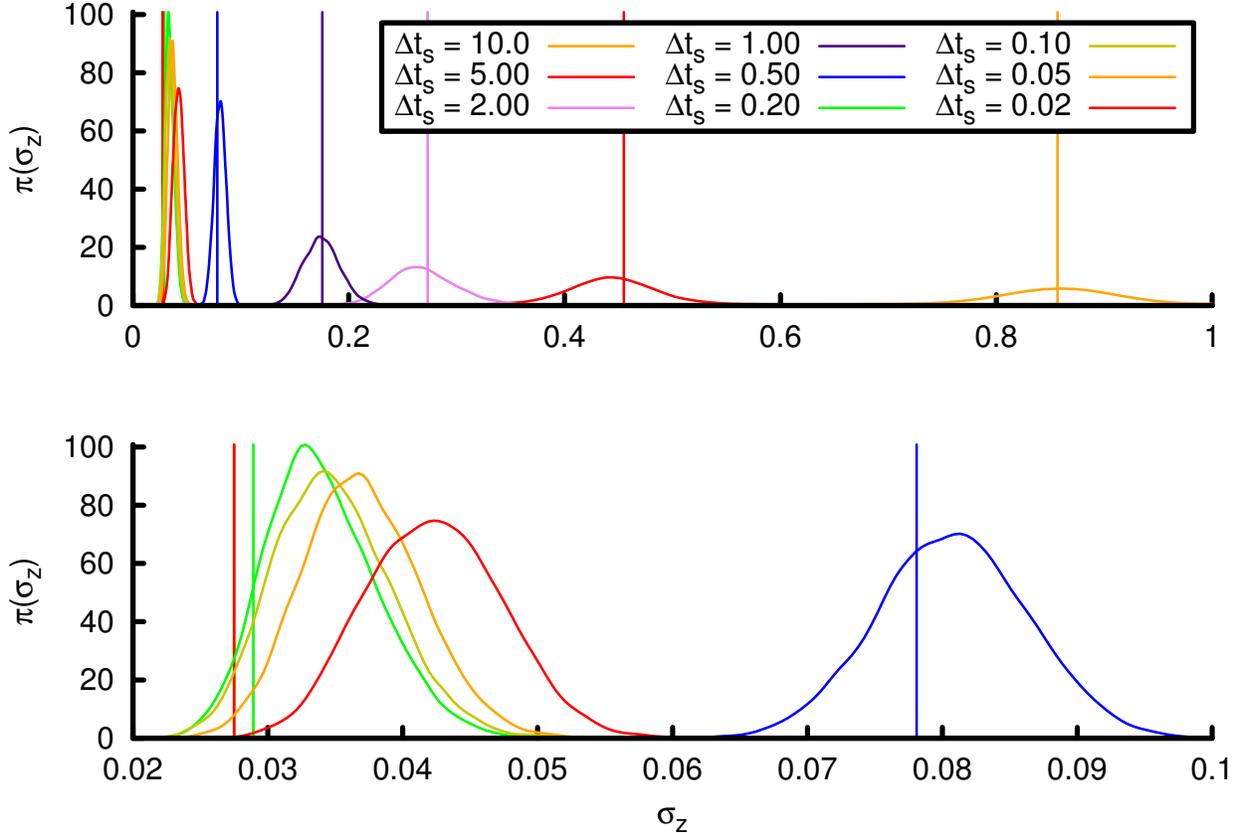}
  \caption{Convergence behavior of autoregressive procedure on Lorenz $z$ data
  gathered from $10,085$ simulations.  Each simulation was of duration $T =
  1000$ with samples $\avg{z_{h}}$ taken every $\Delta{}t_s$ time units.
  Kernel density estimates for the PDF for $\sigma_z$ appear as curves.
  Empirically obtained $\sigma_\text{true}$ from the many
  realizations are marked with vertical lines.
  \label{fig:estimate_convergence}}
\end{figure}

All of the results shown in Figure~\ref{fig:sigma_estimate} and
Table~\ref{tbl:sigma_numbers} were computed using a very high sampling
frequency (sample every third time step).  
However, it is common in DNS to sample
much less frequently.  To examine the impact of coarse sampling and
as well as the behavior as $\Delta t_s \to \Delta t$, the sampling step
was varied while the RK4 time step remained constant at $\Delta t = 0.001$.
Results of this study are shown in 
Figure~\ref{fig:estimate_convergence}.

When the sampling period is
large, the samples are less correlated.  However, information is still
discarded by neglecting even highly correlated samples, leading to
larger uncertainty in the sample average of $z$.  Beginning with large
$\Delta t_s$, as $\Delta t_s$ is decreased, the estimated $\sigma_z$
and $\sigma_{\mathrm{true}}$ both decrease.  However, for $\sigma_z$,
this trend reverses for small enough $\Delta t_s$.  While
$\sigma_\text{true}$ appears to converge, the estimated $\sigma_z$
begins to grow when $\Delta t_s$ become smaller than about 0.20.  
When $\Delta t_s$ is less than about 0.01, the algorithm used to compute the $\sigma_z$ breaks down due to
the effects of round-off error.  We hypothesize that the increase in
$\sigma_z$ with decreasing $\Delta t_s$ below 0.20
is also due to accumulation of double precision round-off error.
Repeating this study using both lower and higher floating point precision (not
shown) produced behavior consistent with that hypothesis.

% nick: I am killing the ulerich et al ref
%
%Further details of
%the algorithms used in this estimation and their sensitivity to
%floating point errors are documented 
%\todo{This is not an acceptable reference, either eliminate or point to
%something else}
%by~\citeauthor{Ulerich2013AR}.
%More robust algorithms for
%computing the estimate are described in Appendix~\ref{app:robust}.
%
% Convergence as the sampling rate increases
% Need to re-run these for the final paper
% Raw data in lorenz/ subdirectory
%

\subsubsection{Bayesian Richardson Extrapolation Results}
Here we explore the performance of the Bayesian Richardson extrapolation
procedure described in (\S\ref{sec:discretization_error}) in three
regimes: small sampling error, medium sampling error, and large sampling
error relative to the discretization error. In DNS, it is expected that
sampling errors will generally be larger than or comparable to the
discretization error. The small sampling error regime is also considered
here for completeness. 

The data input to the Bayesian Richardson extrapolation algorithm is
listed in Table~\ref{tbl:large_sample_uncertainty}, where $\Delta t$ is
the time step used, $T$ is the total simulation time, $\hat{q}$ is the 
observed sample average, and $\sigma_z$ is the estimated standard
deviation of the sample average.  In all cases, the sampling period
was $\Delta t_s = 0.15$.

%
% new table (hows this look? do we want to transpose?)
%
\begin{table}[tp]
\caption{Conditions for cases across large, medium and small sampling
 uncertainty at varying time step.}
\label{tbl:large_sample_uncertainty}
\begin{center}
%\begin{tabular}{|c|c|c|c|c|c|c|c|c|c|}
\begin{tabular}{|>{\centering}p{1.6cm}|>{\centering}p{.9cm}|>{\centering}p{1.6cm}|>{\centering}p{1.4cm}|>{\centering}p{.9cm}|>{\centering}p{1.6cm}|>{\centering}p{1.4cm}|>{\centering}p{.9cm}|>{\centering}p{1.6cm}|>{\centering}p{1.8cm}|}
\hline
  & 
 \multicolumn{3}{c|}{Large}  &
 \multicolumn{3}{c|}{Medium} &
 \multicolumn{3}{c|}{Small}  \tabularnewline
\hline 
%\hline
$\Delta t$ & T & $\hat q$ & $\sigma_z$ & T & $\hat q$ & $\sigma_z$ & T & $\hat q$ & $\sigma_z$ \tabularnewline
\hline
%\noalign{\smallskip}
0.075  & 10  & 23.6081  & 0.457 &  $10^{3}$ & 23.1873 & 0.0290 &
 $10^{6}$ & 23.1911 & 0.000661 \tabularnewline
0.05   & 10  & 23.3747  & 0.546 &  $10^{3}$ & 23.4942 & 0.0325 &
 $10^{6}$ & 23.4874 & 0.000813 \tabularnewline 
0.025  & 10  & 23.3432  & 0.724 & $10^{3}$ & 23.5718 & 0.0382 & $10^{6}$
 & 23.5486  & 0.000884 \tabularnewline 
\hline
\end{tabular}
\end{center}
\end{table}
%
%

% nick : old table
% 
%\todo{Should we roll these into one table?}
%
%
% \begin{table}[htp]
% \caption{Conditions for the case with large sampling uncertainty.}
% \label{tbl:large_sample_uncertainty}
% \begin{center}
% \begin{tabular}{|c|c|c|c|}
% \hline
% $\Delta t$ & T & $\hat q$ & $\sigma_z$ \\
% \hline
% 0.075   & 10  & 23.6081  & 0.457 \\ 
% 0.05  & 10  & 23.3747  & 0.546 \\
% 0.025 & 10  & 23.3432  & 0.724 \\
% \hline
% \end{tabular}
% \end{center}
% \end{table}
% %
% %
% \begin{table}[htp]
% \caption{Conditions for the case with medium sampling uncertainty.}
% \label{tbl:medium_sample_uncertainty}
% \begin{center}
% \begin{tabular}{|c|c|c|c|}
% \hline
% $\Delta t$ & T & $\hat q$ & $\sigma_z$ \\
% \hline
% 0.075 & $10^{3}$ & 23.1873 & 0.0290 \\ 
% 0.05  & $10^{3}$ & 23.4942 & 0.0325   \\
% 0.025 & $10^{3}$ & 23.5718 & 0.0382    \\
% \hline
% \end{tabular}
% \end{center}
% \end{table}
% %
% %
% \begin{table}[htp]
% \caption{Conditions for the case with small sampling uncertainty.}
% \label{tbl:small_sample_uncertainty}
% \begin{center}
% \begin{tabular}{|c|c|c|c|}
% \hline
% $\Delta t$ & T & $\hat q$ & $\sigma_z$ \\
% \hline
% 0.075  & $10^{6}$ & 23.1911 & 0.000661 \\ 
% 0.05   & $10^{6}$ & 23.4874 & 0.000813 \\
% 0.025  & $10^{6}$ & 23.5486  & 0.000884 \\
% \hline
% \end{tabular}
% \end{center}
% \end{table}
%
%

\begin{figure}[htp]
\begin{center}
\subfloat[Large statistical uncertainty.]{
  \includegraphics[height=2.4in]{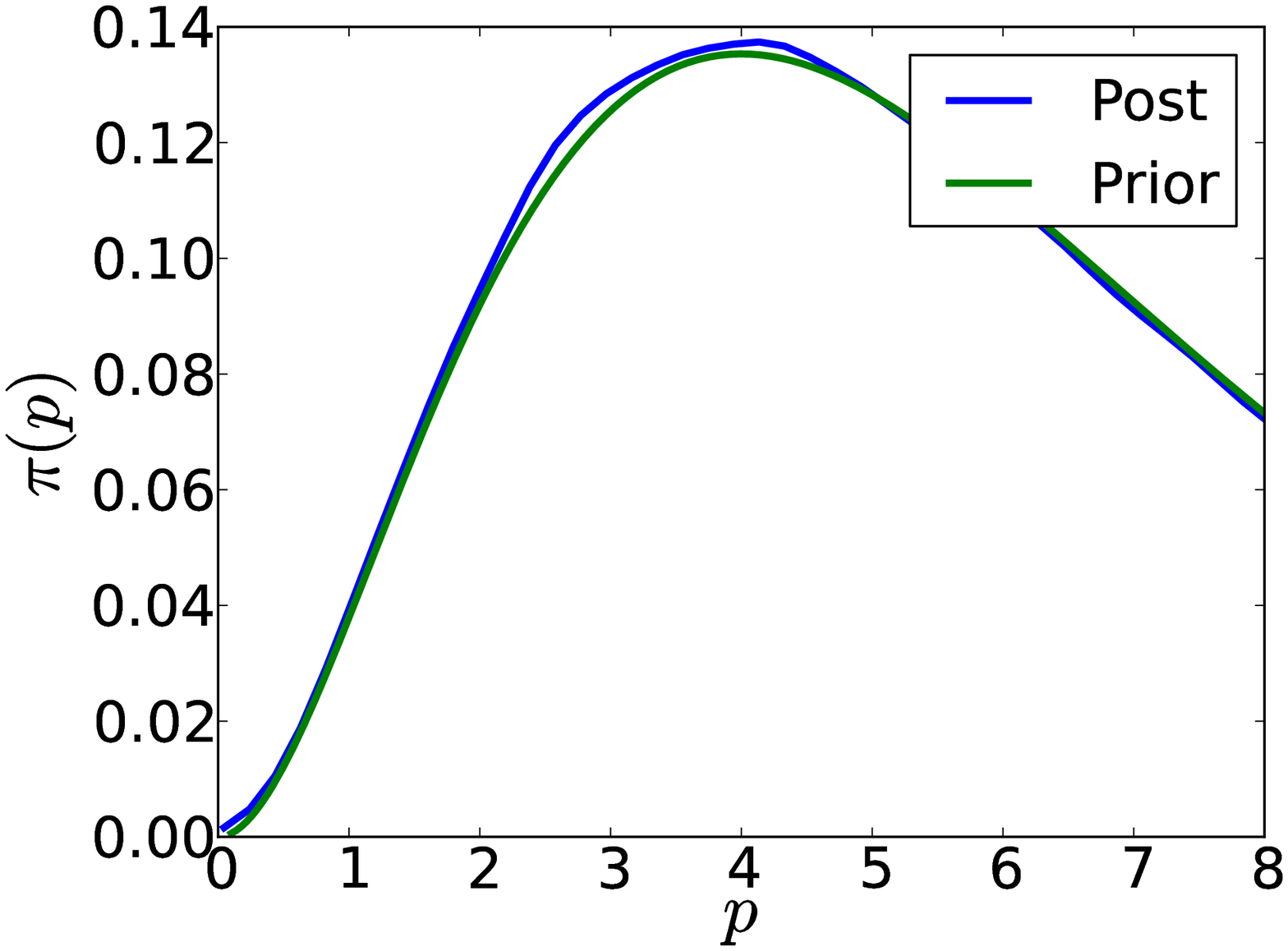}
  \includegraphics[height=2.4in]{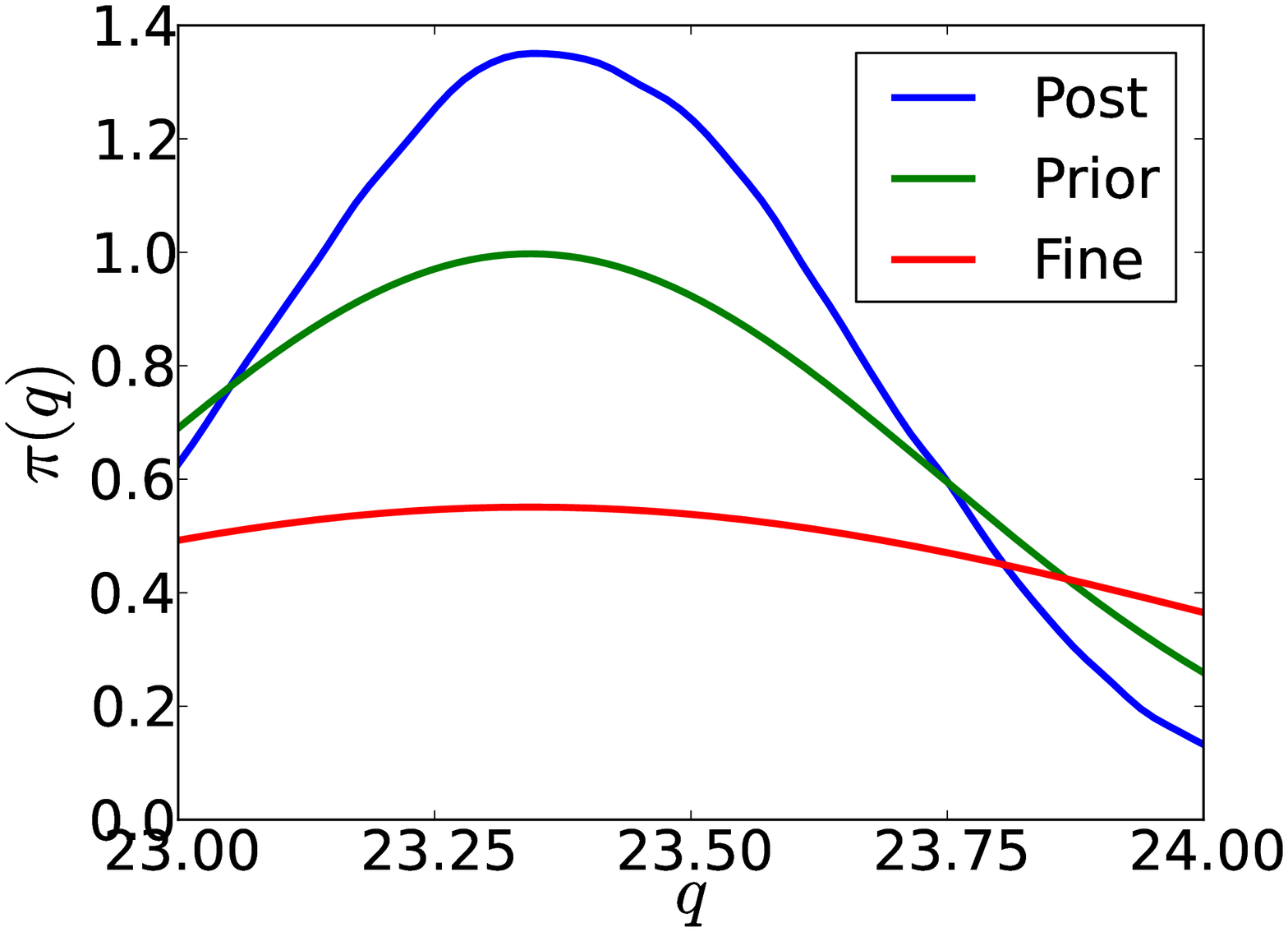}
  \label{fig:noisy}} \\
\subfloat[Medium statistical uncertainty.]{
  \includegraphics[height=2.4in]{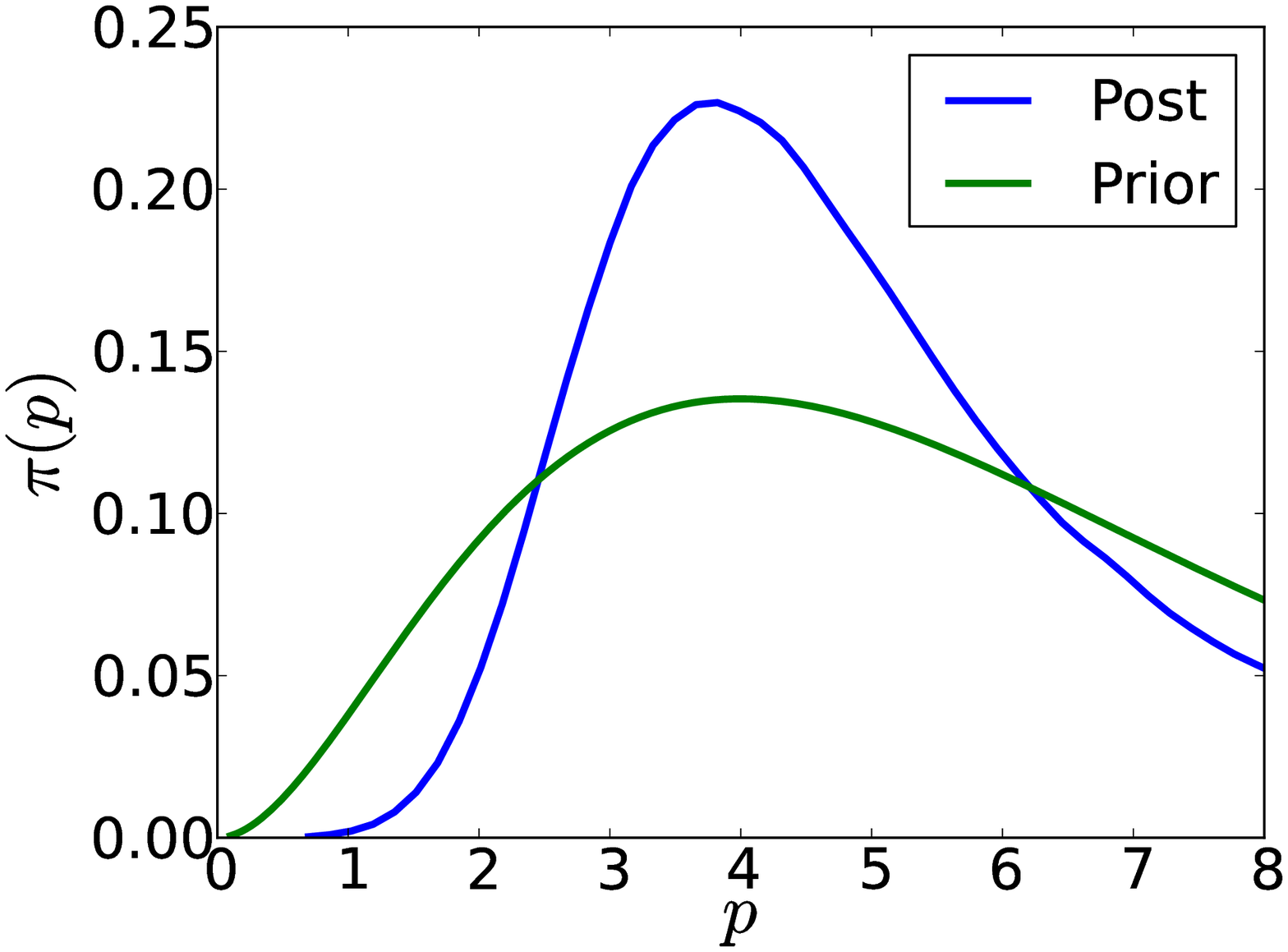}
  \includegraphics[height=2.4in]{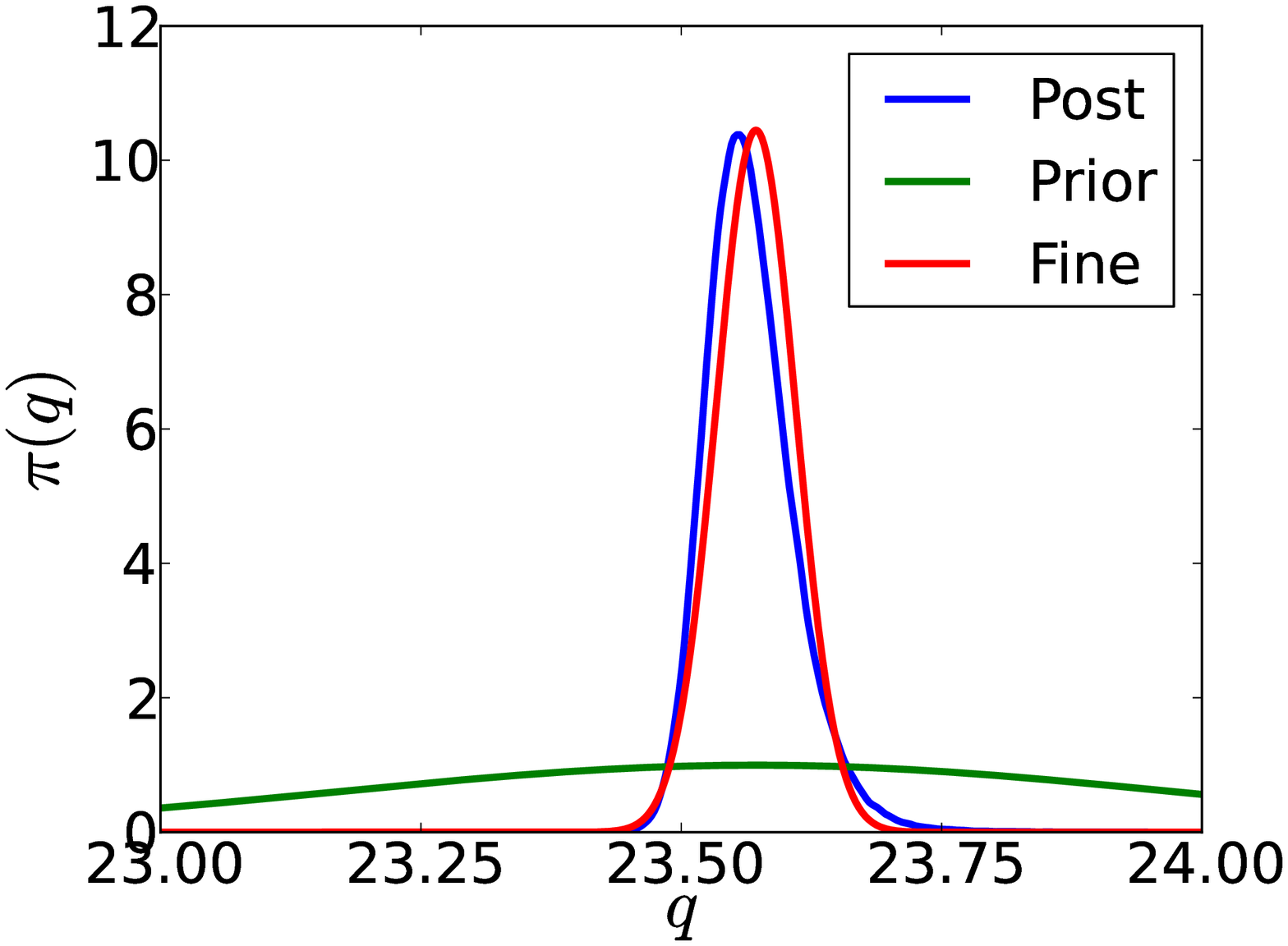}
  \label{fig:mid}} \\
\subfloat[Small statistical uncertainty.]{
  \includegraphics[height=2.4in]{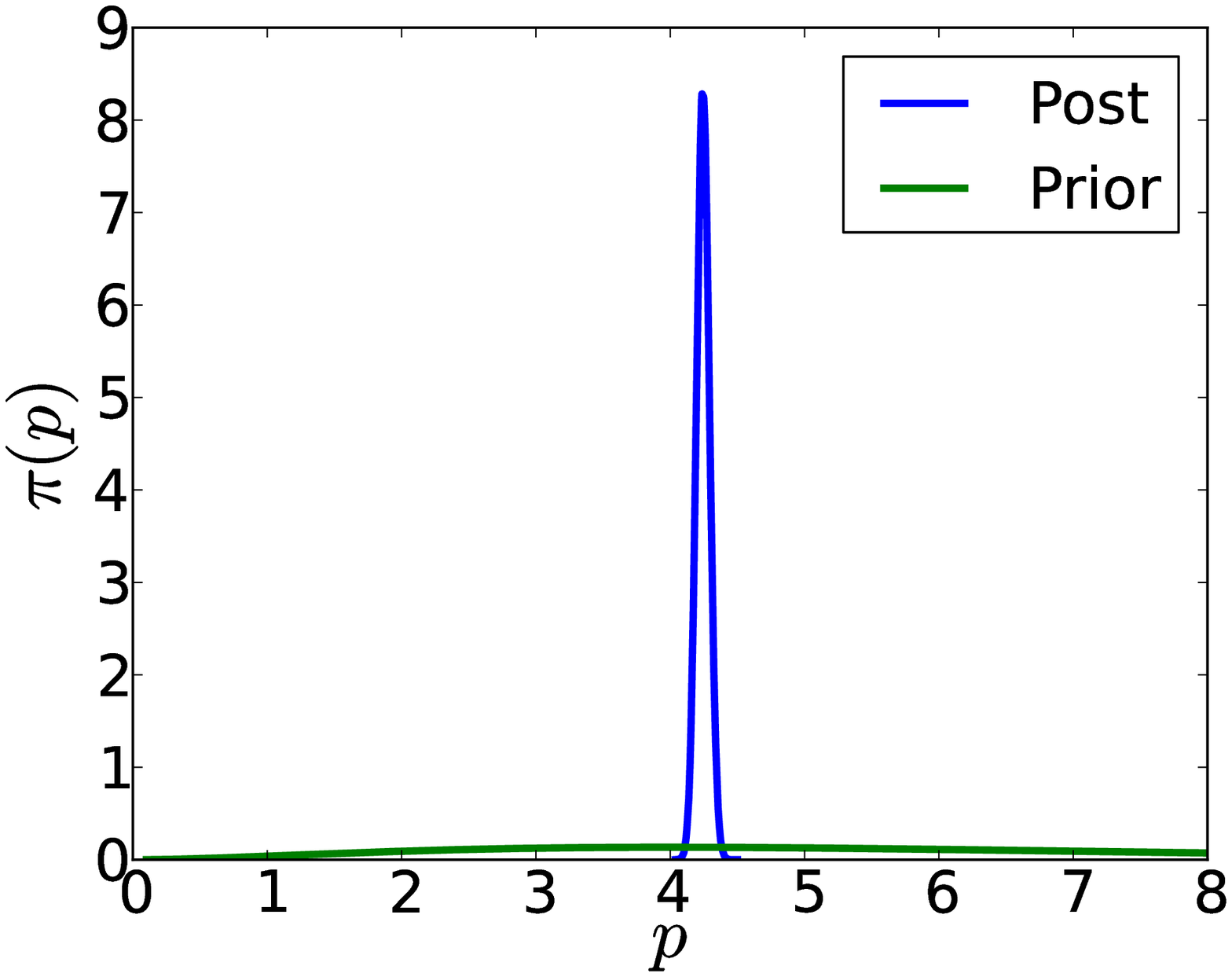}
  \includegraphics[height=2.4in]{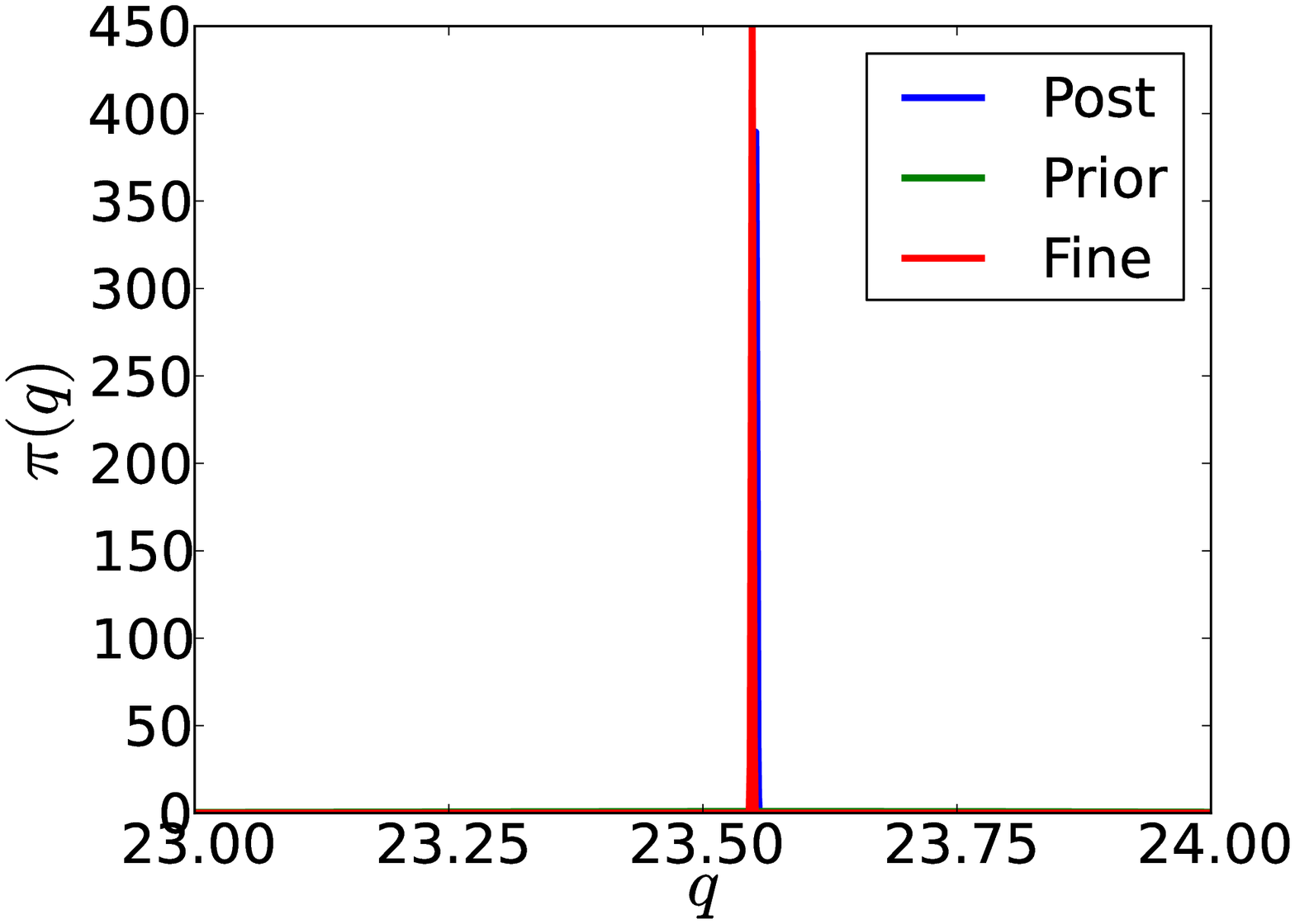}
  \label{fig:sharp}}
\end{center}
\caption{Bayesian Richardson extrapolation results from three different
 regimes: large, medium and small statistical uncertainty in comparison
 to discretization error.} \label{fig:lorenz_posteriors}
\end{figure}

Marginal prior and posterior PDFs for both the order of accuracy $p$ and
the true value of the mean of $z$ (denoted $q$) were obtained using the Bayesian
Richard extrapolation procedure and are shown in
Figure~\ref{fig:lorenz_posteriors}.
%
% posteriors from bayesian richardson inference problem
%            data generated from: lorenz.py
%
%\todo{Need to fix legend in (a) and figure sizes in (c)}
% 
In addition, the plots for $q$ also show the PDF
for the sample average of $z$ with the finest time step (i.e., a Gaussian with
mean equal to the observed sample average and standard deviation of
$\sigma_z$).

When the averaging time duration is sufficiently small so that the
sampling error is large, as shown in Figure~\ref{fig:noisy}, the
sampling error effectively masks the discretization error. In this
case, the data contain little information about the true
discretization error.  Thus, the marginal posterior PDF for $p$ is
essentially the same as the prior PDF.  However, because the prior PDF
for $q$ is so broad, the prior for $p$ constrains the results.  That
is, because we indicate \emph{a priori} that the scheme is convergent,
the data are inconsistent with values of $q$ in the tails of the
prior.  For this reason, the posterior for $q$ is somewhat more peaked
than the prior even though the marginal posterior for $p$ is the same
as the prior.

Figure~\ref{fig:mid} shows the ``medium'' sampling uncertainty level.
For this case, the sampling uncertainty dominates discretization error
at the smallest $\Delta t$, but discretization error dominates at the
largest $\Delta t$.  Some information regarding the order of
accuracy can be learned from the data in this case, leading to a posterior PDF for
$p$ that is significantly different from the prior, unlike the large
sampling uncertainty case.  Note that the peak of the marginal
posterior for $p$ is nearly the formal order of accuracy ($p=4$), but
that there is significant uncertainty associated with this estimate.
Since the discretization error can be estimated with more confidence
and the sampling error is smaller, the posterior PDF for $q$ is much
narrower than in the large sampling uncertainty case.  Further, it is
slightly shifted from the fine resolution result.  This shift is a
correction for discretization error at the fine resolution.

The final case is the small sampling uncertainty case shown
in~\ref{fig:sharp}.  In this case, the sampling error is many times
smaller than the
discretization error, and the Bayesian procedure should reduce to the standard
Richardson extrapolation.
%\todo{worth pointing out standard Richardson only
%holds together because of the artificially-low sampling error?  feels
%necessary since otherwise the incorrect seed is planted that all this fussing
%was somehow unnecessary}.  
This fact is confirmed by the observation
that the PDFs for both $p$ and $q$ are very narrow.  The peak of the
posterior distribution for $p$ occurs at approximately $p = 4.24$,
which is slightly larger than the expected order of accuracy but
agrees with the estimate that would be obtained from standard
Richardson extrapolation.  For comparison, the prior and posterior
mean and standard deviation values for the estimated true mean are
given in Table~\ref{tbl:q_table}.
\begin{table}[htp]
\caption{Prior and posterior mean and standard deviations for the true
  expectation of $z$ as determined from Bayesian inference.}
\label{tbl:q_table}
\begin{tabular}{|c|cc|cc|}
\hline
Case & Prior Mean & Post Mean & Prior Std Dev & Post Std Dev \\
\hline
High Noise & 23.3432 & 23.3672 & 0.4 & 0.29 \\
Med Noise & 23.5718 & 23.5669 & 0.4 & 0.04 \\
Low Noise & 23.5486 & 23.5520 & 0.4 & 0.0010 \\
\hline
\end{tabular}
\end{table}

\section{DNS of $Re_{\tau} = 180$ Channel Flow} \label{sec:channel}

The techniques described in \S\ref{sec:method} have been used to
investigate sampling and discretization errors in DNS of a wall-bounded
turbulent flow. Specifically, DNS of fully-developed incompressible
turbulent channel flow at bulk Reynolds number $Re_{b} = U_b \delta /
\nu = 2925$ has been analyzed , where $U_b$ is the bulk velocity, $\nu$ is the kinematic
viscosity and $\delta$ is the channel half-height.
For this case, the friction Reynolds number is
$Re_{\tau}=u_\tau\delta/\nu \approx 180$, where $u_\tau$ is the friction
velocity, and it has been previously simulated by many
authors~\citep{KMM:87,hoyas}. In the following, quantities are
normalized by $U_b$ and $\delta$, unless otherwise indicated. As is
customary, a superscript $+$ will indicated normalization in wall units;
that is, normalization by $u_\tau$ and $\nu$.

%Fig. \ref{fig:channel_geo} details the channel geometry. \todo{RDM: Do we really need this figure?} 
%This relatively low Reynolds number case is used to allow simulations
%to be performed with higher resolution than is normally used in DNS,
%to help characterize discretization errors, and to allow long
%simulation times to help characterize sampling error. The
%characterization of errors in this simple problem will provide
%insights relevant to DNS of wall bounded flows in general.
This relatively low Reynolds number case has been chosen to enable
testing of the methods developed here because it is computationally
tractable to simulate for times longer than usual, using higher
resolution than usual.  This allows the model predictions to be
tested against observed results, as shown in the small domain case.
Though the physical results are not scientifically new, the
characterization of the two error sources in DNS is novel.
This characterization will provide
insights relevant to DNS of wall-bounded flows in general.  All of the
data used in the Bayesian Richardson extrapolation, including computed
statistics and estimated sampling error, are available from
\url{http://turbulence.ices.utexas.edu}.

\subsection{Discretization and Sampling Details}\label{sec:channel_def}
The incompressible 3D Navier-Stokes equations are solved using the
formulation of \citet*{KMM:87} (KMM), as implemented in the code developed by
Lee {\it et al}\cite{SC13-lee}.
This formulation involves integrating
evolution equations for the wall-normal vorticity, $\omega_y$, and the
Laplacian of the vertical velocity, $\nabla^2 v$. Periodic boundary
conditions are imposed in the streamwise ($x$) and spanwise ($z$) directions,
while in the wall normal direction ($y$), no slip conditions are imposed at
the walls.  A semi-implicit, third-order Runge--Kutta/Crank--Nicholson
scheme is used for the time discretization \cite{spalart_etal_1991}.
The flow is driven by a uniform pressure gradient which is adjusted
continuously to maintain a constant mass flux. In space, a Fourier/Galerkin
method is used in the streamwise and spanwise
directions.  Unlike KMM, here a B-spline/collocation representation is used in the
wall-normal direction because it allows for flexible non-uniform grids 
while retaining spectral-like resolution\citep{Kwok2001Critical}. The
B-spline breakpoints $y_i$ for $i = 0, \ldots, N_b-1$ are set in the
interval $[-1,1]$ according to
\begin{align}
 y_i &=
\frac{ 
\sin \left(\frac{\alpha \pi}{2} \left[-1 + \frac{2 i}{N_b -1} \right] \right)
}{
\sin \left(\frac{\alpha \pi}{2} \right)
},
\end{align}
where $N_b$ is the number of breakpoints and $\alpha$ is a stretching
parameter, which is set to 0.985 for this study.  The Greville
abscissae, also called the Marsden--Schoenberg points, implied by these
breakpoints\citep{Johnson2005Higher,Botella2003Bspline} are used as the
collocation points, of which there are $N_y=N_b+p_{\rm bs}-1$, where
$p_{\rm bs}$ is the B-spline order (7 in the simulations reported here). To develop Richardson extrapolation estimates, a
nominal mesh resolution was defined, along with two uniform
de-refinements (by factors of approximately $\sqrt{2}$ and $2$), labeled
``coarse'' and ``coarsest.''
The nominal mesh was designed to conform to resolution
heuristics typically used in DNS of wall-bounded turbulence. That is, in
$x$- and $z$-directions, $\Delta x^+ \approx 13$, $\Delta
z^+\approx 7$, 
where $\Delta x = L_x/N_x$ and $\Delta_z=L_z/N_z$ with $N_x$ and $N_z$
being the number of Fourier modes in the representation in these
directions. In the $y$-direction, the nominal mesh is required to have
$\Delta y_{\rm wall}^+<1$ at the walls and $\Delta y_{\rm CL}^+\approx \Delta z^+$ at the channel
center, where $\Delta y$ is the spacing between the 
break points.
% I think we should be quoting break points

A constant timestep $\Delta t$ was used in the simulations reported here. This is
slightly different from typical DNS practice, in which variable
timesteps based on a Courant--Friedrichs--Lewy (CFL)
condition\citep{MR1512478} is used. Constant timesteps are used here to ensure
equidistant temporal samples, which simplifies the temporal analysis
required to estimate the sampling uncertainty.  The timestep size for
the nominal mesh was selected by monitoring the timestep in a CFL-based
variable time-step calculation and choosing a step smaller than the
smallest observed timestep.  Accordingly, these simulations are
somewhat better resolved in time than is common practice.

Two sets of DNS simulations were conducted: one using a relatively small
domain with $L_x=4\pi$ and $L_z=2\pi$; the other in a larger domain with
$L_x=12\pi$ and $L_z=4\pi$. The domain size of the latter is identical
to that of the $Re_\tau\approx180$ simulation reported by Hoyas \&
Jim\'{e}nez\citep{hoyas}. The small domain case was studied because the simulations
are less expensive, so that it was practical to perform a simulation
with a finer grid than nominal (by a factor of 2, called finest), to
allow validation of the Bayesian Richardson extrapolations. The large
domain was simulated because the error estimates for this case
will be relevant to the interpretation of the reference simulation
results of Hoyas \& Jim\'{e}nez. For each case, the simulation was run until
a statistically stationary state was reached, and then statistics were
collected over an evolution time $T$, with a sampling period of
$0.1L_x/U_b$ or 10 samples per flow-through. The numerical parameters for
each simulation are given in Table~\ref{tbl:channel_runs}

% nick: fixed the table
% RDM: The $\Delta x$ and $\Delta z$ values are incorrect in the table.

\begin{table}[htbp]
\setlength{\tabcolsep}{5pt}
\renewcommand{\arraystretch}{1.2}
\caption{Numerical and sampling parameters for turbulent channel
 flow simulations conducted at $Re_b=2925$,
 $Re_\tau\approx180$. Variables are as defined in
 section~\ref{sec:channel_def}.}

\begin{tabular}{|l|rr|rrr|rccc|rl|}
\hline
Name & $L_x$ & $L_z$ & $N_x$ & $N_z$ & $N_y$ & $\Delta x^+$ & $\Delta z^+$ & $\Delta
 y^+_{\rm wall}$ & $\Delta y^+_{\rm CL}$ & $T U_b / L_x$ & $\Delta t U_b / \delta$ \\
\hline
%
% e.g. (2 * math.pi )/136. * 186
% 
\multicolumn{12}{|l|}{{\bf Small Domain}}\\
\hline
Coarsest  & $4\pi$  &  $2\pi$  &   96 &  96 &  64 & 24.3 & 12.2 & 0.44 & 9.14 & 2651.0 &0.02\\ 
Coarse    & $4\pi$  &  $2\pi$  &  136 & 136 &  90 & 17.2 &  8.6 & 0.26 & 6.46 &  273.5 &0.01414\\ 
Nominal   & $4\pi$  &  $2\pi$  &  192 & 192 & 128 & 12.2 &  6.1 & 0.16 & 4.53 & 2145.3 &0.01\\ 
% Never use fine mesh result, so leave him out.
%Fine      & $4\pi$  &  $2\pi$  &  268 & 268 & 180 &  5.8 & 2.9 & 0.10 & 3.22 &  337.0 &  3370 &0.0707\\ 
Finest    & $4\pi$  &  $2\pi$  &  384 & 384 & 256 &  6.1 & 3.0 & 0.07 & 2.26 &  709.3 &0.005\\
\hline
\multicolumn{12}{|l|}{{\bf Large Domain}}\\
\hline
Coarsest  &$12\pi$  &  $4\pi$  &  256 & 192 &  64 & 27.4 & 12.2 & 0.44 & 9.14 & 40.0 & 0.02\\
Coarse    &$12\pi$  &  $4\pi$  &  362 & 270 &  90 & 19.4 &  8.7 & 0.26 & 6.46 & 30.0 & 0.01414\\
Nominal   &$12\pi$  &  $4\pi$  &  512 & 384 & 128 & 13.7 &  6.1 & 0.16 & 4.53 & 20.0 & 0.01\\
\hline

\end{tabular}
\label{tbl:channel_runs}
\end{table}

\subsection{Small Domain Results} \label{sec:channel_results_small}
The Bayesian Richardson extrapolation procedure has been applied to a
variety of statistical quantities of particular interest in the
channel flow.  For brevity, we show full results, including details of
the joint posterior PDF for the true value $\avg{q}$, the
discretization error constant $C$, and the order of accuracy $p$, for
only two scalars: the centerline mean velocity and the skin friction
coefficient.  A summary of results for single-point statistics
including the mean velocity, the Reynolds stresses, and the vorticity
correlations at multiple points across the channel is also given.  In
all cases, the inverse problem is formulated using data from the
coarsest, coarse, and nominal mesh resolutions.  Data from the finest
mesh is reserved to provide a validation test of the procedure.

\subsubsection{Centerline Mean Velocity} \label{sec:small_box_cl_results}
The results of Bayesian Richardson extrapolation applied to the
centerline mean velocity $U_{CL}$, normalized by the bulk
velocity $U_b$, are presented here.  First we examine the results of
the inverse problem for the discretization error model.  Then, we test
the calibrated model by using the model to predict the value that
should be observed on the finest mesh.  Finally, we use the model to
examine the discretization error on the nominal mesh.

\begin{figure}[thp]
\begin{center}
\includegraphics[width=0.7\linewidth]{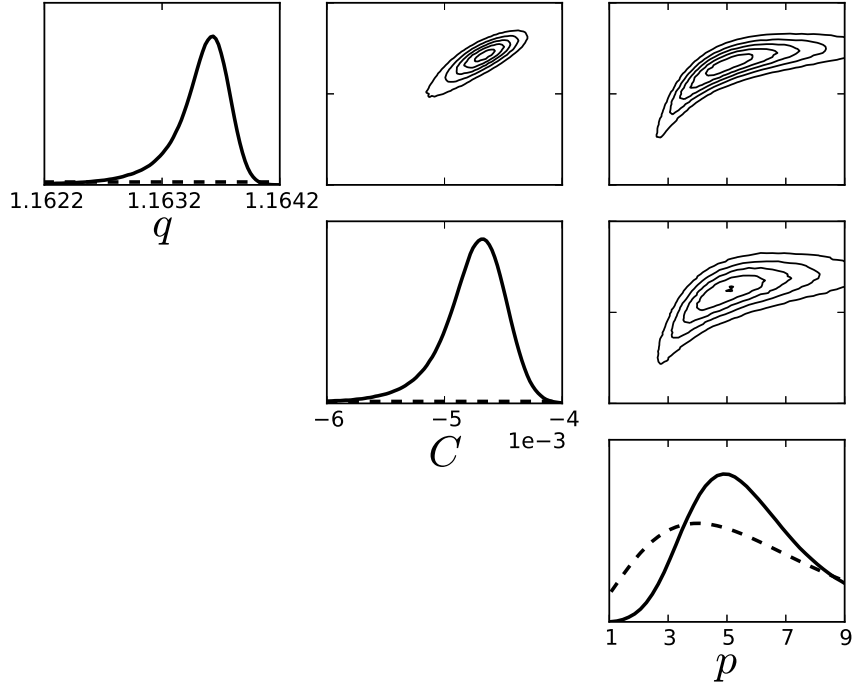}
\end{center}
\caption{Results of the inverse problem for $q=U_{CL}$, $C$, and $p$.  The
input data are $U_{CL}$ data from the coarsest, coarse, and nominal
meshes.  The diagonal shows the marginal PDFs for each of the
parameters while the off-diagonal entries show samples of the joint
posterior PDF projected onto planes in parameter space.}
\label{fig:cl_chp_joint_post}
\end{figure}

Figure~\ref{fig:cl_chp_joint_post} shows the posterior PDF for the
calibration parameters.
The posterior PDF for $p$ is maximum near $p=5$.  While this does
not correspond directly to any of the schemes used here, there is
large uncertainty about the order, with the first and third quartiles
of the marginal distribution for $p$ at approximately $4.4$ and $7.4$,
respectively.  Despite the large uncertainty about the order of
accuracy, the uncertainty regarding the true value is quite small.
For instance, the difference between the $5$th and $95$th percentiles
is less than $0.08\%$ of the mean value.

Given the samples from the posterior PDF represented in
Figure~\ref{fig:cl_chp_joint_post}, one can use the calibrated model
to make predictions of the value of the average centerline velocity that
should be observed for
any value of the resolution parameter $h$, by evaluating
%\todo{We are not consistent with nomenclature, vis $E[q]$ and $<q>$}
%
\begin{equation}
\avg{q_h}_N = E[q] - C_0 h^p - e_{h,N}.
\label{eqn:val_pred_model}
\end{equation}
Here $e_{h,N}$ is the sampling uncertainty for the simulation from
which the observed average velocity is obtained. Thus the distribution
for $\avg{q_h}_N$ obtained from~\eqref{eqn:val_pred_model} includes
uncertainty from two sources.  First, the calibrated discretization
error model parameters ($E[q]$, $C_0$ and $p$) are uncertain.  Second,
the observation is contaminated by sampling error. The consistency of
the model with the actual observation can be assessed by simply
examining whether the observed value is a plausible draw from the
prediction distribution generated according
to~\eqref{eqn:val_pred_model}. If the observed value is highly
unlikely according to the prediction, the model is declared invalid.

\begin{figure}[thp]
\begin{center}
\includegraphics[width=0.7\linewidth]{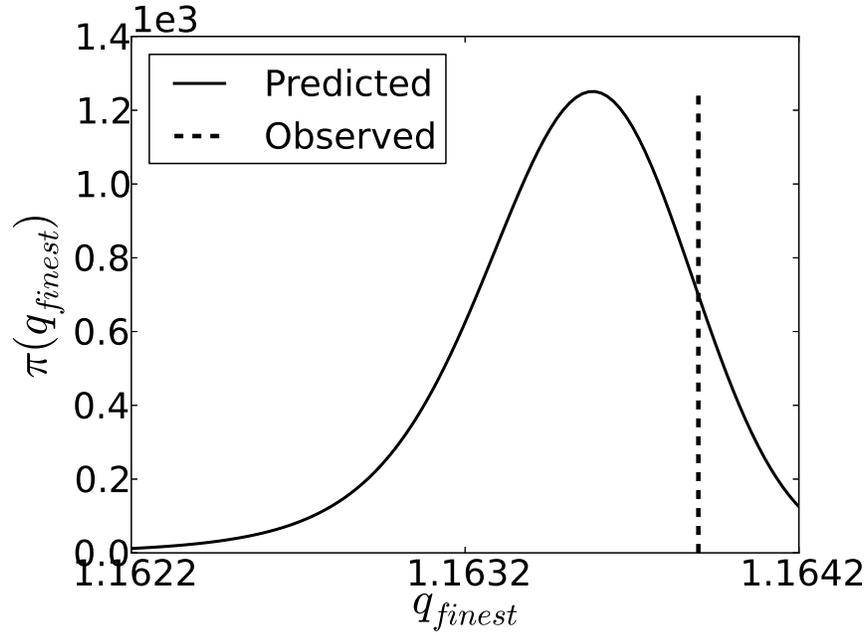}
\end{center}
\caption{
PDF of the mean centerline velocity for the finest mesh predicted
according to~\eqref{eqn:val_pred_model} (blue) and the observed mean
centerline velocity on the finest mesh (green).}
\label{fig:cl_chp_validation}
\end{figure}

Results of this validation check for the centerline velocity are shown in
Figure~\ref{fig:cl_chp_validation}.
Clearly, the observed value is not near the tail of the prediction
distribution, indicating that there is no reason to believe the model is
invalid.

\begin{figure}[thp]
\begin{center}
\includegraphics[width=0.7\linewidth]{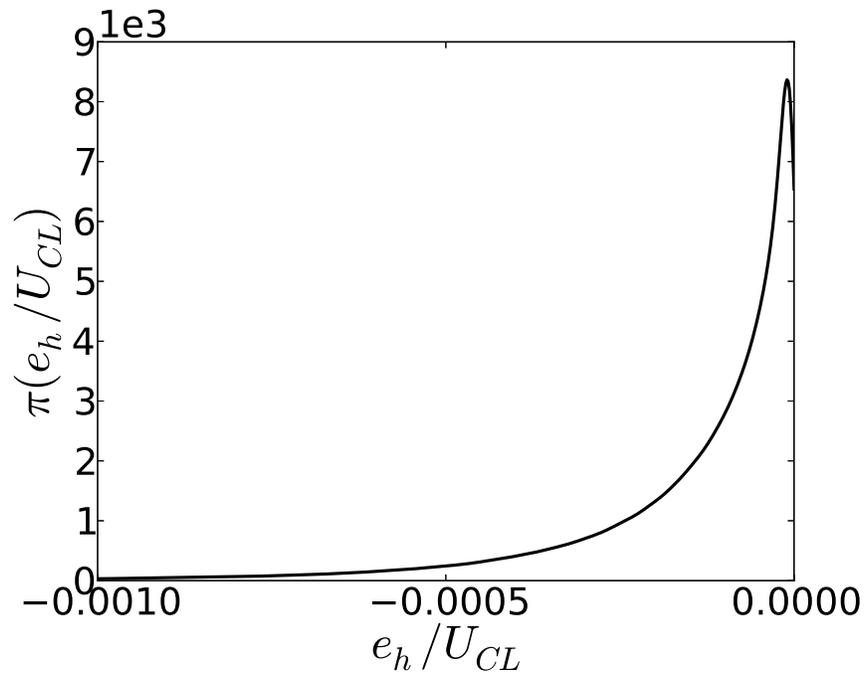}
\end{center}
\caption{
Discretization error, as computed by the calibrated model, for the
centerline mean velocity on the nominal mesh.  }
\label{fig:cl_chp_disc_error}
\end{figure}

Finally, Figure~\ref{fig:cl_chp_disc_error} shows the estimated
discretization error on the nominal mesh, normalized by the observed
mean value.
Note that the discretization error is very small.  Essentially all of
the probability is assigned to values of less than $0.1\%$, and half
is assigned to values less than $0.008\%$.  For comparison, the
standard deviation of the sampling error was estimated as $0.011\%$
for this mesh.  Thus, even after more than 2000 flow-throughs, sampling
uncertainty is still significant for this quantity.

Note that the discretization error distribution in
Figure~\ref{fig:cl_chp_disc_error}
has an odd shape, with high
probability assigned to negative values very close to zero, but essentially zero
probability to positive values. Similar distributions are 
observed in the results shown in subsequent sections
for other quantities as well. This feature can be understood by
examining the posterior distribution shown in
Figure~\ref{fig:cl_chp_joint_post}.  Specifically, the value of $C$ is
bounded away from zero.  Since $\epsilon_h = Ch^p$, $C$ is the only
parameter that can change the sign of $\epsilon_h$.  Thus, since it is
bounded away from zero, the the model is completely sure of the sign
of the discretization error.  Also, this result for $C$ is entirely
consistent with monotonic data with sampling uncertainty that is small
relative to the changes observed between different resolution
simulations.  In this case, one should be able to determine the sign
of the discretization error with very high confidence.

This explains why the discretization error tends to have all its
probability on one side of zero, but we also observe that the
probability density is highest near zero.  This feature results from
the fact that $p$ is not well-informed.  In particular, large values
of $p$, which lead to small $\epsilon_h$ are not ruled out by the
data.  Since increasingly larger values of $p$ lead to increasingly
smaller values of $\epsilon$, the probability clusters near zero.

\subsubsection{Skin Friction} \label{sec:small_box_cf_results}
\begin{figure}[thp]
\begin{center}
\includegraphics[width=0.7\linewidth]{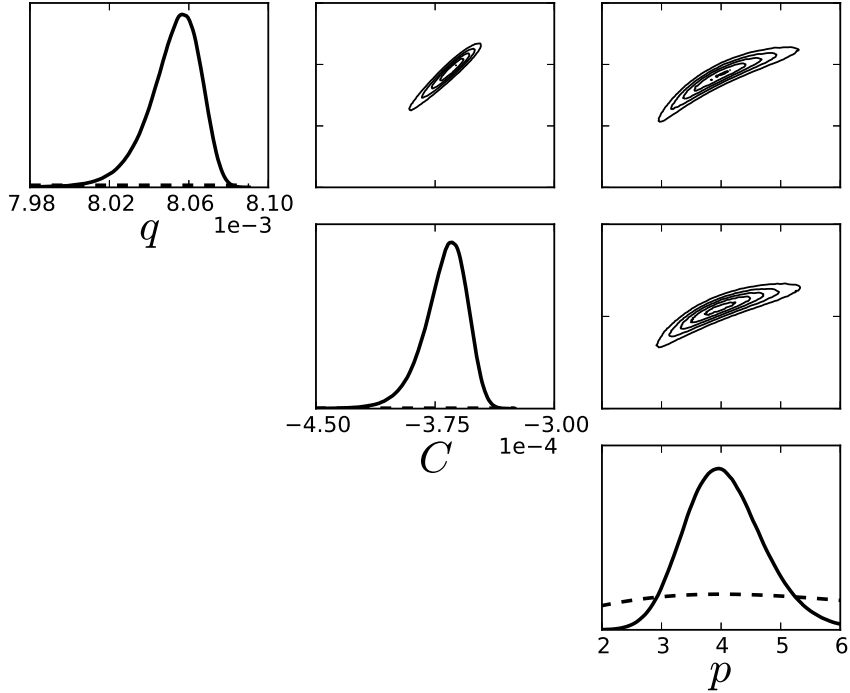}
\end{center}
\caption{Results of the inverse problem for $q=C_f= 2 \tau_w / \rho U_b^2$, $C$, and $p$.  The
input data are $U_{CL}$ data from the coarsest, coarse, and nominal
meshes.  The diagonal shows the marginal PDFs for each of the
parameters while the off-diagonal entries show samples of the joint
posterior PDF projected onto planes in parameter space.}
\label{fig:tau_chp_joint_post}
\end{figure}

Results for the skin friction coefficient are analyzed here in a
series of figures analogous to those shown for the centerline mean
velocity.  To begin, Figure~\ref{fig:tau_chp_joint_post} shows the
joint posterior PDF for the parameters of the discretization error
model.
While the order of accuracy appears somewhat better informed than for
the centerline velocity, there is still significant uncertainty, with
the $5$th and $95$th percentiles at 3.06 and 5.28, respectively.
However, the marginal posterior for the true value of $C_f$ is again
quite narrow, with the difference between the $5$th and $95$th
percentiles being only $0.54\%$ of the mean value.

\begin{figure}[thp]
\begin{center}
\includegraphics[width=0.7\linewidth]{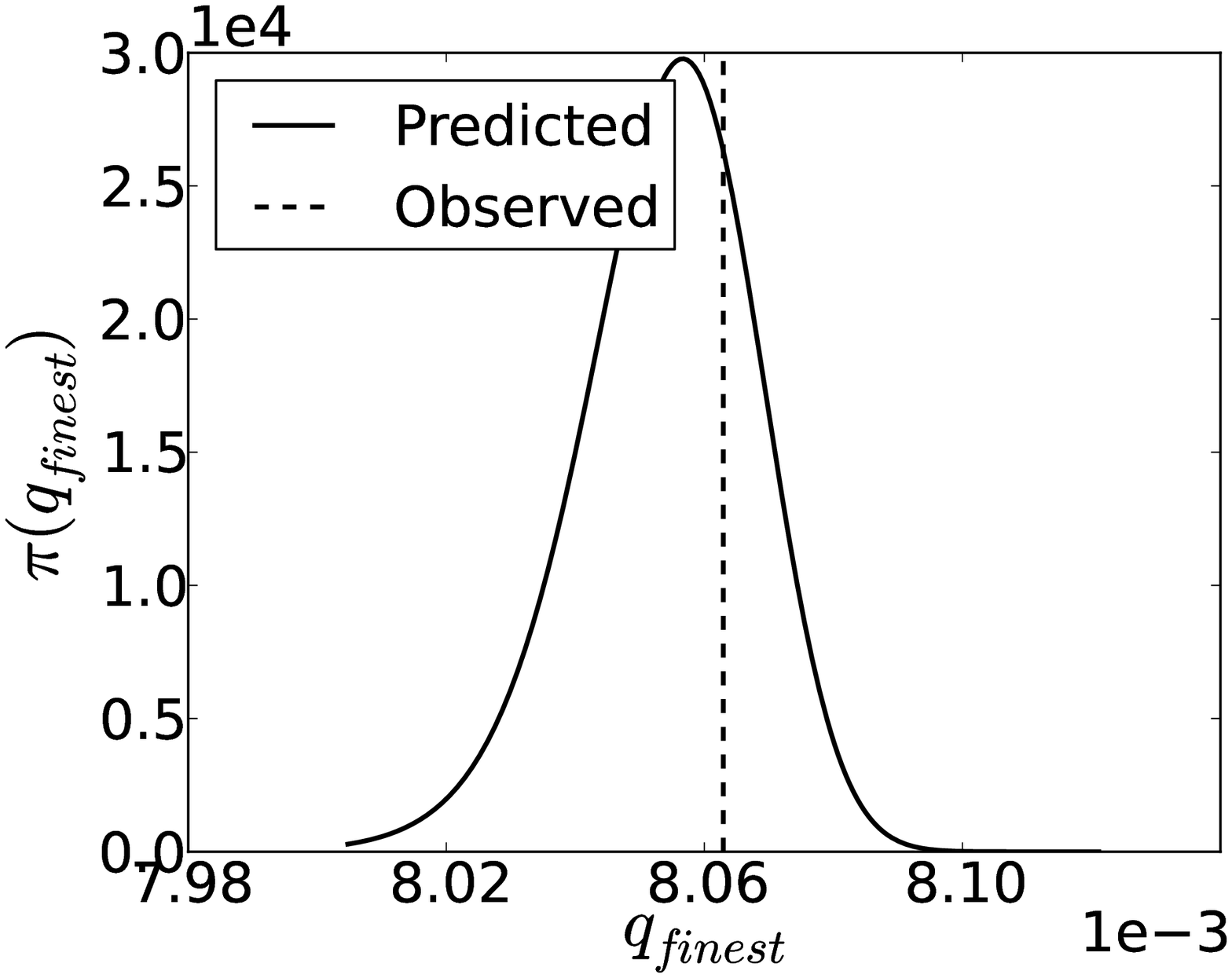}
\end{center}
\caption{
PDF of the mean skin friction coefficient for the finest mesh predicted
according to~\eqref{eqn:val_pred_model} (blue) and the observed mean
centerline velocity on the finest mesh (green).}
\label{fig:tau_chp_validation}
\end{figure}

Figure~\ref{fig:tau_chp_validation} compares the model prediction of
the skin friction on the finest mesh, computed from the calibrated
results and estimated sampling uncertainty in the finest mesh result
using~\eqref{eqn:val_pred_model}, and the observed results.
Clearly there is good agreement between the prediction PDF and the
observation.  As with the centerline velocity, there is no reason to
question the discretization error model in this case.

\begin{figure}[thp]
\begin{center}
\includegraphics[width=0.7\linewidth]{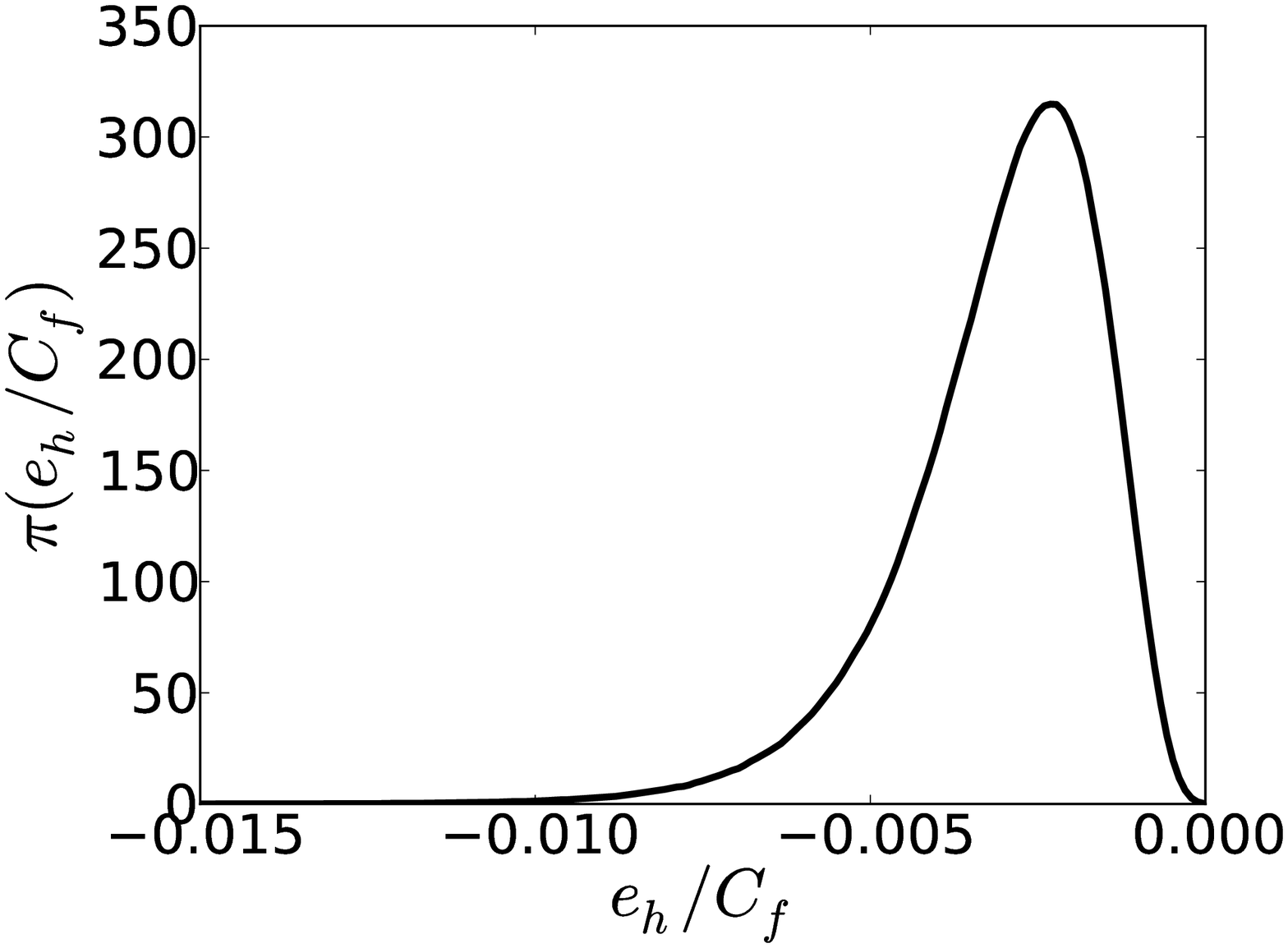}
\end{center}
\caption{
Discretization error, as computed by the calibrated model, for the
skin friction on the nominal mesh.  }
\label{fig:tau_chp_disc_error}
\end{figure}
Finally, the estimated discretization error on the nominal mesh is
shown in Figure~\ref{fig:tau_chp_disc_error}.
As with the centerline velocity, the discretization error is quite
small.  The mean discretization error is only $0.3\%$ of the mean
value.  Unlike the centerline velocity, the discretization error is
large relative to the estimated sampling error standard deviation,
which is less than $0.05\%$.

\subsubsection{Summary of Results for Single-Point Statistics}
\label{sec:single-pt}

Uncertainties in a number of single-point statistics that are generally
of interest in DNS are presented here, including the mean velocity,
Reynolds stresses, and vorticity variances, as functions of the
wall-normal location.  As shown for the skin friction and centerline
velocity, the first step after performing the Bayesian update to
calibrate the discretization error model is to assess the predictions of
the model relative to the finest mesh results.  Here, this assessment is
performed by evaluating the cumulative distribution function (CDF)
corresponding to the prediction for the finest mesh value, as given
by~\eqref{eqn:val_pred_model}), at the observed result for the finest
mesh.  This value is important because, if it is close to zero or close
to one, then the observed value corresponds to a draw from one of the
tails of the prediction distribution.

\begin{figure}[htp]
\centering
\begin{tabular}{ccc}
\subfloat[$\avg{u}$]{\includegraphics[width=0.32\linewidth]{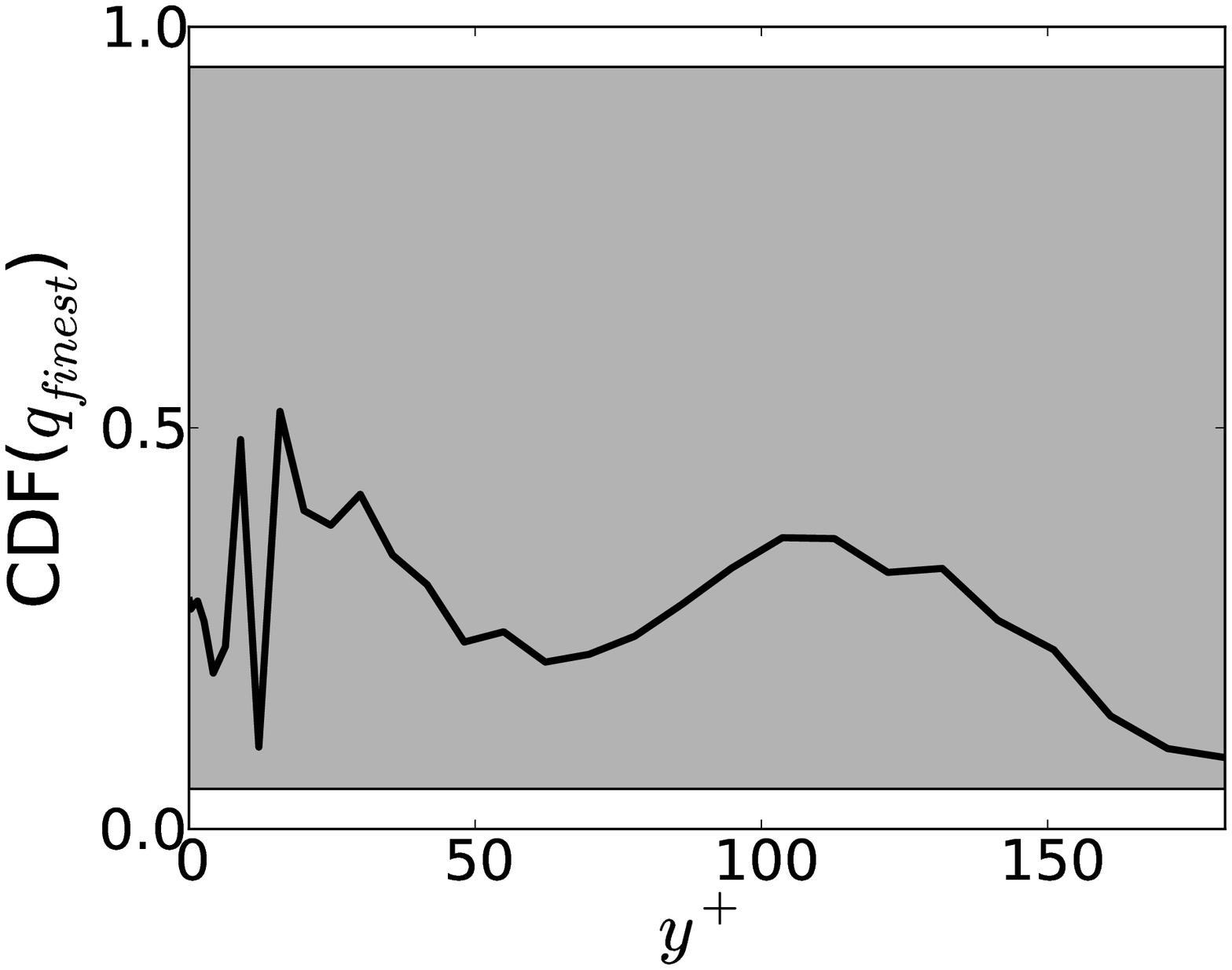}} &
\subfloat[$\nu d\avg{u}/dy$]{\includegraphics[width=0.32\linewidth]{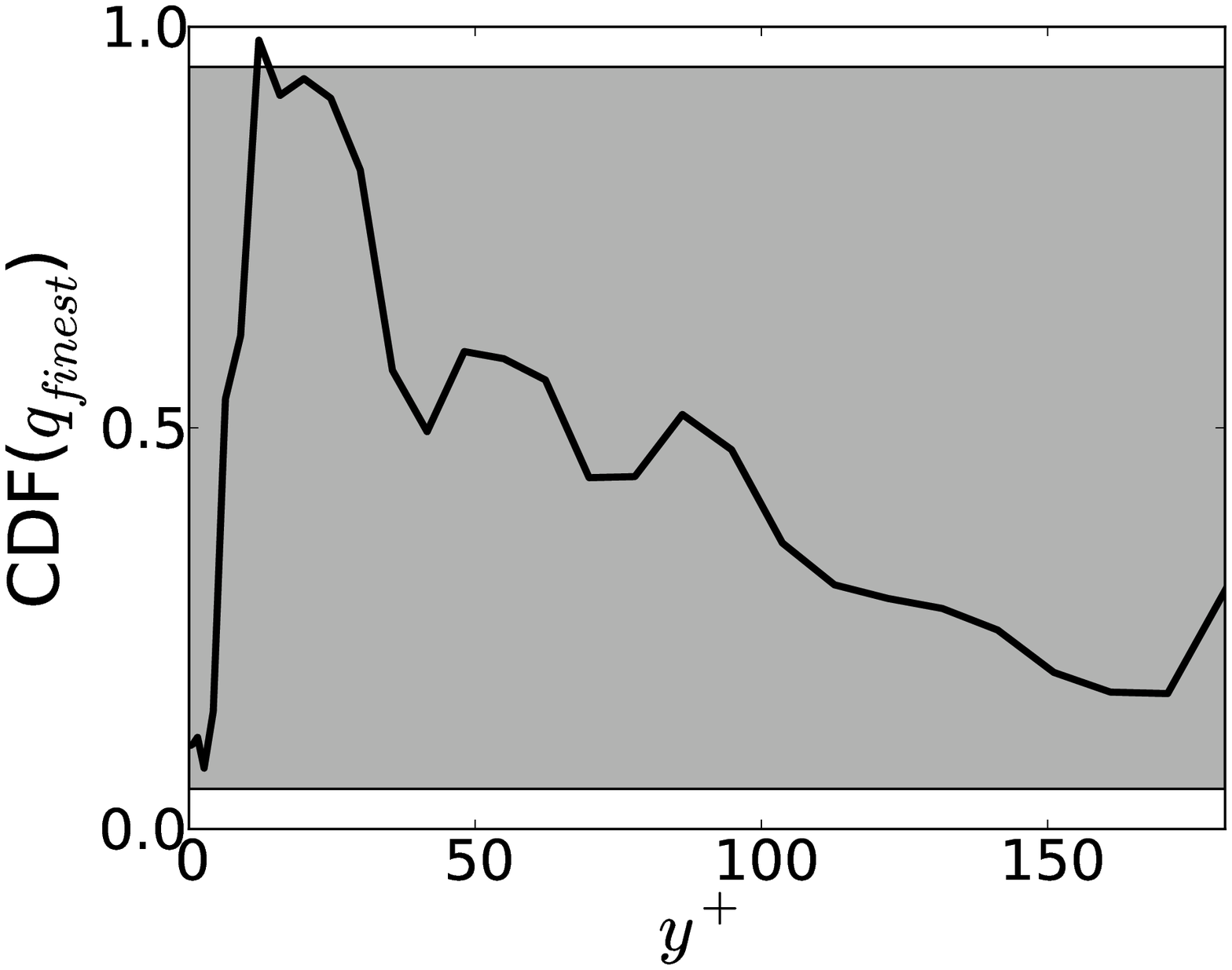}} &
\subfloat[$\avg{u'v'}$]{\includegraphics[width=0.32\linewidth]{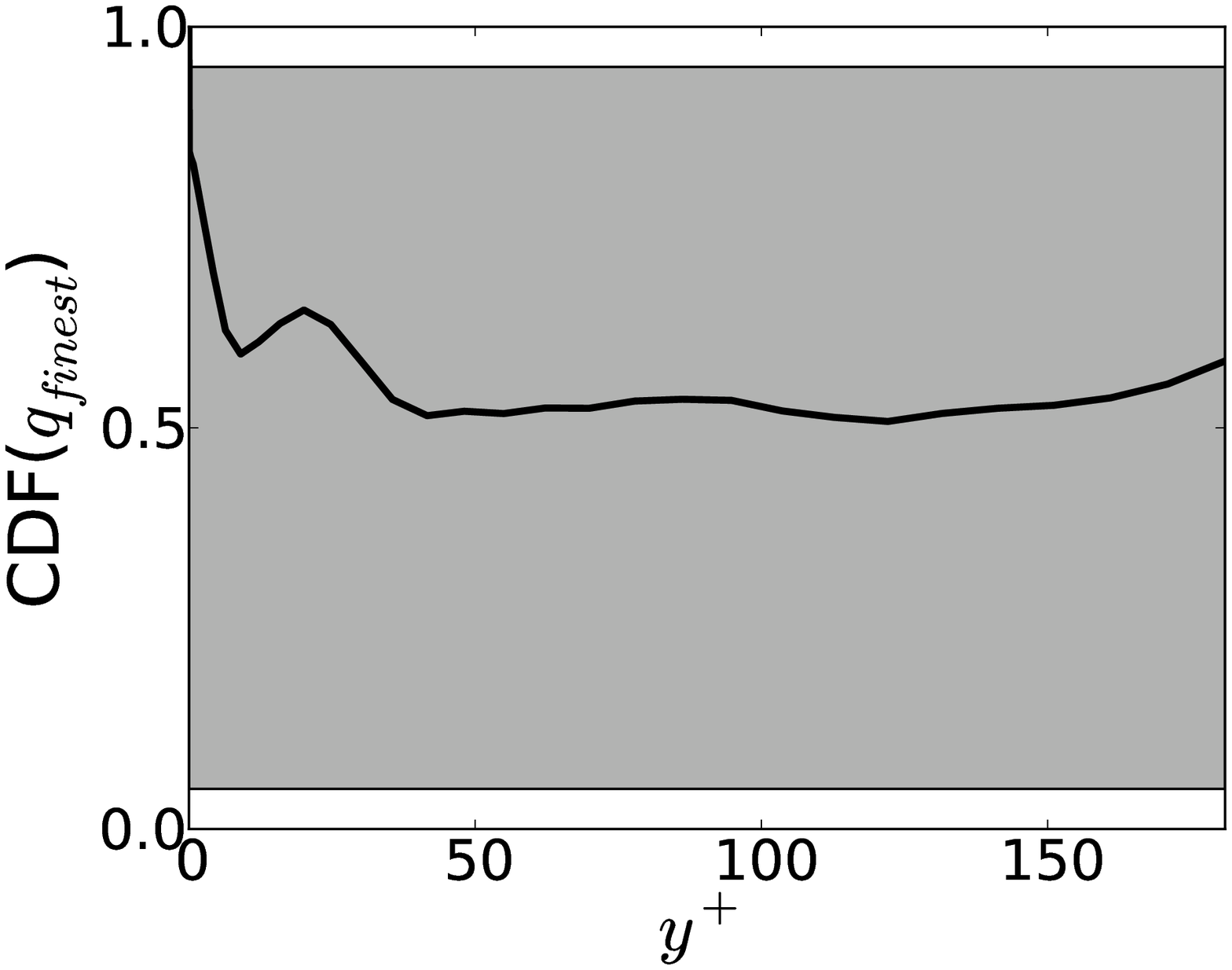}} \\
\subfloat[$\avg{u'u'}$]{\includegraphics[width=0.32\linewidth]{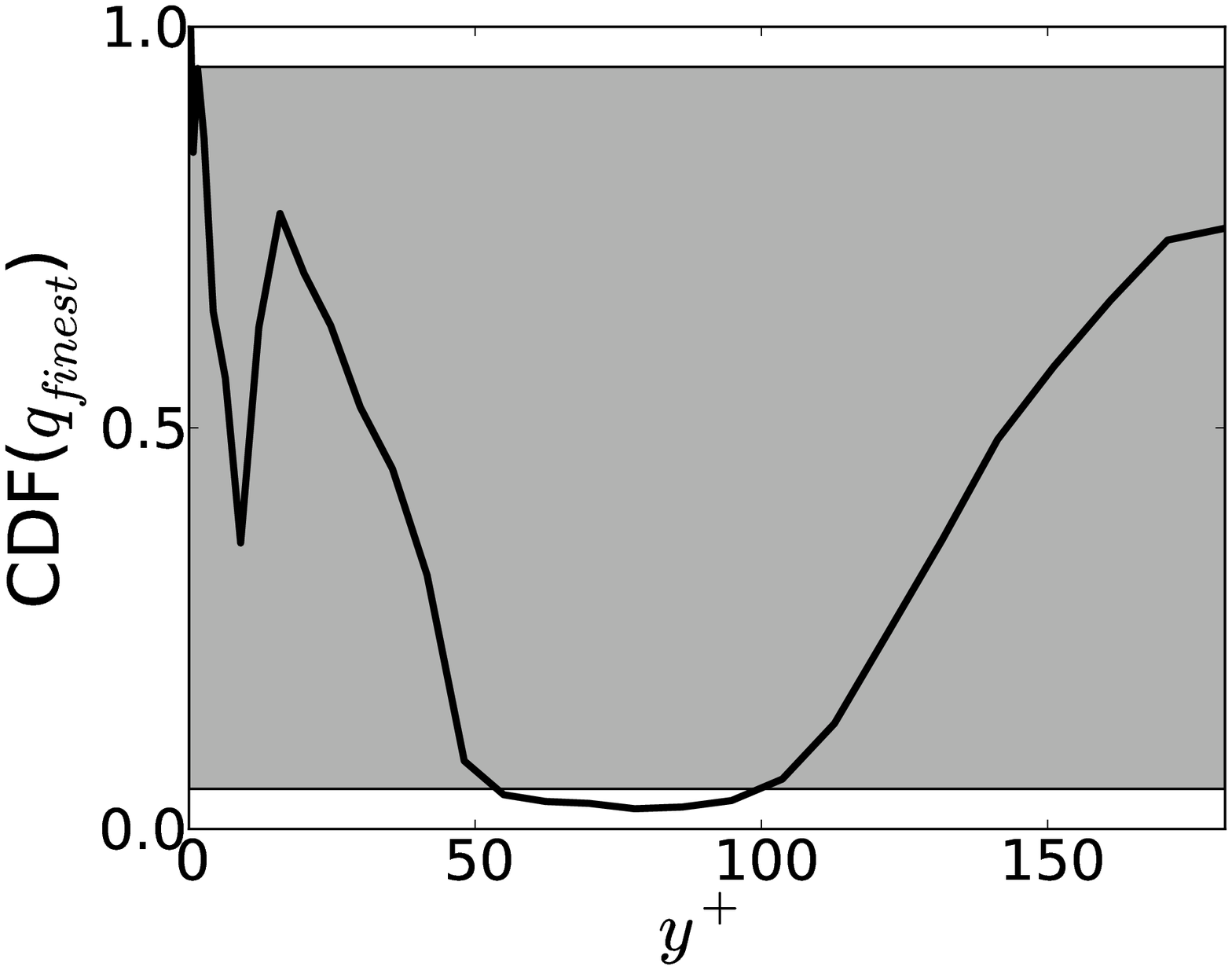}} &
\subfloat[$\avg{v'v'}$]{\includegraphics[width=0.32\linewidth]{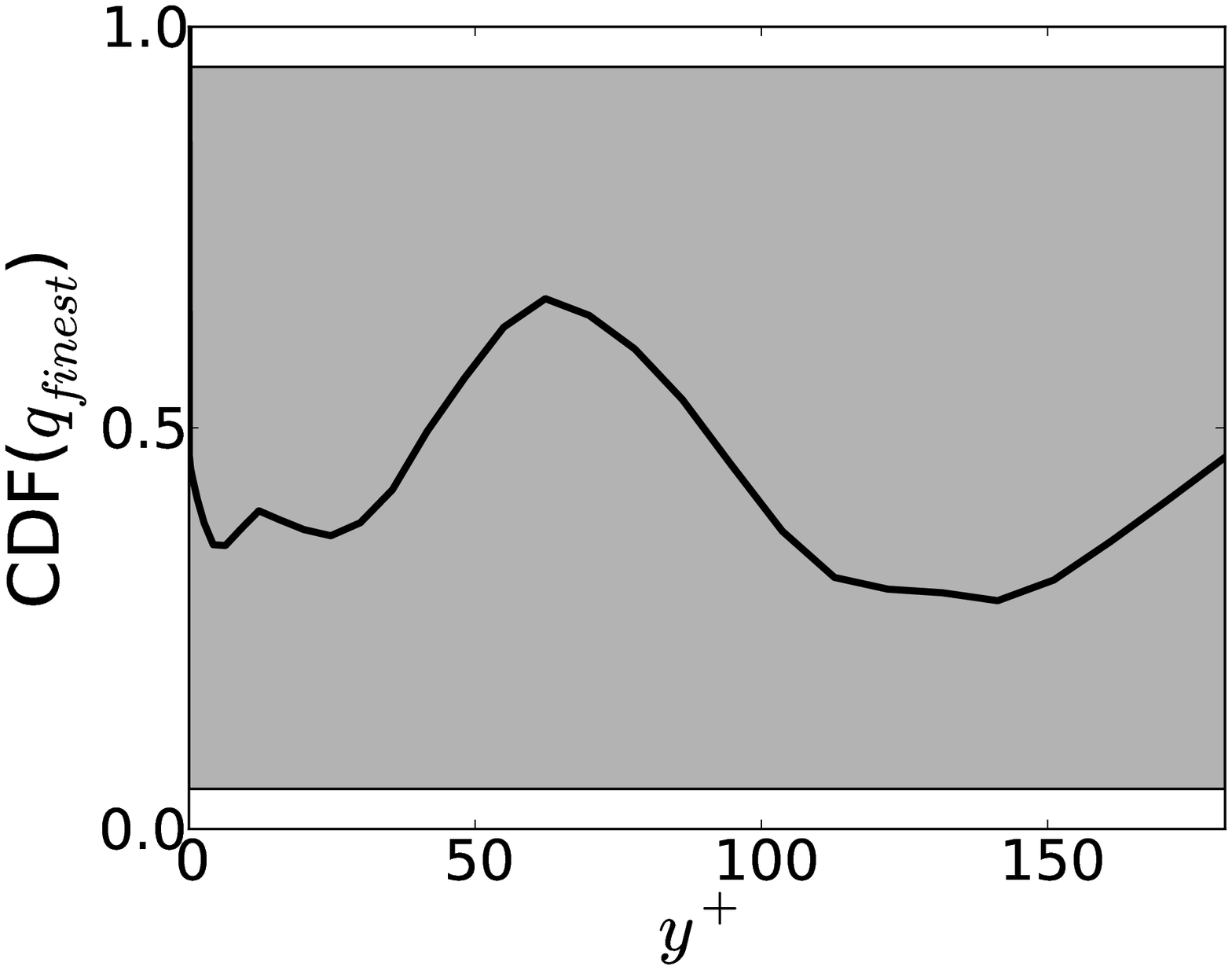}} &
\subfloat[$\avg{w'w'}$]{\includegraphics[width=0.32\linewidth]{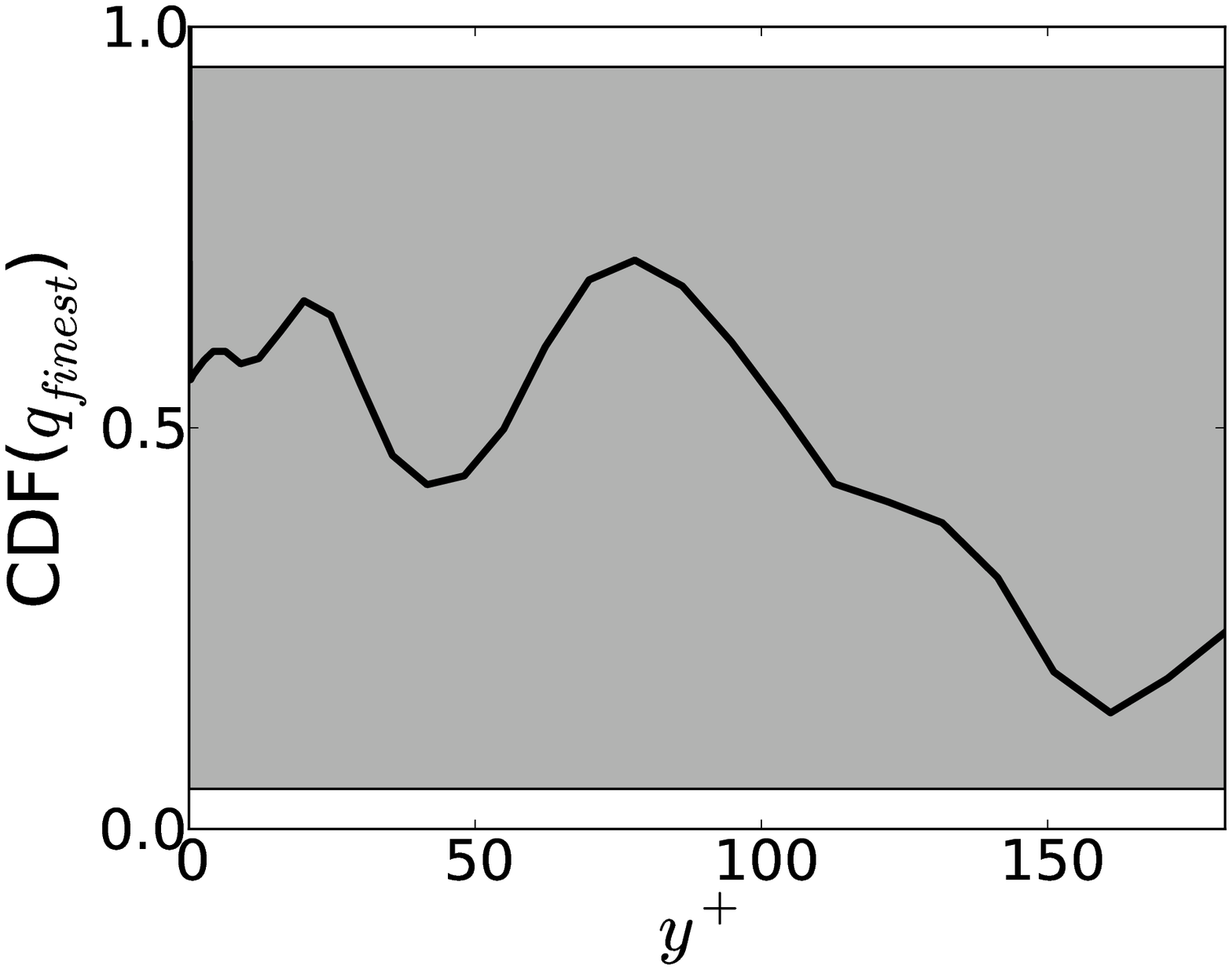}} \\
\subfloat[$\avg{\omega_x' \omega_x'}$]{\includegraphics[width=0.32\linewidth]{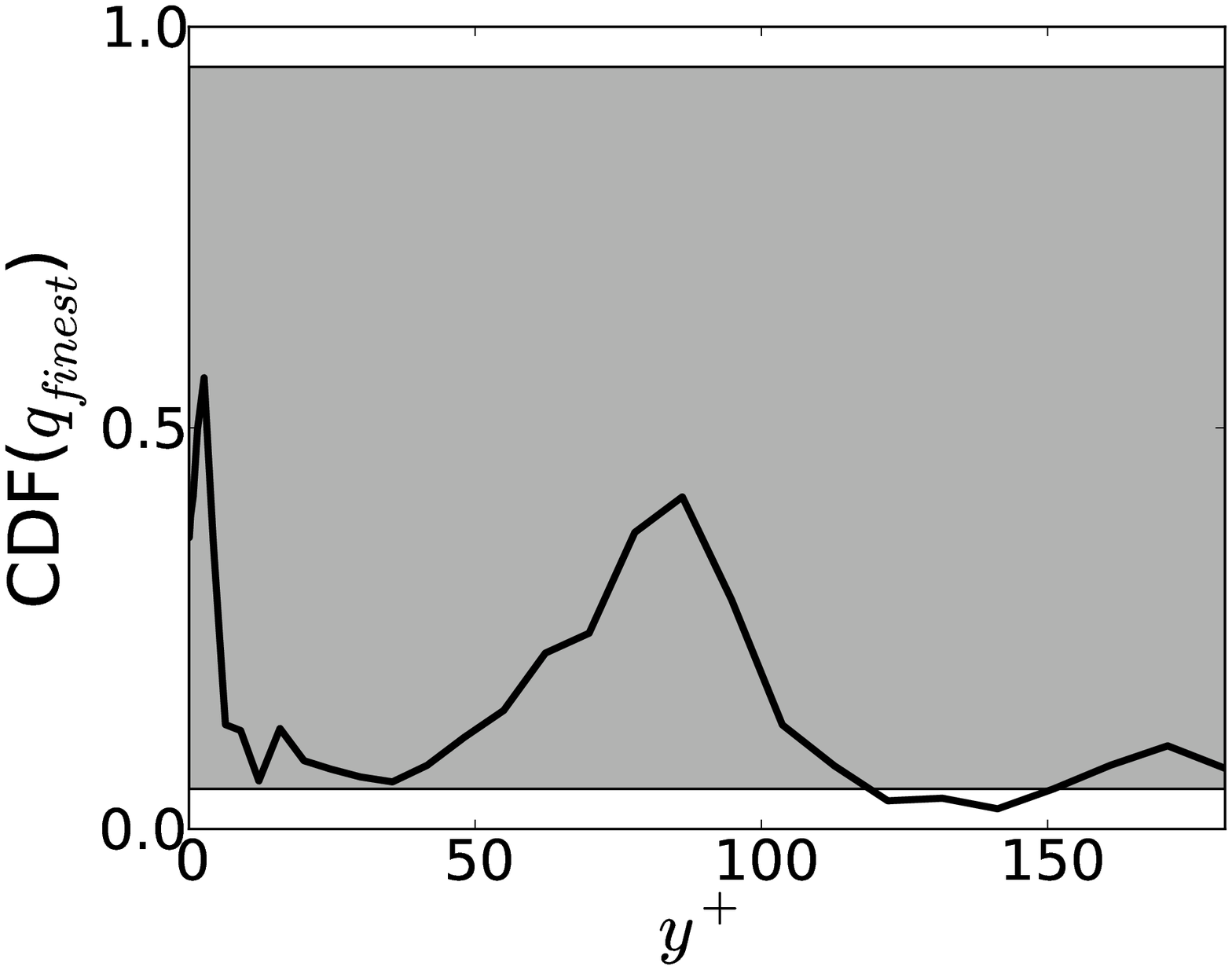}} &
\subfloat[$\avg{\omega_y' \omega_y'}$]{\includegraphics[width=0.32\linewidth]{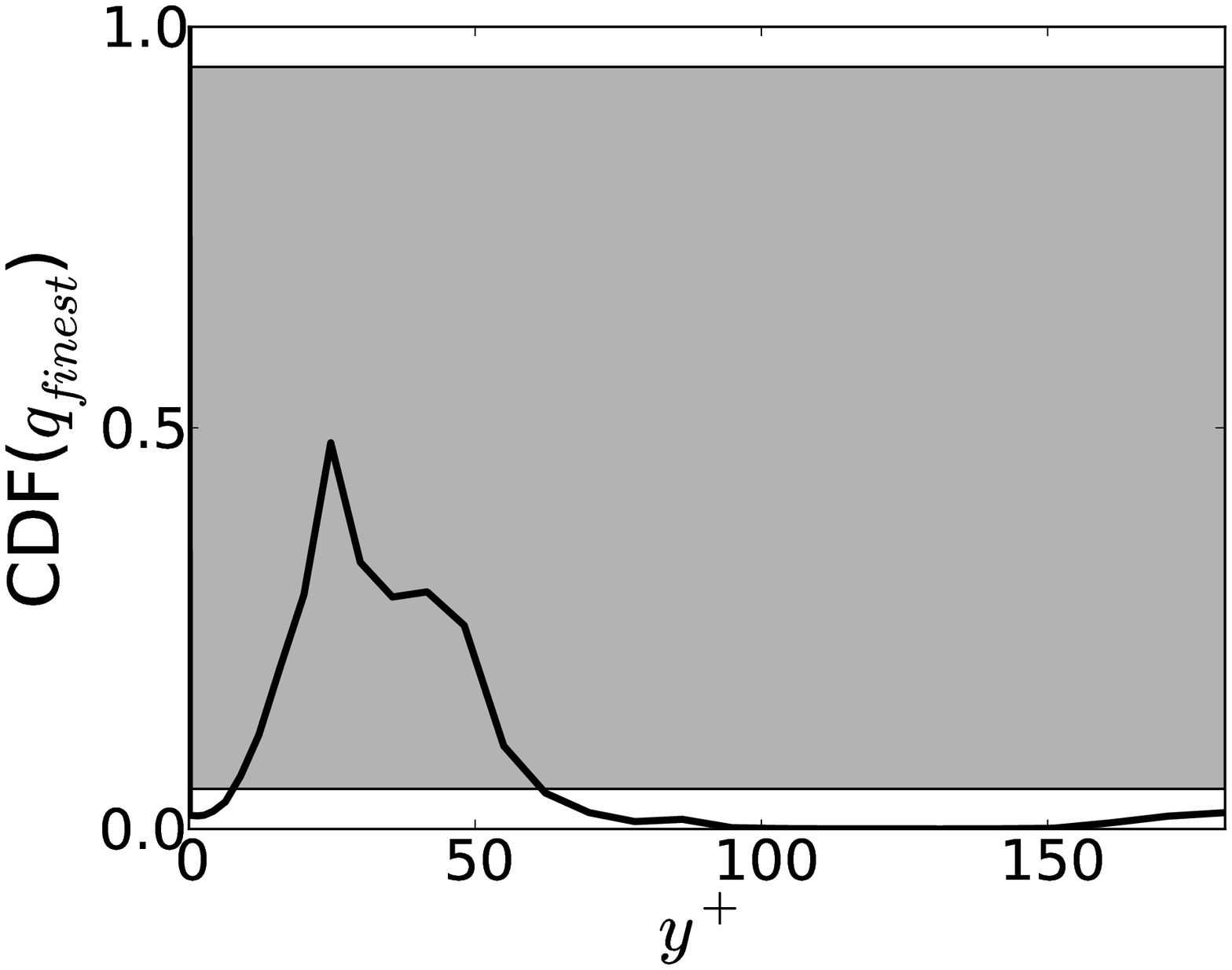}} &
\subfloat[$\avg{\omega_z' \omega_z'}$]{\includegraphics[width=0.32\linewidth]{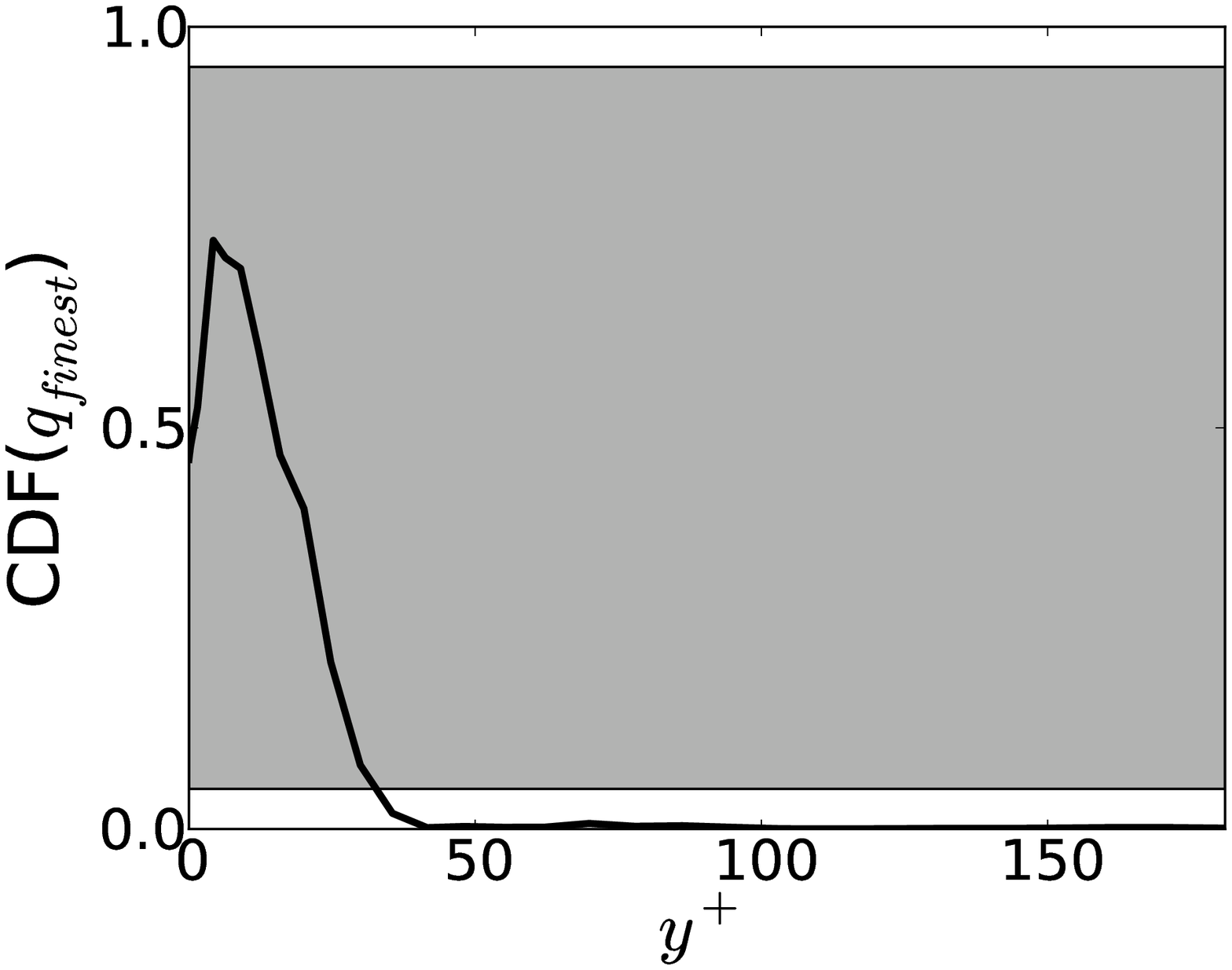}} \\
\end{tabular}
\caption{Comparison of median predicted change between the nominal and finest
mesh results (dashed line) and the observed change (solid line), both
with $90\%$ credibility intervals.}
\label{fig:validation}
\end{figure}

The results are presented in Figures~\ref{fig:validation}.  In each
figure, the solid line is the computed value of the CDF at the
observed value.  The grey shows the region between $0.05$ and $0.95$,
which is the $90\%$ credibility interval.  When the observed
results give a CDF value that falls outside of this region, the model
and the observation are in poor agreement.  In this case, we cannot
have confidence in the model, and it is declared invalid for our
purposes.  When the values are in the grey region, the model passes
this validation check.

For the mean velocity, viscous stress, Reynolds shear stress,
wall-normal velocity variance, and spanwise velocity variance, there
is reasonable to excellent agreement between the model predictions and
the observations.  For these quantities, the model is not invalidated
by this assessment.  Alternatively, for the streamwise velocity
variance and the vorticity variances, there are large regions of the
channel where the observed value falls outside of the $90\%$
credibility interval.  For example, examining $\avg{u'u'}$, for
$50 \lesssim y^+ \lesssim 100$, the percentile of the observed value
is less then $5\%$, meaning that the model assigns probability greater
than $0.95$ to values larger than the observed value.  This level of
disagreement means that the model cannot be used with confidence.  The
model for the vorticity variances also appears to be invalid based on
this assessment.

A closer examination of the $\avg{u'u'}$ data at these $y^+$
locations shows the problem.  The results are not converging
monotonically with increasing mesh resolution. For example, at
$y^+ \approx 62$, the value of $\avg{u'u'}/U_b^2$ on the coarsest,
coarse and nominal meshes was 0.010343, 0.010042, and 0.00997,
respectively. However, the value observed on the finest mesh was
0.010010, an increase in magnitude compared to the nominal mesh.  This
non-monotonic behavior cannot be captured by the simple model used
here.  Further, even if a model capable of producing non-monotonic
convergence were used, it would be unlikely to produce an accurate
prediction given that the calibration data (i.e., the three coarser
mesh results) are monotonic.  The vorticity variance data also show
non-monotonic behavior with increasing resolution.

Regardless, it is clear that the simple model used here is
insufficient for some quantities.  Given that the complete
discretization is a mix of spectral, high-order B-spline, and 2nd and
3rd order time marching schemes, it is not necessarily surprising that
the convergence behavior is complex, and it is clear that none of the
results are consistent with the final asymptotic behavior of the
scheme, which must be 2nd order due to the temporal discretization of
the viscous terms.  More importantly, the invalidity of the model
does not imply that the errors are large.  For example, the change
between the nominal and finest mesh results for $\avg{u'u'}/U_b^2$
is less than $0.5\%$.  However, for quantities where the model is
invalid, we clearly cannot use it to make reliable statements about
the discretization error.  For this reason, no additional results are
shown for the streamwise velocity variance or vorticity variances.

\begin{figure}
\centering
\begin{tabular}{cc}
\subfloat[$\avg{u}$]{\includegraphics[width=0.49\linewidth]{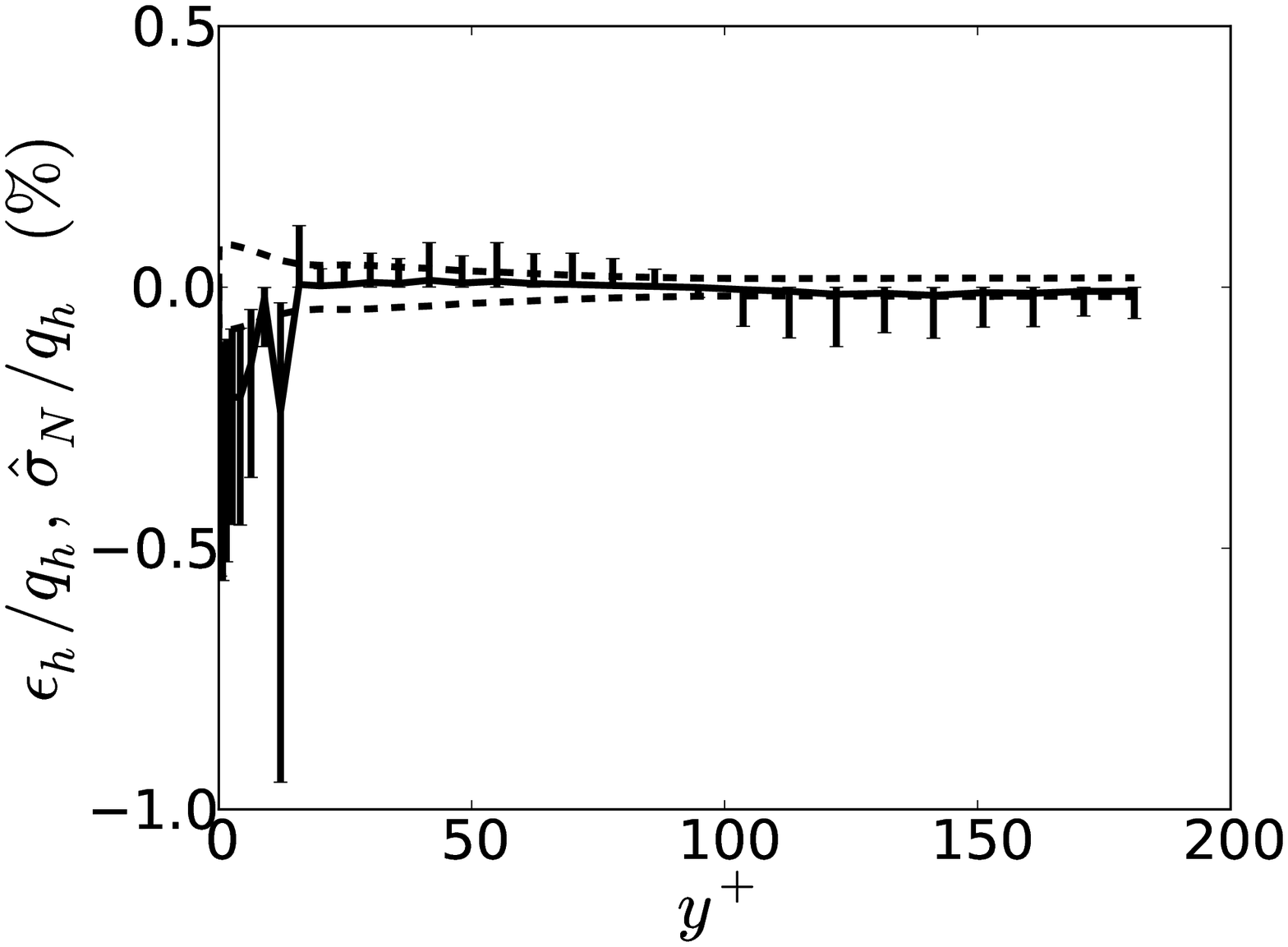}} &
\subfloat[$\nu d\avg{u}/dy$]{\includegraphics[width=0.49\linewidth]{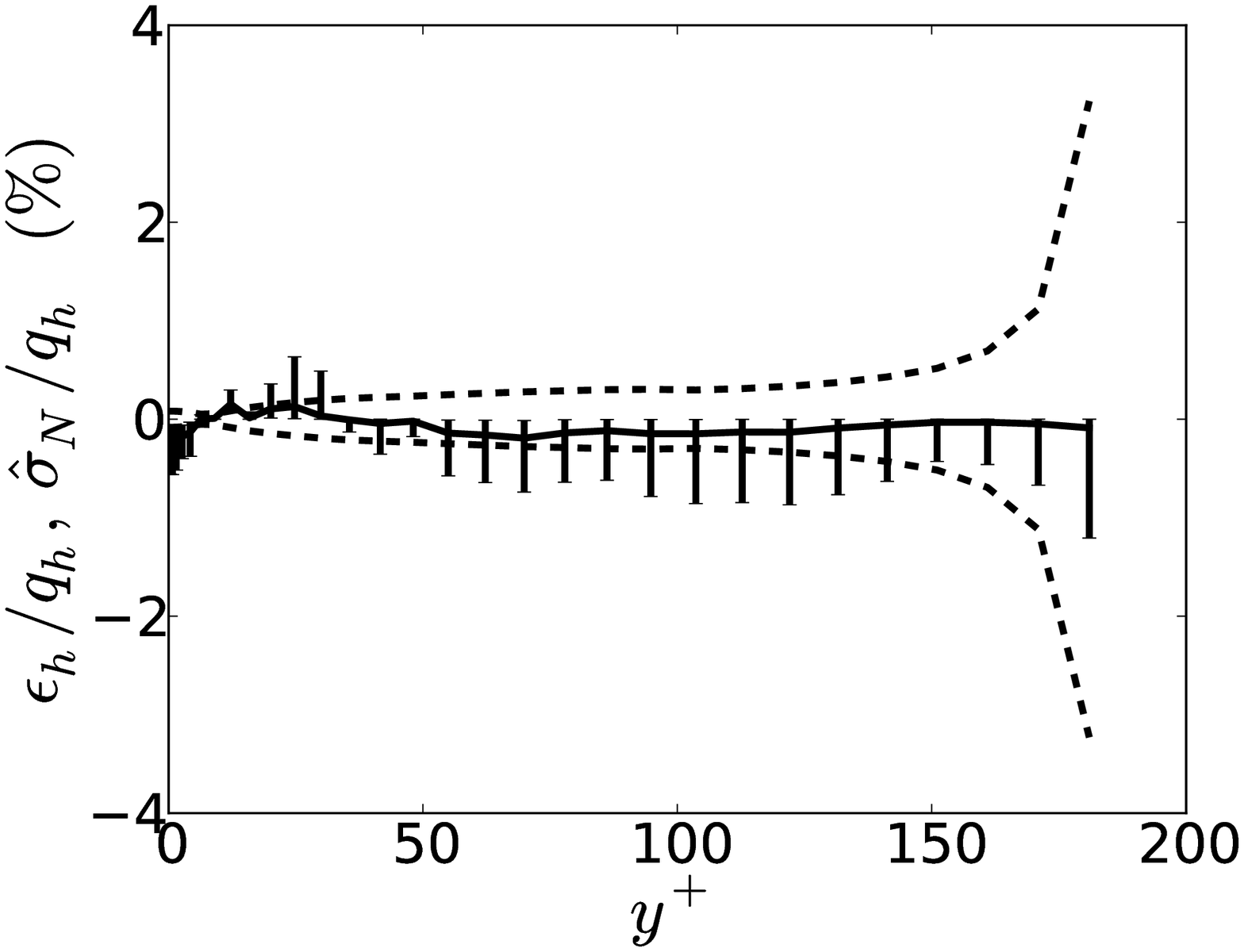}} \\
\end{tabular}
 \caption{Estimated discretization error (solid) and sampling
uncertainty (dashed) and their $90\%$ credibility intervals.}
\label{fig:mean_disc}
\end{figure}

\begin{figure}
\centering
\begin{tabular}{ccc}
\subfloat[$\avg{u'v'}$]{\includegraphics[width=0.32\linewidth]{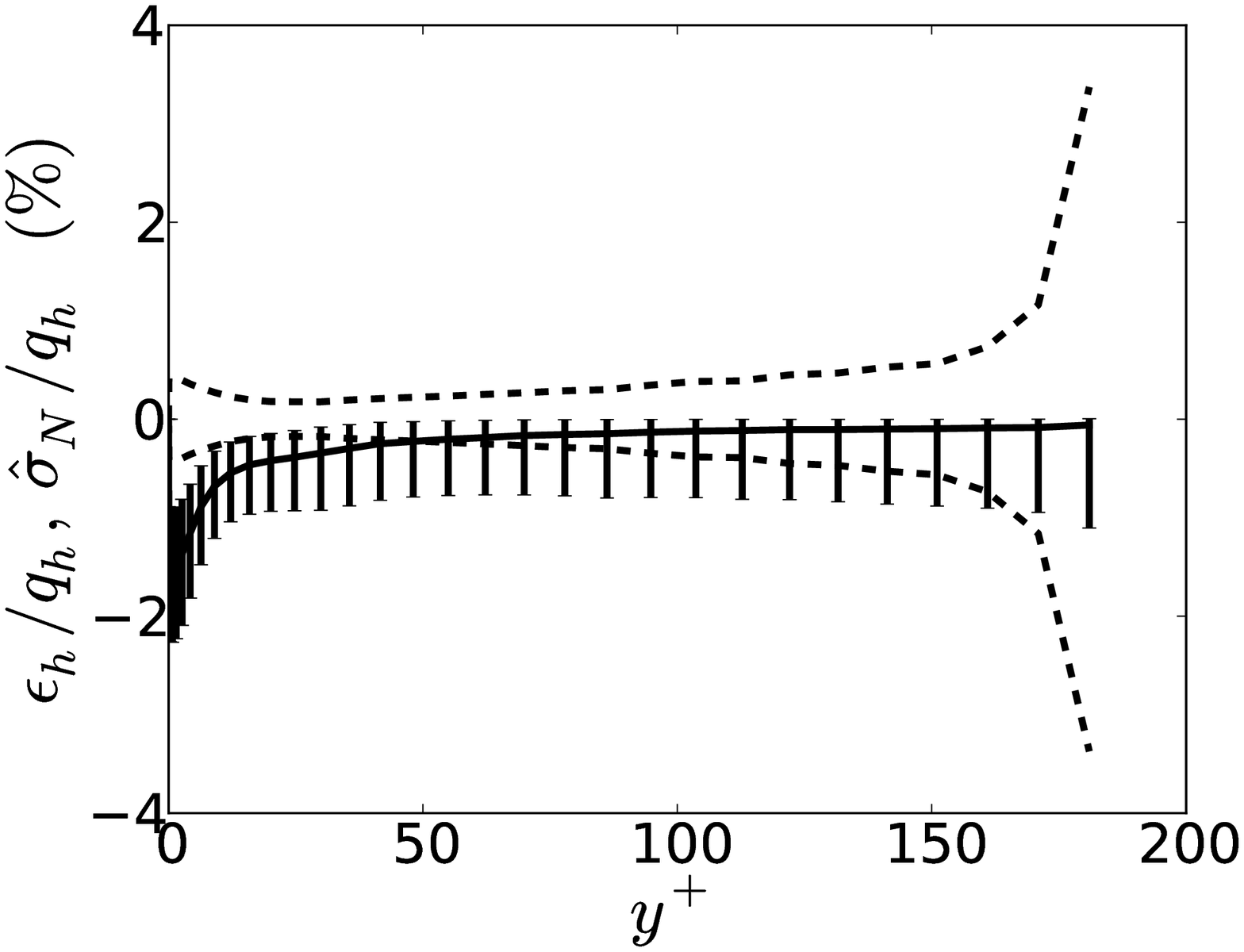}} &
\subfloat[$\avg{v'v'}$]{\includegraphics[width=0.32\linewidth]{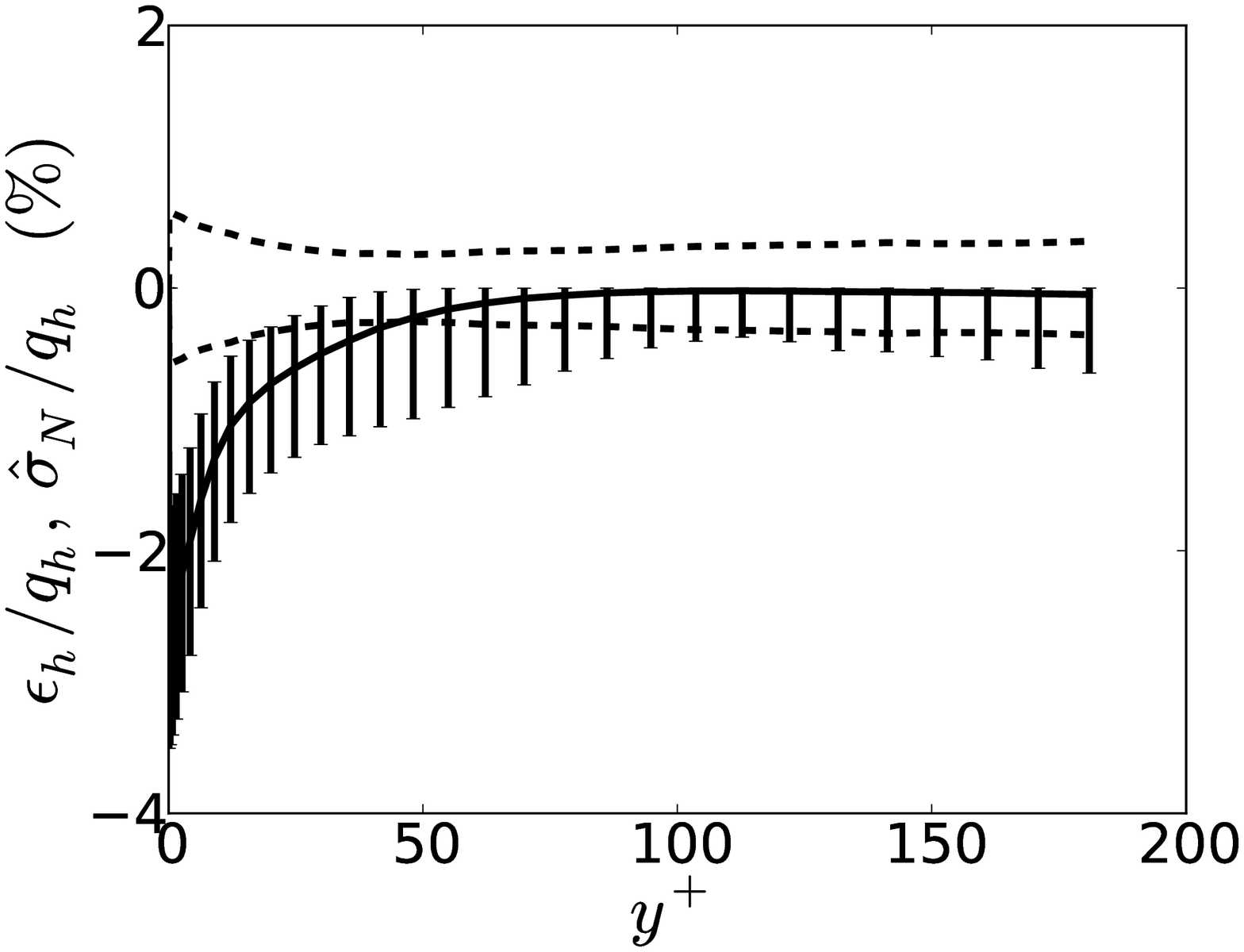}} &
\subfloat[$\avg{w'w'}$]{\includegraphics[width=0.32\linewidth]{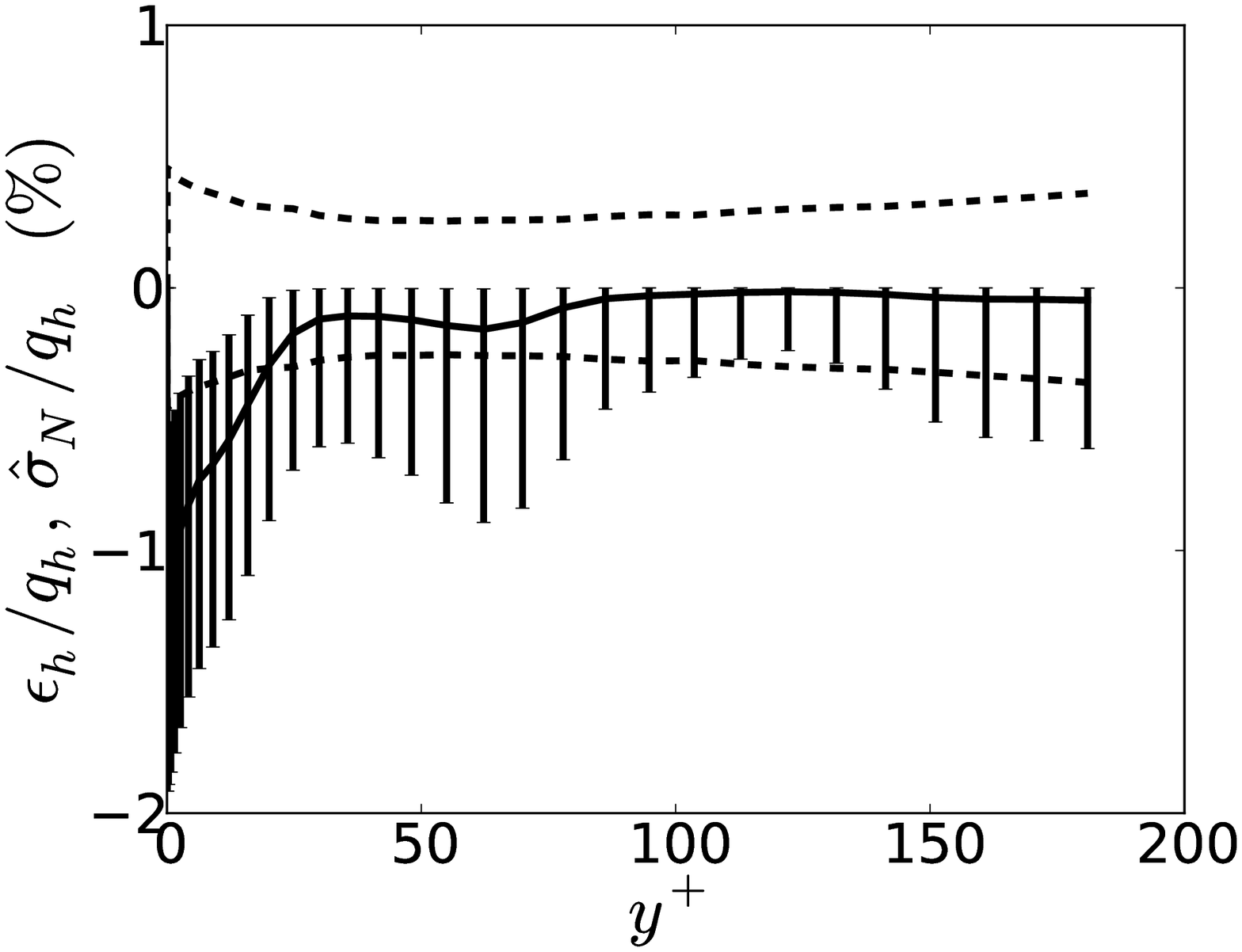}} \\
\end{tabular}
 \caption{Estimated discretization error (solid) and sampling
uncertainty (dashed) and their $90\%$ credibility intervals.}
\label{fig:reynolds_disc}
\end{figure}

%
% discretization error
%
For the quantities where the model is not invalidated, we use it to
predict the discretization error for the nominal mesh result.
Figures~\ref{fig:mean_disc} and~\ref{fig:reynolds_disc} show these
predictions.  The median prediction is shown as a solid line with
error bars indicating the $90\%$ credibility interval.  For
comparison, the estimated sampling error is indicated by the dashed
lines, which correspond to the $90\%$ credibility interval of the
sampling error model.  For all quantities, the errors are presented as
percentage values.

For the mean velocity and viscous shear stress, both the estimated
discretization error and sampling errors are less than 1\% in
magnitude everywhere across the channel.  In fact, for most points,
the median error in the mean velocity is less than one quarter of a
percent, with nearly all the 90\% confidence intervals at less than
one half of a percent.

Very near the wall, the discretization errors are estimated to be
larger than the sampling error. For $y^+ \gtrsim 15$, the median
of the discretization error lies within the $90\%$ credibility
interval for the sampling error, but generally there is some
probability that the discretization error is larger.  On the whole, it
appears that neither error is dominant.

Qualitatively similar conclusions can be drawn for $\avg{u'v'}$,
$\avg{v'v'}$, and $\avg{w'w'}$, but the errors are somewhat larger.
The median discretization error is less than $2\%$ everywhere and is
largest near the wall.  Near the wall, the discretization error is
larger than sampling error.  Near the center of the channel, the
situation is reversed.

\subsection{Large Domain Results}
Turbulent channel flow simulations at $Re_\tau\approx 180$ have been
performed many times. Currently, one of the most useful simulations is
that of Hoyas \& Jim\'{e}nez\citep{hoyas}, because of its large spatial
domain, because statistical data is easily accessible online, and
because it is part of a series of simulations with Reynolds numbers
ranging over an order of magnitude. The large domain simulations
reported in this subsection were performed in the same domain size
($L_x=12\pi$, $L_z=4\pi$) as Hoyas \& Jim\'{e}nez so that a direct
comparison can be made to those results, and so that the uncertainty
estimates developed here will be indicative of the uncertainties in this
commonly referenced work.

Unlike the smaller box case, only three meshes were used, so it is not
possible to test the validity of the calibrated discretization error
model against a higher resolution result.  However, since the the
Reynolds number and mesh resolution are the same or similar to the
small domain case, we expect that the model is valid for the same
quantities.  Further, consistent with typical DNS practice, statistics
were gathered over only a modest simulation time (10s of
flow-throughs), although each flow-through with the large box
represents significantly more data than the small box case.  Full
details of the simulation are given in Table~\ref{tbl:channel_runs}.

\subsubsection{Centerline Mean Velocity and Skin Friction} \label{sec:big_box_cl_results}
As in sections \ref{sec:small_box_cl_results} and
\ref{sec:small_box_cf_results}, we present detailed results for the centerline mean
velocity and the skin friction.  The posterior PDFs for the
calibration parameters are qualitatively
similar to the results for the small domain
(see Figures~\ref{fig:cl_chp_joint_post}
and~\ref{fig:tau_chp_joint_post}), and are therefore not shown. The
posterior PDFs are marginally less well informed due to the somewhat
larger sampling uncertainty in the large domain results, but do not
differ materially.  For example, for the centerline velocity, the true
value shifts slightly to the left to a mean of approximately 1.162769,
and there is still significant uncertainty about the value of $p$.
The mean is 4.84, but the $5$th and $95$th percentiles lie at 3.19 and
7.10, respectively.  As in the small domain case, the order of
accuracy for the skin friction is somewhat better informed than that
for the centerline velocity, but there is still significant
uncertainty, with the $5$th and $95$th percentiles at 3.06 and 5.28,
respectively.  However, the marginal posterior for the true value of
$C_f$ is again quite narrow, with the difference between the $5$th and
$95$th percentiles being only $0.54\%$ of the mean value.

\begin{figure}[thp]
\begin{center}
\includegraphics[width=0.7\linewidth]{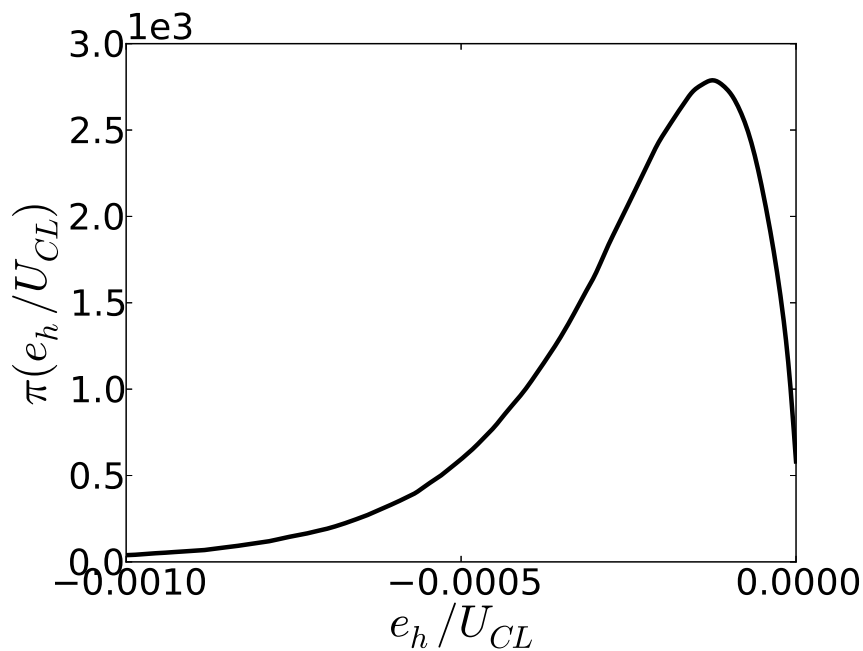}
\end{center}
\caption{
Discretization error, as computed by the calibrated model, for the
centerline mean velocity on the nominal mesh.  }
\label{fig:big_cl_chp_disc_error}
\end{figure}

Figure~\ref{fig:big_cl_chp_disc_error} shows the estimated
discretization error for the centerline velocity on the nominal mesh,
normalized by the observed mean value.
As in the small box case, it is almost certain that the
discretization error is less than $0.1\%$, and the mean is only
$0.011\%$.  The $50$th percentile lies at approximately $0.021\%$,
which is very close to the estimated standard deviation of the
sampling error.

\begin{figure}[thp]
\begin{center}
\includegraphics[width=0.7\linewidth]{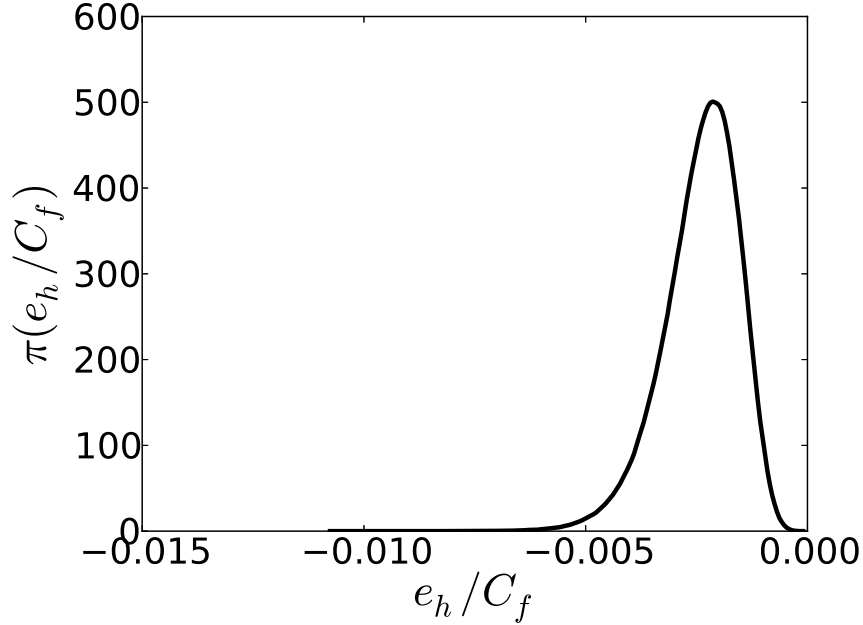}
\end{center}
\caption{
Discretization error, as computed by the calibrated model, for the
skin friction on the nominal mesh.  }
\label{fig:big_tau_chp_disc_error}
\end{figure}

The estimated discretization error in the skin friction on the nominal
mesh is shown in Figure~\ref{fig:big_tau_chp_disc_error}.
As in the small box case, the discretization error is small, with a
mean of approximately $0.24\%$.  For comparison, the estimated
standard deviation of the sampling error is approximately $0.092\%$.

% jimenez comparison here
% should truncate some of these digits, do we have a rule for this?
As mentioned earlier, this domain size is identical to that of a
previous $Re_\tau\approx180$ simulation reported by Hoyas \&
Jim\'{e}nez\citep{hoyas}. For the purposes of verification, a direct
comparison between that simulation and the nominal mesh of this study is
performed. The centerline velocity on the nominal mesh for our study is
% $1.163032943 \pm 0.000247770114486$
$1.16303 \pm 0.00024$ %7770$, % Space present for formatting
where the quoted uncertainty
estimate is one standard deviation of the sampling error. This is within
0.031\% of the value of 1.16267
%1.1626736365304358
quoted by Hoyas \& Jim\'{e}nez. While small, these values differ by more
than a standard deviation of the estimated sampling error. However, these two simulations, while run with
similar resolution, did differ in both their choice
of wall-normal numerics (B-splines vs. Chebychev polynomials) as well
as the number of points in $y$ (128 vs. 97). It is therefore plausible
that the discrepancy between the values of the centerline
velocity in these simulations is a combined result of discretization
error and sampling error. Indeed, the observed difference of 0.00036 is
in the range of plausible discretization errors $e_h$, as shown in
Figure~\ref{fig:big_cl_chp_disc_error}. Similarly, our value of the skin friction
coefficient from the nominal mesh ($0.00807834 \pm 7.49 \times 10^{-6}$) differs by
$\approx 0.4\%$ from the value quoted by Jim\'{e}nez ($0.00811666 \pm
3.4 \times 10^{-7}$). In absolute terms, this is a very small
difference, but it is significantly larger than the estimated 
sampling error.  Recalling the aforementioned differences between the present
wall-normal numerics and those of Hoyas \& Jim\'{e}nez, it is plausible that the
discrepancy is due to discretization
error. Indeed, the $\approx 0.4\%$ discrepancy is plausible as a value of
the discretization error as shown in Figure~\ref{fig:big_tau_chp_disc_error}.

% ---------------------------------------------
% juans response to my inquiry:
% 
% I did not prepare this data, so I am not really sure. Most probably, as
% we do from time to time, the ch180 has been run longer afterwards and
% updated in the website, but not in the main page were the main
% parameters of the simulations are listed (nu,utau,Lx,Lz).

% If I were you, I would infer any information from the actual profiles at
% *.prof
% ---------------------------------------------
% I'm going to take this as true, and use the information from *.prof (file)
%
% tw (ours) = .0040391798259 +- 3.74657*10-e6
%           = 
%
% tw(readme)=  .004043696
%
% tw(file)  =  .004058332     +- 1.732307e-7
%           = 0.008116664     +- 3.464614e-7
%
% >>> (.00807834- 0.008116664)/0.008116664
% -0.004721644261731301
% or, ~ .4% disparity
%
%
%
% Sergio's response: 
%
% Sorry for the delay, I've been out of the office. The correct value is
% the last one. We rerun some of the old files to solve a small bug which
% appeared very close to the wall and also extended the simulation. This
% changed a little bit the u_\tau. I didn't notice this small discrepancy,
% but you are right. I agree with Juan that the best idea is to use all
% the data coming from the .prof files.
%

\subsubsection{Summary of Results for Single-Point Statistics}

This section shows the estimated discretization and sampling errors
for the mean velocity $\avg{u}$, viscous shear stress $\nu
d\avg{u}/dy$, Reynolds shear stress $\avg{u'v'}$, wall-normal velocity
variance $\avg{v'v'}$, and spanwise velocity variance $\avg{w'w'}$.
Recall that the discretization error model for these quantities passed
the validation assessment for the small domain case, as shown
in \S\ref{sec:single-pt}.

%We present a summary of discretization error and sampling error results
%for the large domain size over several low-order statistical quantities
%typically of interest in DNS. Due to the lack of a mesh at significantly
%finer resolution, we are unable to perform a similar validation test. As
%a result, we utilize the small domain results presented in section
%\ref{sec:single-pt} and only consider quantities where the
%discretization error model was not invalidated in this case. These
%quantities were the mean velocity, the $vv$ and $ww$ velocity variances,
%and $uv$. These results were generated by performing the Bayesian
%Richardson Extrapolation procedure on the coarsest, coarse and nominal
%meshes.

\begin{figure}
\centering
\begin{tabular}{cc}
\subfloat[$\avg{u}$]{\includegraphics[width=0.49\linewidth]{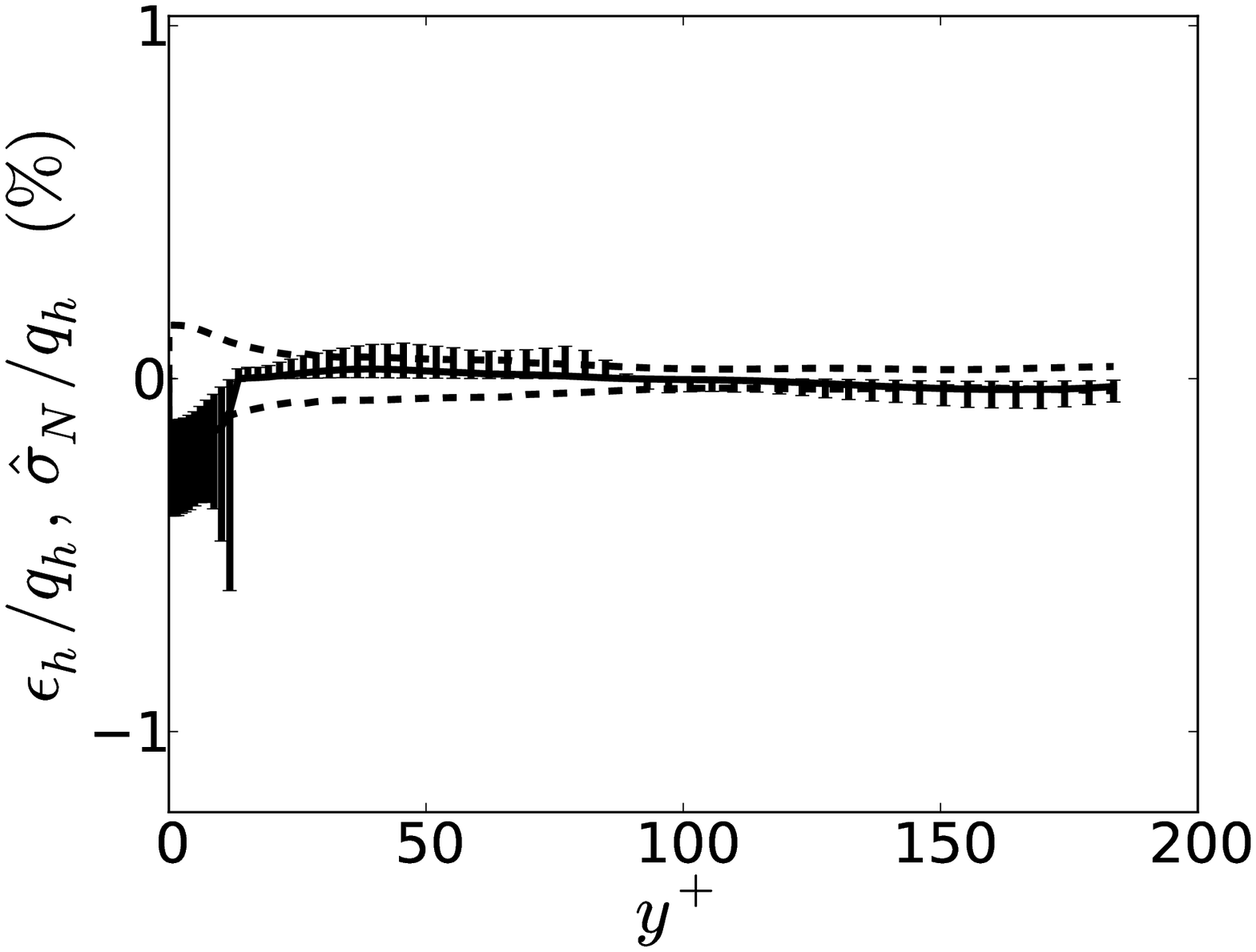}} &
\subfloat[$\nu d\avg{u}/dy$]{\includegraphics[width=0.49\linewidth]{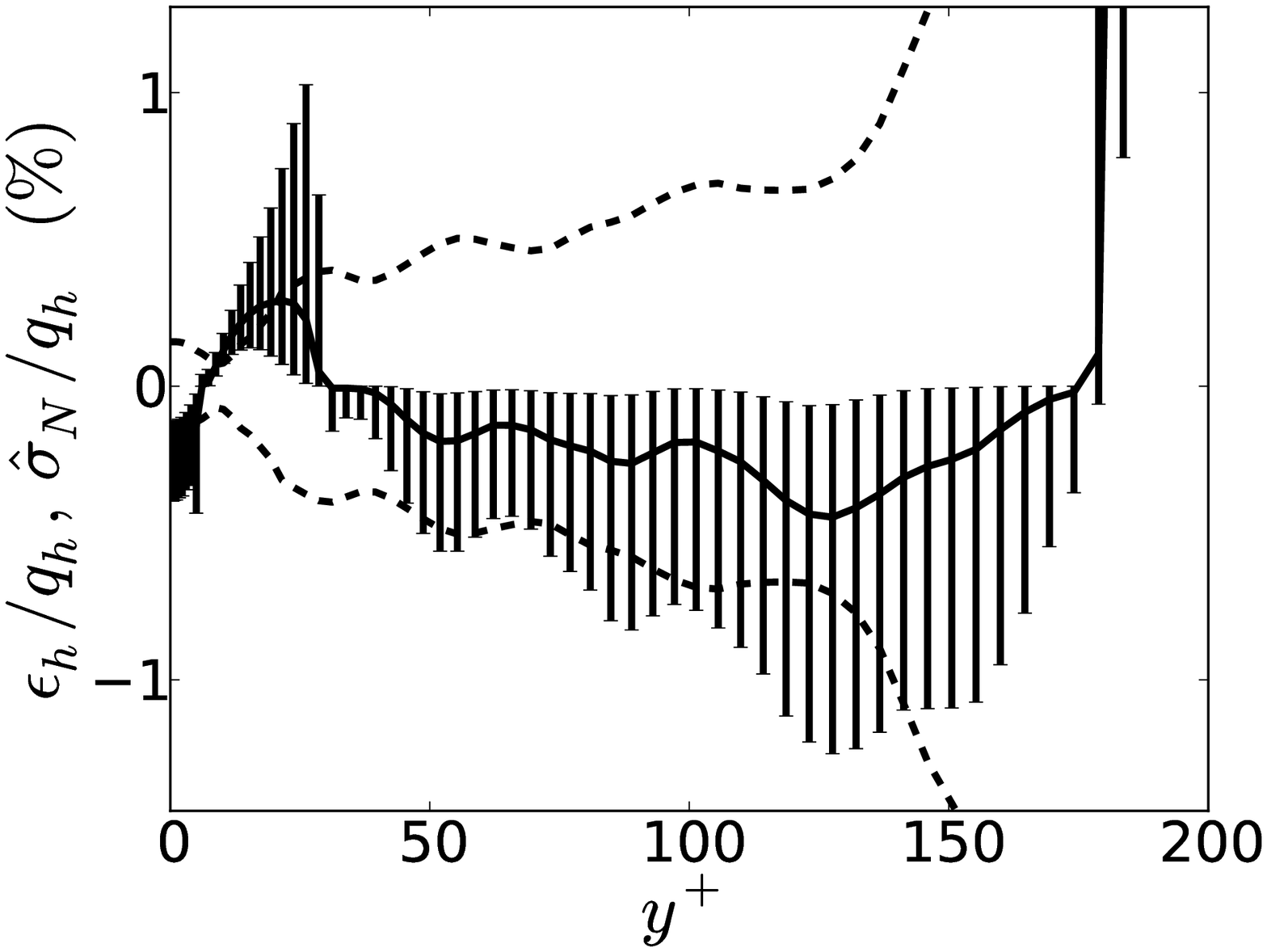}} \\
\end{tabular}
 \caption{Estimated discretization error (solid) and sampling
uncertainty (dashed) and their $90\%$ credibility intervals.}
\label{fig:bb_mean_disc}
\end{figure}

\begin{figure}
\centering
\begin{tabular}{ccc}
\subfloat[$\avg{u'v'}$]{\includegraphics[width=0.32\linewidth]{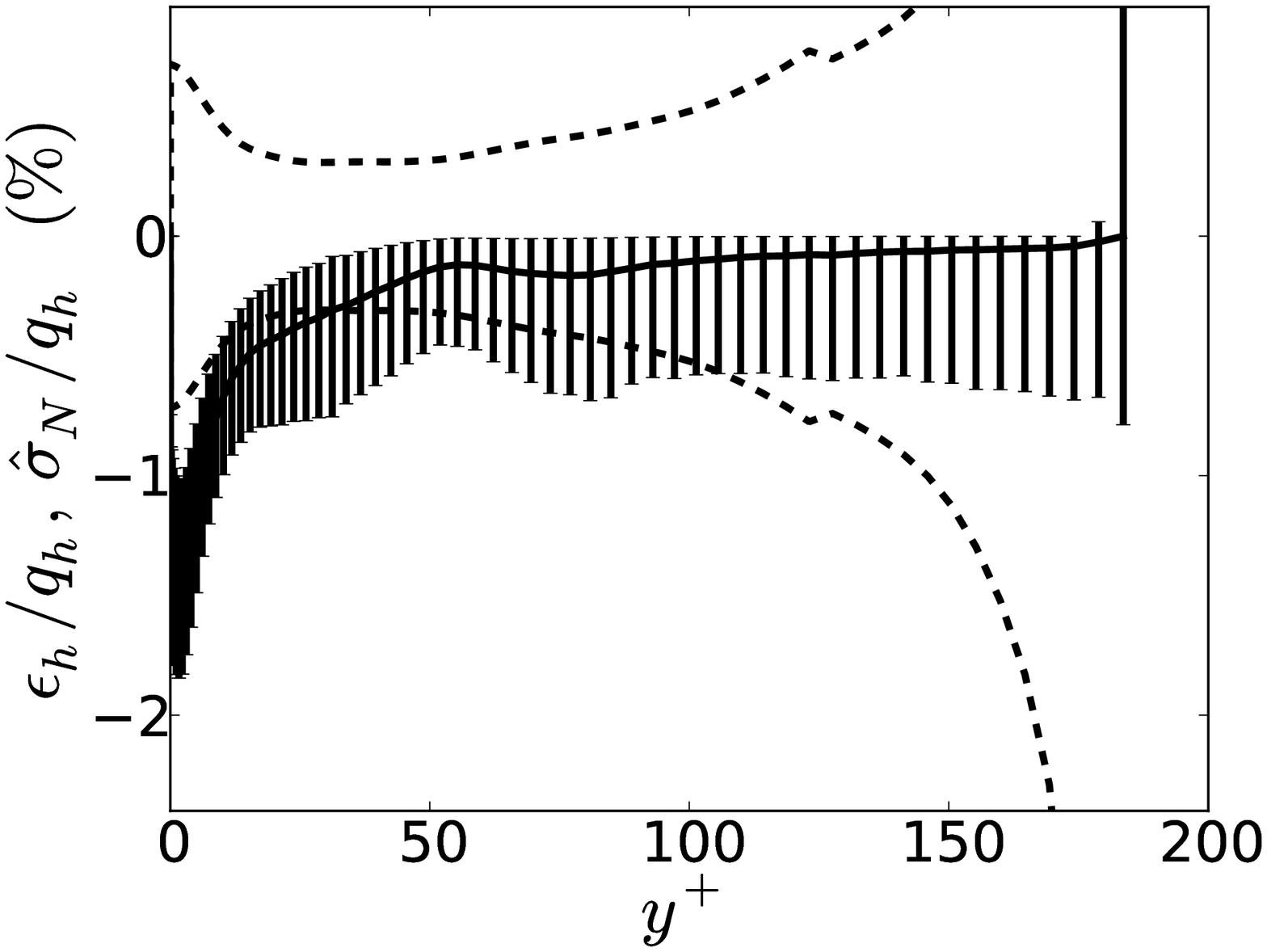}} &
\subfloat[$\avg{v'v'}$]{\includegraphics[width=0.32\linewidth]{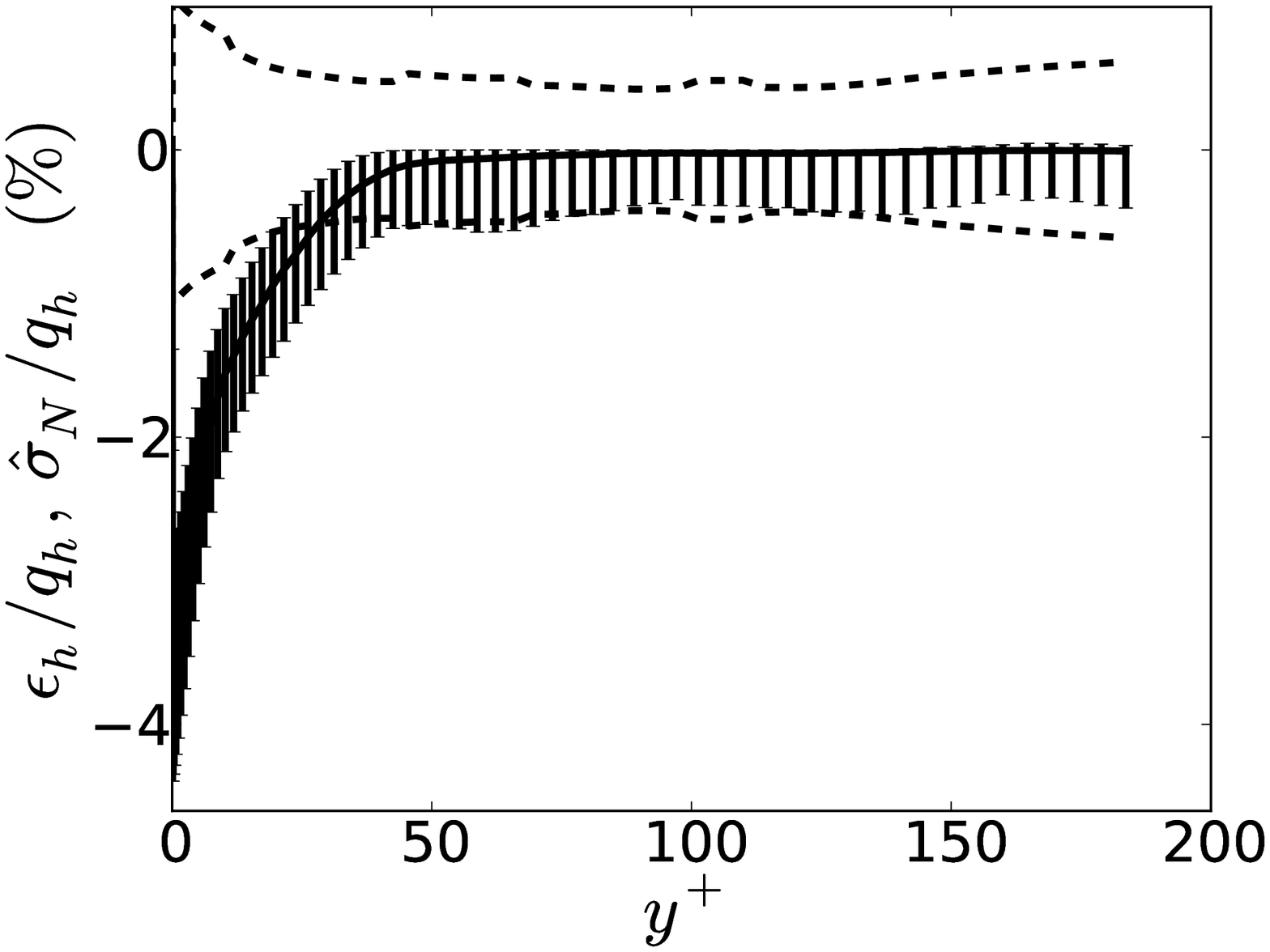}} &
\subfloat[$\avg{w'w'}$]{\includegraphics[width=0.32\linewidth]{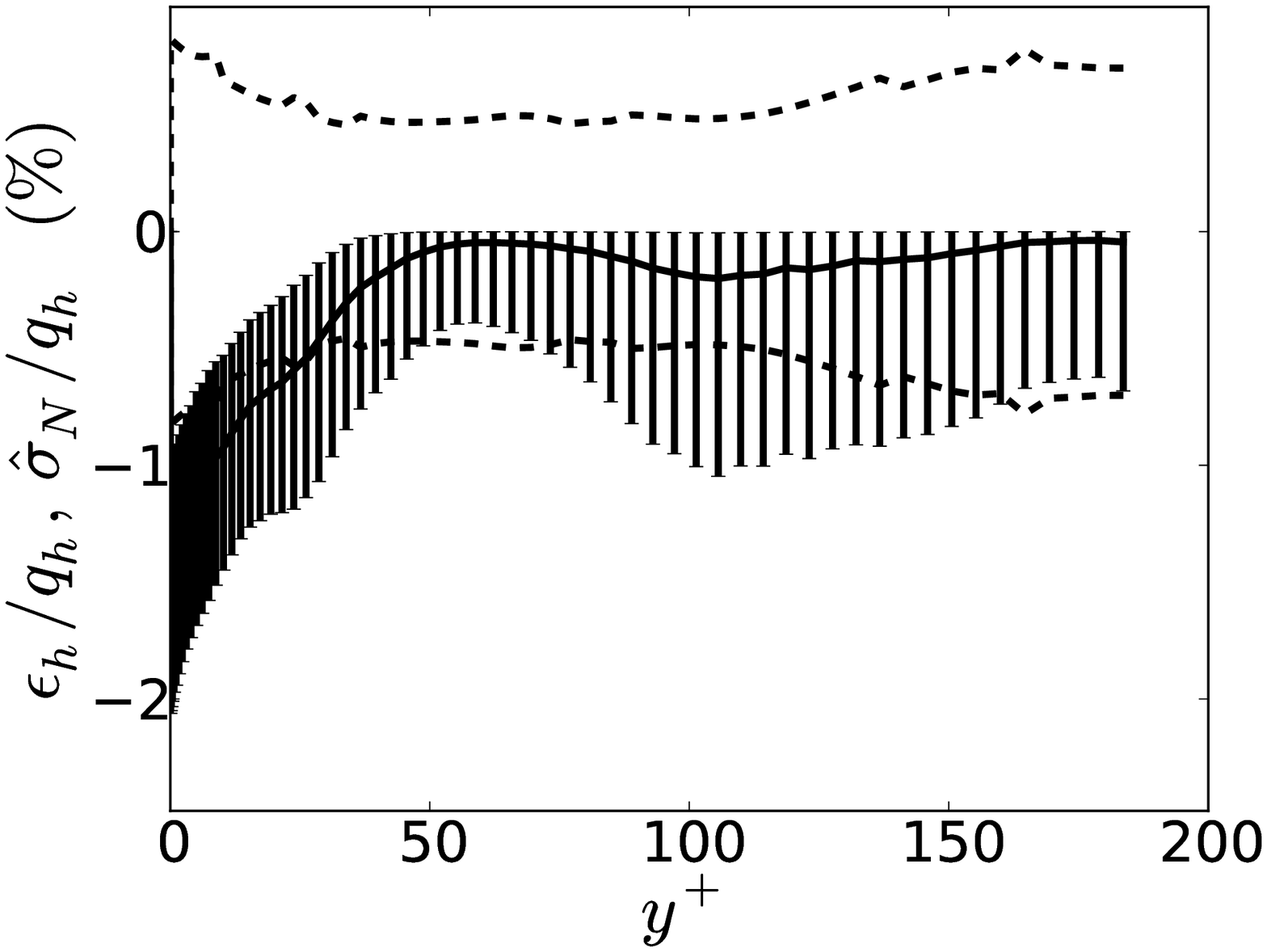}} \\
\end{tabular}
 \caption{Estimated discretization error (solid) and sampling
uncertainty (dashed) and their $90\%$ credibility intervals.}
\label{fig:bb_reynolds_disc}
\end{figure}

The results are shown in Figures~\ref{fig:bb_mean_disc}
and~\ref{fig:bb_reynolds_disc}, which are analogous to
Figures~\ref{fig:mean_disc} and~\ref{fig:reynolds_disc} for the small
domain results.
As in the small domain case, the estimated discretization errors in
the mean velocity and viscous stress are less than one percent nearly
everywhere.  The larger percentage errors in the viscous stress near
the centerline are an artifact of the local viscous stress going to
zero at the center of the channel.  The largest percent error observed
anywhere aside from the centerline is roughly four percent in
$\avg{v'v'}$ very near the wall.  Of course, $\avg{v'v'} \propto y^4$
as $y \rarrow 0$, meaning that this error is still very small.
Finally, in general, the discretization errors observed are largest
near the wall.  In this region, they tend to be larger than the
sampling error.  In the center of the channel, the sampling error is
generally larger.

\begin{figure}
\begin{center}
\subfloat[$\avg{u}$]{\includegraphics[width=0.49\linewidth]{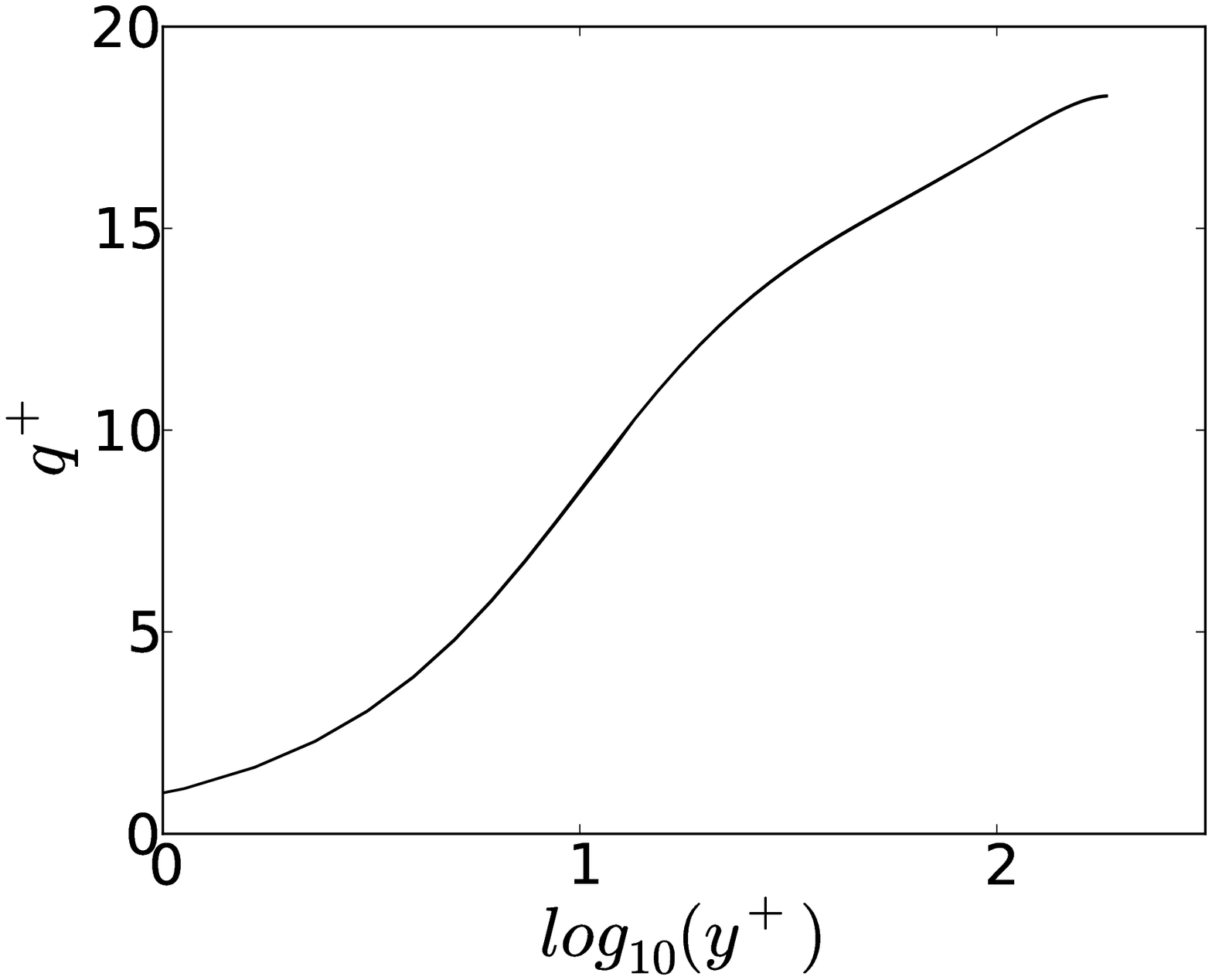}} \\
\subfloat[$\nu d\avg{u}/dy$]{\includegraphics[width=0.49\linewidth]{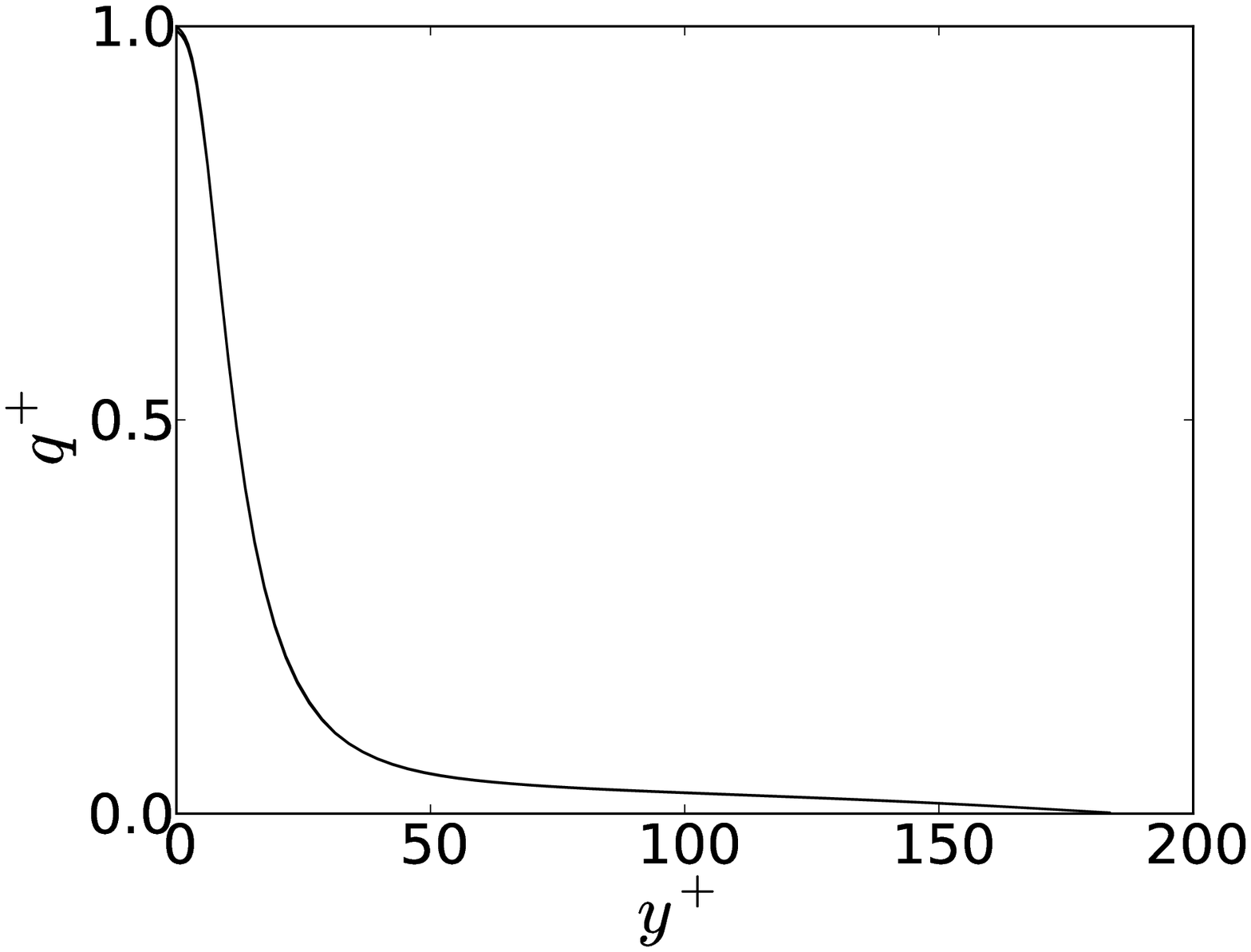}} 
\subfloat[$-\avg{u'v'}$]{\includegraphics[width=0.49\linewidth]{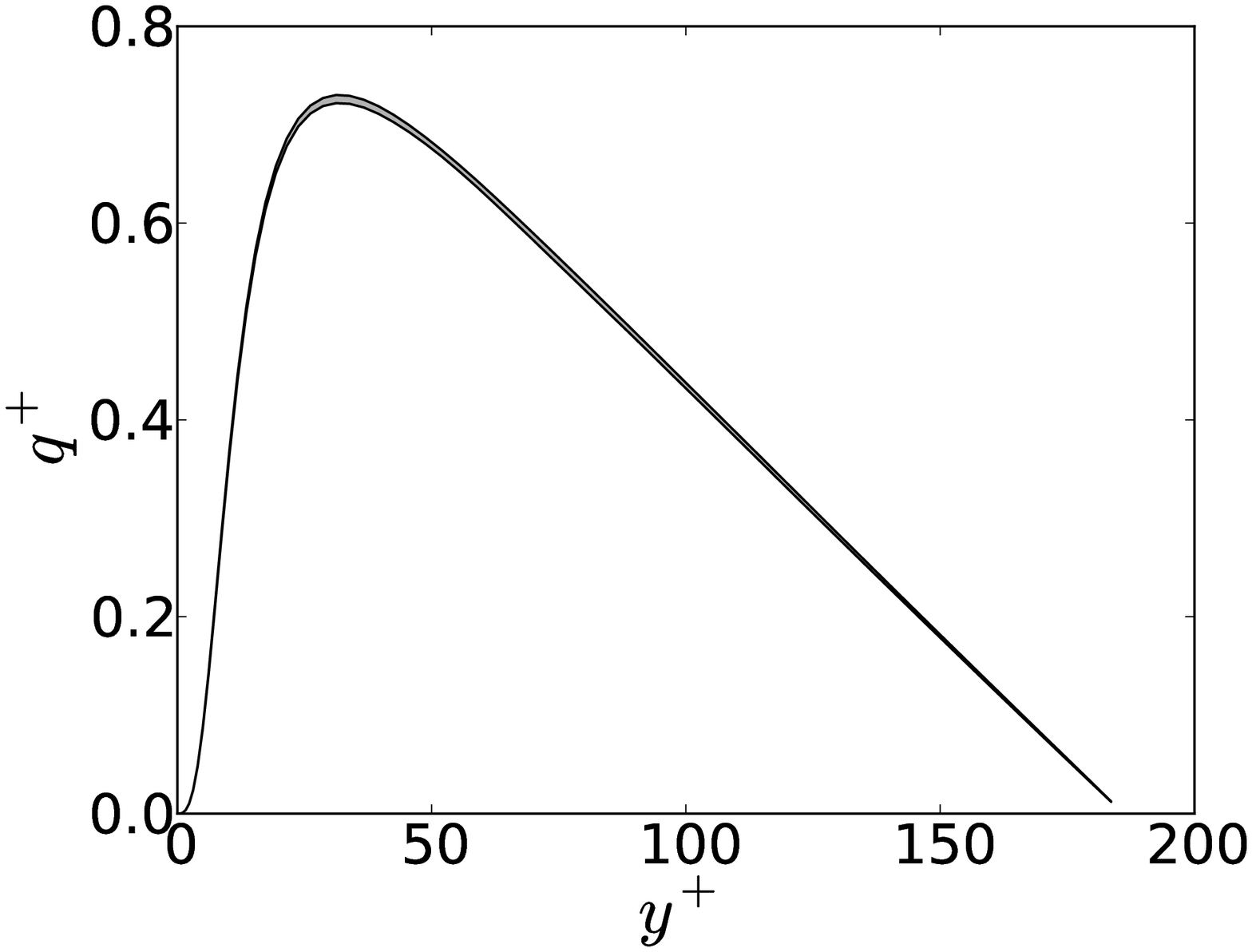}} \\
\subfloat[$\avg{v'v'}$]{\includegraphics[width=0.49\linewidth]{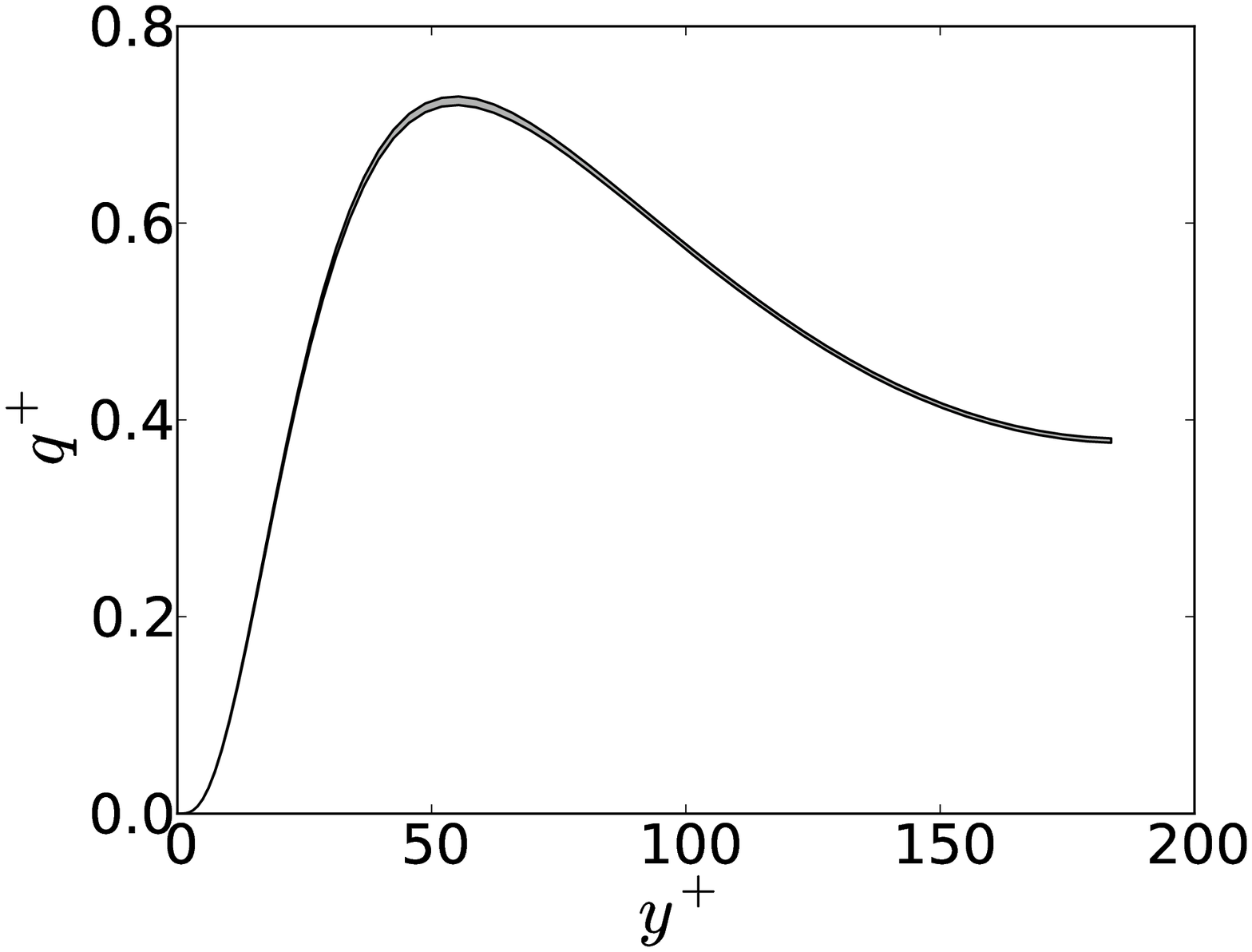}} 
\subfloat[$\avg{w'w'}$]{\includegraphics[width=0.49\linewidth]{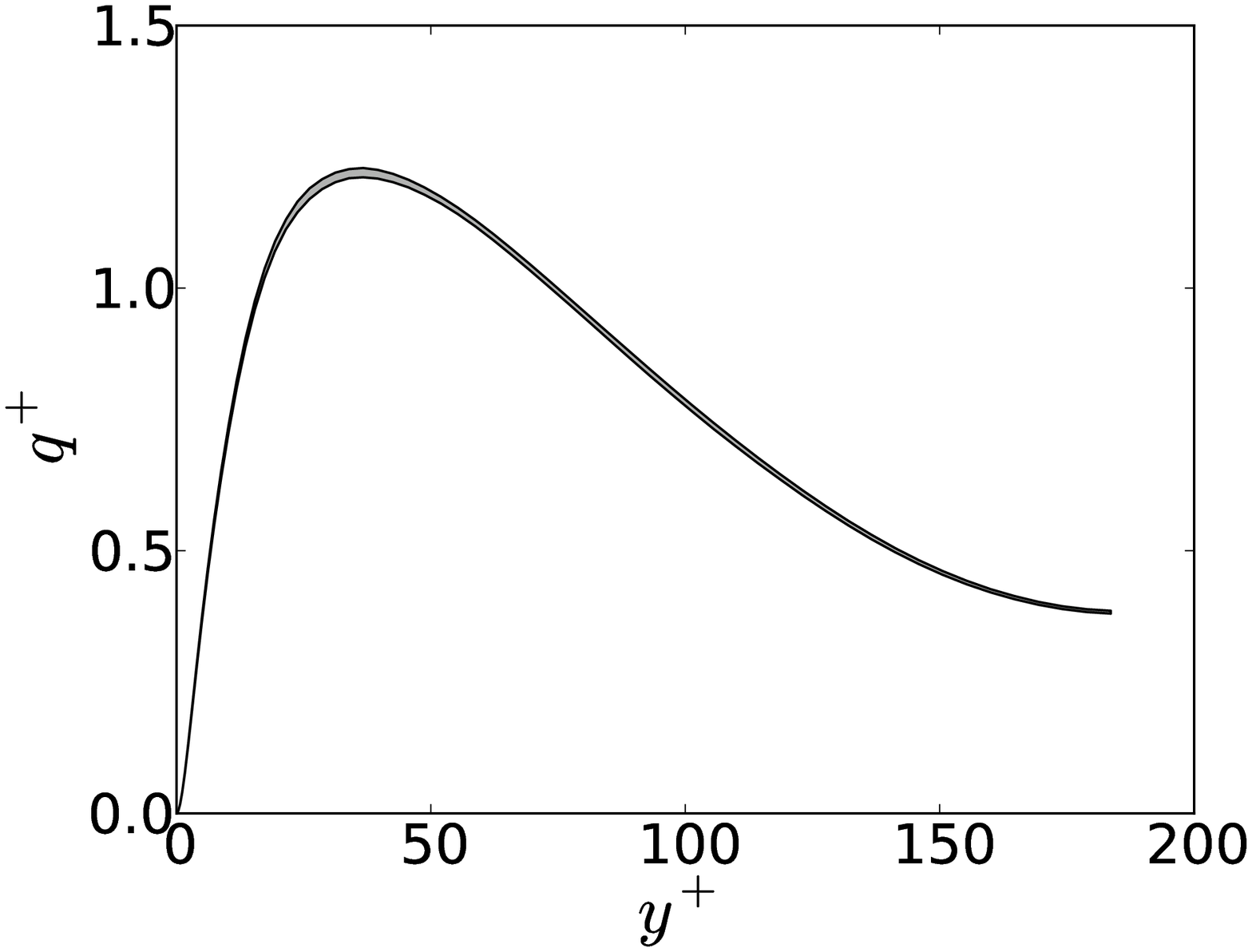}}
\end{center}
 \caption{Estimated true value obtained from the posterior 
 distribution for $q$.  The grey region shows the $90\%$ credibility
 interval between the $5$th and $95$th percentiles of the posterior.
 All quantities are normalized using the nominal value of $u_{\tau}$.}
\label{fig:bb_true_profiles}
\end{figure}

In addition to providing an assessment of the discretization error on
the nominal mesh, the Bayesian procedure provides an estimate of the
true value in the limit of infinite resolution ($h \rarrow 0$).  This
estimate is provided by the posterior distribution for $q$ that is
obtained from the Bayesian update that is performed to calibrate the
discretization error model.  These posterior estimates for the true
profiles are plotted in Figure~\ref{fig:bb_true_profiles} for the
quantities for which the discretization error model was found valid for
the small domain results.
%

%\todo{In figure: fix mean velocity plot so vertical axis starts at 0.}

All quantities are plotted using wall normalization, but, to avoid
introducing additional uncertainty due to the fact that $u_{\tau}$ is
an uncertain quantity, the nominal value for $u_{\tau}$ is used.  The
resulting intervals are small.  For mean velocity and viscous shear
stress, the uncertainty is small enough that the $90\%$ credibility
interval appears as just a thick line.  In the Reynolds stress and
wall-normal and spanwise variances, the effect of the uncertainty is
more visible, particularly near the peak values, but still quite
small.

\clearpage

\section{Conclusions} \label{sec:conclusions}
DNS data is crucial to advancing understanding of turbulent flow
physics and to calibration of engineering models of turbulent flow.
Given these uses, it is important to fully understand and characterize
the errors and uncertainties in computed statistical outputs.
%
%Rigorous assessments of DNS uncertainty are crucial to the utility of
%DNS results.  
However, because of complications due to sampling error, systematic
studies of discretization error are not standard for DNS.  In this
work, two enabling utilities have been developed and applied: a
sampling error estimator that accounts for correlation in the data
used to compute statistics and a Bayesian extension of Richardson
extrapolation that can be used to estimate discretization error in the
presence of uncertainty due to finite sampling.  These tools enable
systematic estimation of both sampling and discretization errors in
statistical quantities computed from simulations of chaotic systems.

The results for the Lorenz equations demonstrate that these tools
perform well in a simple, well-understood setting.  However, the
results for DNS of $Re_{\tau} = 180$ channel flow indicate that their
usage in a complex setting is more difficult.  One obvious
complication is that discretization errors resulting from practical
simulations may not be in the asymptotic regime.  The simple
discretization error representation used here was found to be adequate
for many important quantities, including mean velocity and Reynolds
shear stress.  Further, the estimated errors in these quantities are
small, indicating that the usual heuristics used to design meshes for
DNS of wall-bounded turbulence are reasonable.  Thus, we conclude that
simulations of channel flow based on these resolution heuristics with
similar sampling time, such as those reported by Jim\'{e}nez and
co-authors~\cite{DelAlamo2003Spectra, DelAlamo2004Scaling,
Hoyas2006Scaling, re2000}, can be expected to have errors of the same
magnitude as those reported here.

However, for other quantities, most notably the streamwise velocity
variance, the discretization error model is invalidated by comparison
against higher resolution simulations than those used to calibrate the
model.  Due to this failure, we are unable to quantify the
discretization error in these quantities with confidence. None-the-less,
the errors appear to be small because the observed change
from the nominal to finest resolution results is quite small.

It may be possible to solve this problem by posing a more complex
discretization error model, which could be based on retaining
additional terms in a Taylor series expansion of the discretization
error.  Future work should focus on investigating such models as well
as applying this technique to investigate resolution heuristics used
for other numerical schemes and classes of flow.  While it will likely
not be practical to apply the full Bayesian Richardson extrapolation technique for
each new simulation, by assessing the relevant resolution
heuristics in a computationally tractable setting, as done for low
$Re$ channel flow here, one can develop estimates of the expected numerical
accuracy of the results of more demanding simulations. This can then be
combined with sampling error estimates, which are tractable for even
expensive DNS, to obtain a complete characterization of DNS uncertainties.

\section*{Acknowledgments}
The work presented here was supported by the Department of Energy
[National Nuclear Security Administration] under Award Number
[DE-FC52-08NA28615]. 
% standard ALCF blurb
%This research used resources of the Argonne Leadership Computing
%Facility at Argonne National Laboratory, which is supported by the
%Office of Science of the U.S. Department of Energy under contract
%DE-AC02-06CH11357. 

The authors acknowledge the Texas Advanced Computing Center (TACC) at The 
University of Texas at Austin for providing HPC resources that have contributed 
to the research results reported here.
Finally, the authors wish to thank Mr. Myoungkyu Lee for
the use of his simulation code as well as his assistance in
generating several of the DNS runs.

\appendix

\section{Motivation for the Sampling Error Estimator}

\label{app:background_sample_error}
If the samples $X_i$ were independent, the classical central limit theorem (CLT)
states that
\begin{equation*}
e_N \overset{d}{\rarrow} \mcal{N}(0, \sigma^2/N),
\end{equation*}
%
%where $\mcal{N}(0, \sigma^2/N)$ denotes the Gaussian distribution with mean
%zero and variance $\sigma^2/N$ 
where $\sigma^2 = \variance X = E[(X-\mu)^2]$.
However, the samples resulting from a DNS calculation have \emph{a
priori} unknown correlation structure and, at least for small temporal
or spatial separation, are certainly not independent.
%Calculation of finite sampling errors in a statistically stationary
%DNS is complicated by the fact that the samples possess an \emph{a
%priori} unknown temporal autocorrelation structure.

To avoid the complications of correlated samples,
%As any well-defined numerical experiment must cause the
%autocorrelation to ultimately decay to zero,
many authors downsample
instantaneous measurements until the retained samples are arguably
uncorrelated and then use an estimate based on the classical CLT.  However,
optimally downsampling autocorrelated samples requires coarsening the
data ``just enough'' to decorrelate the signal but not ``too much'' to
avoid discarding useful data~\citep{Zwiers1995Taking}.  As increasing
the number of independent samples is computationally expensive in DNS,
it is imperative to extract all possible information from the data.
Thus, we seek a method to estimate the uncertainty in statistics
computed from correlated samples.

The method developed in \S\ref{sec:sampling_error} is motivated
by an extension of the CLT from a sequence of independent, identically
distributed random variables to an $\alpha$-mixing sequence.  For the
precise statement of the theorem see~\citet[Theorem 27.4]{Billingsley}
or~\citet[Theorem 3.2.1]{Zhengyan}.  Loosely speaking, the theorem
states that, if random variables ``far'' apart in the sequence are
nearly independent, which is expected for data resulting from DNS,
then as $N \rarrow \infty$,
\begin{equation*}
e_N \overset{d}{\rarrow} \mcal{N}(0, s^2/N),
\end{equation*}
where
\begin{equation*}
s^2 \equiv E[(X_0 - \mu)^2] + 2 \sum_{k=1}^{\infty} E[(X_0 - \mu)(X_k - \mu)].
\end{equation*}
Thus,
\begin{equation*}
\variance e_N \rarrow \frac{s^2}{N} = \frac{\sigma^2}{N} \left( 1 + 2 \sum_{k=1}^{\infty} \rho(k) \right),
\end{equation*}
where
\begin{equation*}
\rho(k) = \frac{E[(X_0 - \mu)(X_{k} - \mu)]}{E[(X_0 - \mu)^2]},
\end{equation*}
is the autocorrelation at separation $k$.  Thus, the extension for
weak dependence simply modifies the effective number of samples from
the classical CLT.  That is,
\begin{equation*}
\variance e_N \rarrow \frac{\sigma^2}{N_\text{eff}},
\end{equation*}
where
\begin{equation*}
N_\text{eff} = \frac{N}{1 + 2 \sum_{k=1}^{\infty} \rho(k)}.
\end{equation*}
%

%\section{Data}
%\input{appendix_data}
\clearpage % force all floats to be place before references

\bibliography{references}

\end{document}